\documentclass[twocolumn]{aastex}

\usepackage{graphicx}
\usepackage{enumitem}
\setitemize{noitemsep,topsep=0pt,parsep=0pt,partopsep=0pt}
\usepackage{booktabs}
\usepackage{amsmath}
\usepackage{hyperref}

\shortauthors{Wu et al}

\begin{document}

\title{ExoEcho: Exoplanet Ephemerides Change Observations. III. A JWST Survey for Transit Timing Variations of 37 Exoplanets}

\correspondingauthor{Bo Ma}
\email{mabo8@mail.sysu.edu.cn}

\author{Yunke Wu}
\affil{School of Physics and Astronomy, Sun Yat-sen University, Zhuhai 519082, China; {\it mabo8@mail.sysu.edu.cn}}
\affil{CSST Science Center for the Guangdong-Hong Kong-Macau Great Bay Area, Sun Yat-sen University, Zhuhai 519082, China}

\author{Zixin Zhang}
\affil{School of Physics and Astronomy, Sun Yat-sen University, Zhuhai 519082, China; {\it mabo8@mail.sysu.edu.cn}}
\affil{CSST Science Center for the Guangdong-Hong Kong-Macau Great Bay Area, Sun Yat-sen University, Zhuhai 519082, China}

\author{Xinyue Ma}
\affil{School of Physics and Astronomy, Sun Yat-sen University, Zhuhai 519082, China; {\it mabo8@mail.sysu.edu.cn}}
\affil{CSST Science Center for the Guangdong-Hong Kong-Macau Great Bay Area, Sun Yat-sen University, Zhuhai 519082, China}

\author{Wenqin Wang}
\affil{School of Physics and Astronomy, Sun Yat-sen University, Zhuhai 519082, China; {\it mabo8@mail.sysu.edu.cn}}
\affil{CSST Science Center for the Guangdong-Hong Kong-Macau Great Bay Area, Sun Yat-sen University, Zhuhai 519082, China}

\author{Siqi Guo}
\affil{School of Physics and Astronomy, Sun Yat-sen University, Zhuhai 519082, China; {\it mabo8@mail.sysu.edu.cn}}
\affil{CSST Science Center for the Guangdong-Hong Kong-Macau Great Bay Area, Sun Yat-sen University, Zhuhai 519082, China}

\author{Cong Yu}
\affil{School of Physics and Astronomy, Sun Yat-sen University, Zhuhai 519082, China; {\it mabo8@mail.sysu.edu.cn}}
\affil{CSST Science Center for the Guangdong-Hong Kong-Macau Great Bay Area, Sun Yat-sen University, Zhuhai 519082, China}

\author{Shangfei Liu}
\affil{School of Physics and Astronomy, Sun Yat-sen University, Zhuhai 519082, China; {\it mabo8@mail.sysu.edu.cn}}
\affil{CSST Science Center for the Guangdong-Hong Kong-Macau Great Bay Area, Sun Yat-sen University, Zhuhai 519082, China}

\author{Dichang Chen}
\affil{School of Physics and Astronomy, Sun Yat-sen University, Zhuhai 519082, China; {\it mabo8@mail.sysu.edu.cn}}
\affil{CSST Science Center for the Guangdong-Hong Kong-Macau Great Bay Area, Sun Yat-sen University, Zhuhai 519082, China}

\author{Bo Ma}
\affil{School of Physics and Astronomy, Sun Yat-sen University, Zhuhai 519082, China; {\it mabo8@mail.sysu.edu.cn}}
\affil{CSST Science Center for the Guangdong-Hong Kong-Macau Great Bay Area, Sun Yat-sen University, Zhuhai 519082, China}

\nocollaboration

\begin{abstract}

Transit timing variations (TTVs) provide a powerful probe of planetary dynamics and orbital evolution. We extend the ExoEcho project by incorporating \textit{James Webb Space Telescope} (JWST) transit observations and analyze 84 light curves of 37 exoplanets, including 20 hot Jupiters, 11 Neptune-like planets, and 6 super-Earths. We derive updated transit ephemerides and conduct a systematic search for long-term and periodic TTV signals.
We identify two strong TTV candidates, including HATS-6~b and HAT-P-18~b. The hot Jupiter HATS-6~b exhibits a long-term trend with $\dot{P} = 230.5 \pm 46.8\ \mathrm{ms\ yr^{-1}}$, potentially driven by tidal interactions with its host star. 
In addition, the Neptune-like planet HAT-P-18~b shows periodic TTV signal with amplitude of $92.9 \pm 13.7$~s. 
We also show that the previously claimed positive period derivative in WASP-121~b is inconsistent with our new transit timing data.
Although continued monitoring is required to confirm these findings, our results demonstrate the potential of JWST for high-precision transit timing and extend TTV studies beyond the traditional hot-Jupiter population to lower-mass exoplanets.

\end{abstract}


\section{Introduction}
Since the discovery of the first exoplanet orbiting a Sun-like star, the number of confirmed exoplanets has grown to over 6,000\footnote{\href{https://exoplanetarchive.ipac.caltech.edu}{https://exoplanetarchive.ipac.caltech.edu}}. Early discoveries were heavily biased toward hot Jupiters: gas giant planets characterized by large radii and short orbital periods. These characteristics made them the most detectable targets for the transit method with low-precision instruments, establishing them as primary subjects for initial atmospheric characterization studies. The commissioning of the James Webb Space Telescope (JWST) has ushered in a new era for exoplanet science. With its superior sensitivity and photometric stability, JWST enables detailed studies of more diverse planet types, such as Neptune-like exoplanets and super-Earths \citep{kempton_reflective_2023,beatty_sulfur_2024,dyrek_so2_2024,powell_sulfur_2024}. The growing archive of JWST transit observations provides high-precision data that are ideal for updating planetary ephemerides and significantly extending the baseline for transit timing variation (TTV) analysis.

Transit timing variations (TTVs), defined as deviations from a strictly linear ephemeris, offer a powerful tool for probing the dynamical properties of exoplanet systems. These variations can be categorized by their timescales and physical origins. Long-term TTVs can arise from several mechanisms, including orbital decay due to tidal dissipation \citep{yee_orbit_2019, turner_decaying_2021, alvarado-montes_orbital_2025, basturk_orbit_2025}, planetary mass loss \citep{valsecchi_tidally_2015, jackson_tidal_2016}, apsidal precession \citep{jordan_observability_2008, ragozzine_probing_2009}, Rømer effect \citep{yee_orbit_2019,bouma_wasp-4_2020,harre_apparent_2023,winn_orbital_2025} and the Applegate mechanism \citep{applegate_mechanism_1992, watson_orbital_2010}. Short-term, periodic TTVs, on the other hand, are often indicative of gravitational perturbations from additional, unseen bodies in the system, such as non-transiting exoplanets, exo-trojans, and exo-moons \citep{agol_detecting_2005, holman_use_2005, zhang_constraining_2024, vitkova_toi-4504_2025, nabbie_high_2025, Zhangzi2026, Cui2026}. The high-precision data from JWST are particularly valuable for extending TTV studies to a wider range of planetary types.

This study is part of the Exoplanet Ephemerides CHange Observations (ExoEcho) project \citep{wu_evidence_2026, wang_long-term_2024, wang_exoplanet_2025, ma_exoplanet_2025}, which aims to investigate the photodynamics of exoplanets using high-precision transit timing from ground- and space-based telescopes. Here, we present a uniform analysis of JWST transit observations for a sample of 37 exoplanets. This sample, comprising 20 hot Jupiters, 11 Neptune-like planets, and 6 super-Earths, significantly expands the diversity of planets subjected to high-precision TTV analysis, allowing us to probe dynamical phenomena across different planetary architectures and physical conditions. We extract light curves, measure mid-transit times, and perform a systematic TTV analysis for each target, obtaining 84 high-precision transit times from the JWST archive. We identify targets exhibiting significant TTVs and assess their robustness to select the most prominent candidates.

The paper is structured as follows: Section \ref{sec:light curve analysis} describes our target selection, the JWST data used, and our procedures for extracting light curves and measuring transit times. Section \ref{sec:transit timing variation analysis} presents the results of the TTV analysis for each target. In Section \ref{sec:discussion}, we discuss the prominent targets with possible TTVs. Finally, Section \ref{sec:conclusion} summarizes our conclusions.

\section{JWST Light Curve Analysis \label{sec:light curve analysis}}

\subsection{Data Sample}
\begin{figure}[h!]
    \centering
    \includegraphics[width=0.45\textwidth]{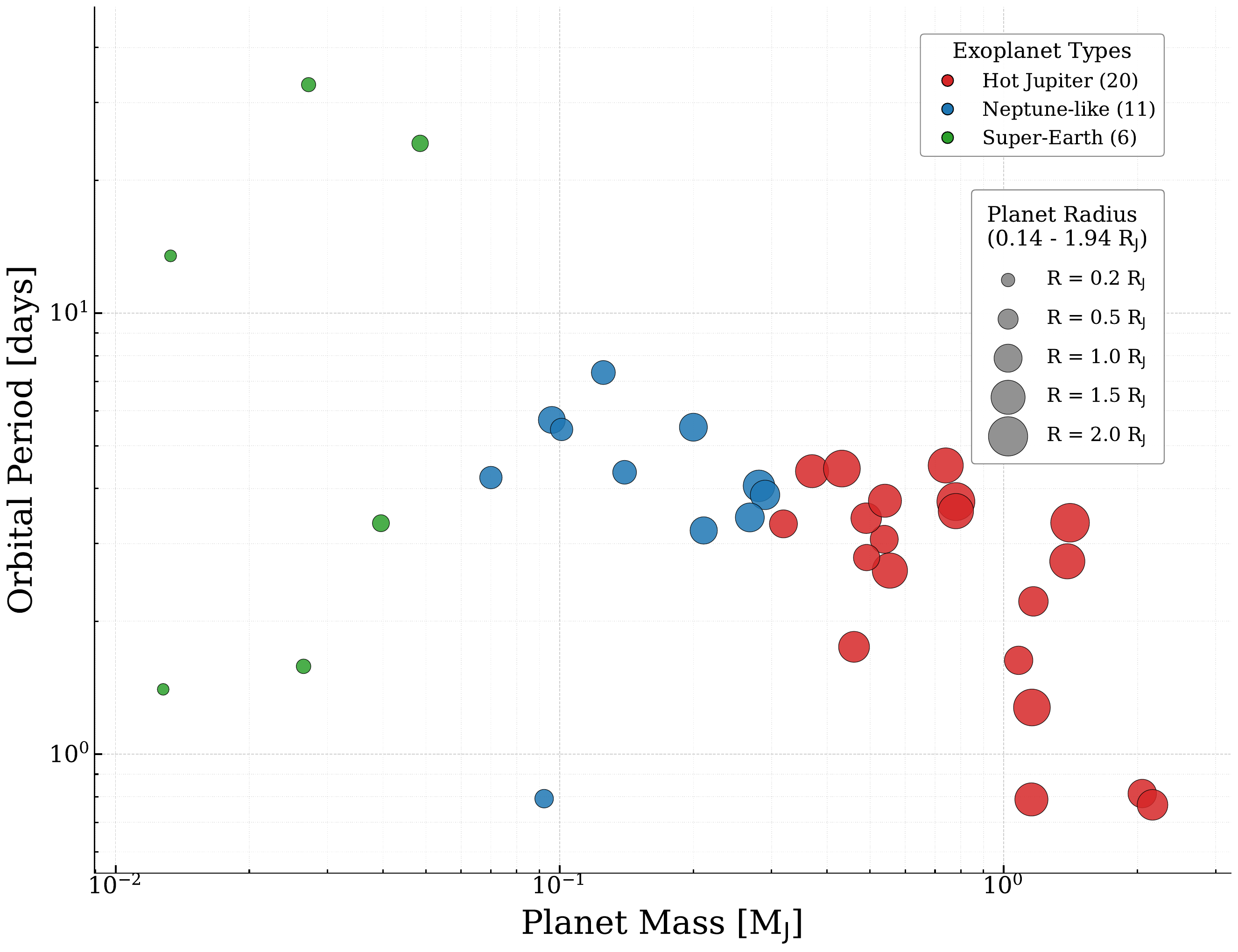}
    \caption{Exoplanet Sample Distribution: Mass versus orbital period for the 37 exoplanets in our sample. Point sizes are proportional to planetary radius. Colors indicate different exoplanet types: red for Hot Jupiters (20), blue for Neptune-like planets (11), and green for Super-Earths (6).}
    \label{fig:exoplanet ditribution}
\end{figure}

Our sample comprises 37 exoplanets, including 20 hot Jupiters, 11 Neptune-like planets, and 6 super-Earths. The corresponding JWST observing programs are summarized in Table~\ref{tab:data proposal}, while the distribution of planetary properties is shown in Figure~\ref{fig:exoplanet ditribution}. In total, we have analyzed 84 JWST transit observations obtained between 2022 June and 2025 June. All data were downloaded from the MAST archive\footnote{\href{https://mast.stsci.edu}{https://mast.stsci.edu}}.

\begin{deluxetable}{ccc}[!h]
\tablecaption{Observation Proposal Information Used in This Work \label{tab:data proposal}}
\tabletypesize{\scriptsize}
\tablewidth{0pt}
\tablehead{
\colhead{Planet} & \colhead{IDs\tablenotemark{a}} & \colhead{Transits\tablenotemark{b}}
}
\startdata
GJ 1214~b   & 1185o18,1185o19,1803o1        & 3 \\
GJ 3470~b   & 1185o16,1185o17            & 2 \\
HAT-P-12~b   & 1281o2,1281o1,1281o3,6219o1      & 4 \\
HAT-P-18~b   & 2734o1                & 1 \\
HAT-P-26~b   & 1177o6,1312o3,1312o1         & 3 \\
HAT-P-65~b   & 3969o4,3969o6             & 2 \\
HATS-6~b    & 3171o6,3171o7             & 2 \\
HATS-72~b   & 5924o10                & 1 \\
HATS-75~b   & 3171o9,3171o10,3171o19        & 3 \\
HD 189733~b  & 1633o5,1633o4,2001o1         & 3 \\
K2-18~b    & 2722o1,2722o3,2372o1,2722o2,2372o7  & 5 \\
KELT-7~b    & 3838o2,5924o7             & 2 \\
Kepler-12~b  & 3969o1,3969o5             & 2 \\
LTT 9779~b   & 1201o2                & 1 \\
NGTS-2~b    & 3838o3,5924o5             & 2 \\
NGTS-10~b   & 2158o1                & 1 \\
TOI-134~b   & 2512o13,2512o11,2512o12        & 3 \\
TOI-260~b   & 2512o1,2512o2             & 2 \\
TOI-1231~b   & 3557o2                & 1 \\
TOI-3757~b   & 3171o4,3171o5             & 2 \\
TOI-3984~Ab   & 3171o1,3171o2,3171o3         & 3 \\
TOI-5205~b   & 3171o16,3171o17,3171o18        & 3 \\
TrES-4~b    & 3838o1                & 1 \\
WASP-15~b   & 3838o4                & 1 \\
WASP-17~b   & 1353o5,1353o101,1353o3        & 3 \\
WASP-19~b   & 5924o12                & 1 \\
WASP-39~b   & 1366o4,1366o2,1366o1,1366o3,2783o1  & 5 \\
WASP-43~b   & 1366o11,1224o2            & 2 \\
WASP-52~b   & 1201o501,3969o3               & 2 \\
WASP-63~b   & 2113o5,5924o9             & 2 \\
WASP-69~b   & 5924o2                & 1 \\
WASP-80~b   & 1177o1,1185o2,1185o103,5924o3             & 4 \\
WASP-96~b   & 2734o2                & 1 \\
WASP-107~b\tablenotemark{c}   & 1185o8,1280o1,1201o8,1224o3,1185o9  & 5 \\
WASP-121~b   & 1729o1,1201o1             & 2 \\
WASP-166~b   & 2062o4,2062o3             & 2 \\
WASP-178~b   & 2055o1                & 1 \\
\enddata
\tablecomments{
\tablenotetext{a}{This column contains the proposal IDs corresponding to the observations used for each target. For example, 1185o18 for GJ 1214~b indicates proposal 1185, observation 18.}
\tablenotetext{b}{This column denotes the total number of transits used for each target.}
\tablenotetext{c}{The analysis of these transit observations for WASP-107~b was published in our previous paper \citep{wu_evidence_2026}.}
}
\end{deluxetable}

\subsection{White Light Curve Extraction}

We reduce the downloaded time-series spectroscopic data from JWST and generate broadband white light curves for TTV analysis. Two independent pipelines, \texttt{Eureka!} \citep{bell_eureka_2022,ahrer_early_2023,rustamkulov_early_2023,alderson_early_2023,howard_characterizing_2023,powell_sulfur_2024} and \texttt{exoTEDRF} \citep{radica_exotedrf_2024,radica_awesome_2023,feinstein_early_2023,howard_characterizing_2023,lim_atmospheric_2023}, are employed for this preprocessing. Although tailored to different instruments, both begin with raw uncalibrated data and produce white light curves that are subsequently analyzed with a uniform fitting framework to ensure consistency in the derived transit parameters and mid-transit times.

The choice of pipeline is dictated by the spectral trace characteristics of each instrument. \texttt{Eureka!} is used for NIRSpec, NIRCam, and MIRI, which produce single or spatially well-separated traces. \texttt{exoTEDRF} is adopted for NIRISS, whose slitless disperser projects partially overlapping diffraction orders onto the detector; \texttt{exoTEDRF} deblends overlapping traces and extracts the orders independently. To optimize the signal-to-noise ratio, we retain only the first-order (order~1) NIRISS spectra and use exclusively the NRS1 detector for NIRSpec to avoid inter-detector systematics.

Following the extraction process, we obtain 84 broadband white light curves suitable for high-precision transit timing measurements. These light curves are subsequently modeled to determine the transit parameters and mid-transit times.

\subsection{Transit Time Measurements}\label{sec:transit measurements}

The 84 white light curves are fitted independently to determine 84 mid-transit times. 
Transit models are generated using the \texttt{batman} package \citep{kreidberg_batman_2015}. We fix the orbital period and adopt a circular orbit ($e=0$), while allowing all remaining transit parameters to vary freely with gaussian shaped priors based on literature values (Table~\ref{tab:planet parameters}). Limb darkening was described using the reparameterized quadratic law, with coefficients $u_1$ and $u_2$ assigned uniform priors between 0 and 1 \citep{kipping_efficient_2013}. Detailed descriptions of the extraction procedures, fitting configurations, prior distributions, posterior analyses, and diagnostic products for all 84 transits are provided online.

Instrumental systematics were modeled using combinations of several commonly adopted components, including a linear trend in time, linear decorrelations with the spectral position and point-spread-function full width at half maximum (FWHM), exponential ramps to account for detector settling effects, and step functions to correct discontinuities such as mirror segment tilt events. Additional terms were included when warranted by individual datasets. We also incorporated a flux uncertainty scaling factor to account for residual noise not captured by the nominal photometric uncertainties \citep{bell_nightside_2024,valentine_jwst-tst_2024}. The exact systematics model was selected on a transit-by-transit basis according to the characteristics of each observation. 

Two additional mechanisms that may affect transit timing are non-zero orbital eccentricity and spot-crossing events. 
We adopt a circular orbit ($e=0$) for all fits to ensure a homogeneous treatment of the sample, as the majority of our targets are short-period planets with nearly circular orbits.
Even for systems with comparatively larger reported eccentricities, such as HAT-P-26~b and TOI-3984~b, the light curves are adequately described by circular-orbit models. Transit timing shifts induced by non-zero eccentricity depend on both $e$ and the argument of pericentre \citep{kipping_transiting_2008}. To evaluate their impact, we repeated the fits while allowing eccentricity and the position of pericentre to vary freely. The resulting transit times are indistinguishable from those obtained under the circular-orbit assumption, consistent with the findings of \citet{wu_evidence_2026}.
We therefore retain the simpler circular model for all targets.

For the second effect, spot-crossing events, the vast majority of the light curves show no evidence of such features. 
The few light curves that do exhibit possible spot-crossing signatures show only low-level anomalies close to the transit bottom rather than near ingress or egress, as in the case of HAT-P-18~b \citep{fournier-tondreau_near-infrared_2024}. 
Since spot-crossing anomalies can bias the measured mid-transit time, we further assessed whether the features identified in our light curves could produce a significant timing offset. We quantified their impact using two independent approaches: a first-order estimate derived directly from the best-fit solution and direct refitting of the light curves after masking the affected data points (see Section~\ref{sec:hat-p-18b} for a detailed discussion). Both approaches yield timing shifts well below the corresponding timing uncertainties. We therefore conclude that the identified spot-crossing events have a negligible impact on the measured transit times and do not require special treatment in our timing analysis.

\begin{figure}[!h]
    \centering
    \includegraphics[width=0.48\textwidth]{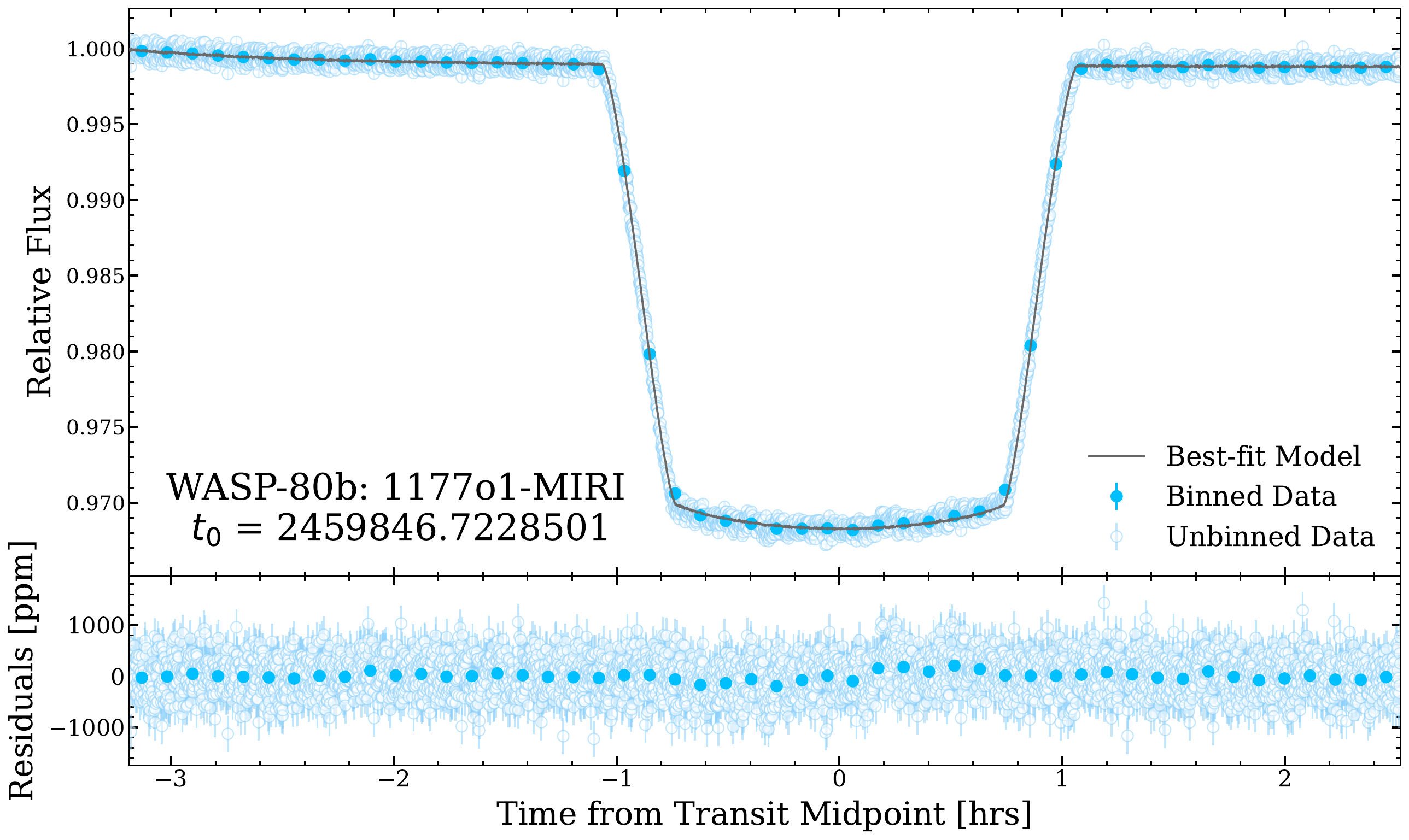}
    \caption{White Light Curve Example: Reduced white light curves and fitting results for WASP-80~b (ID: 1177o1-MIRI, indicating proposal 1177, observation 1, observed with the JWST MIRI instrument). The top panel shows the relative flux light curve, while the bottom panel displays the corresponding residuals. Unfilled blue points represent unbinned data, while solid blue points show 50-point binned data. The black line indicates the best-fitting model.}
    \label{fig:white light curve example}
\end{figure}

\begin{figure}[!h]
    \centering
    \includegraphics[width=0.45\textwidth]{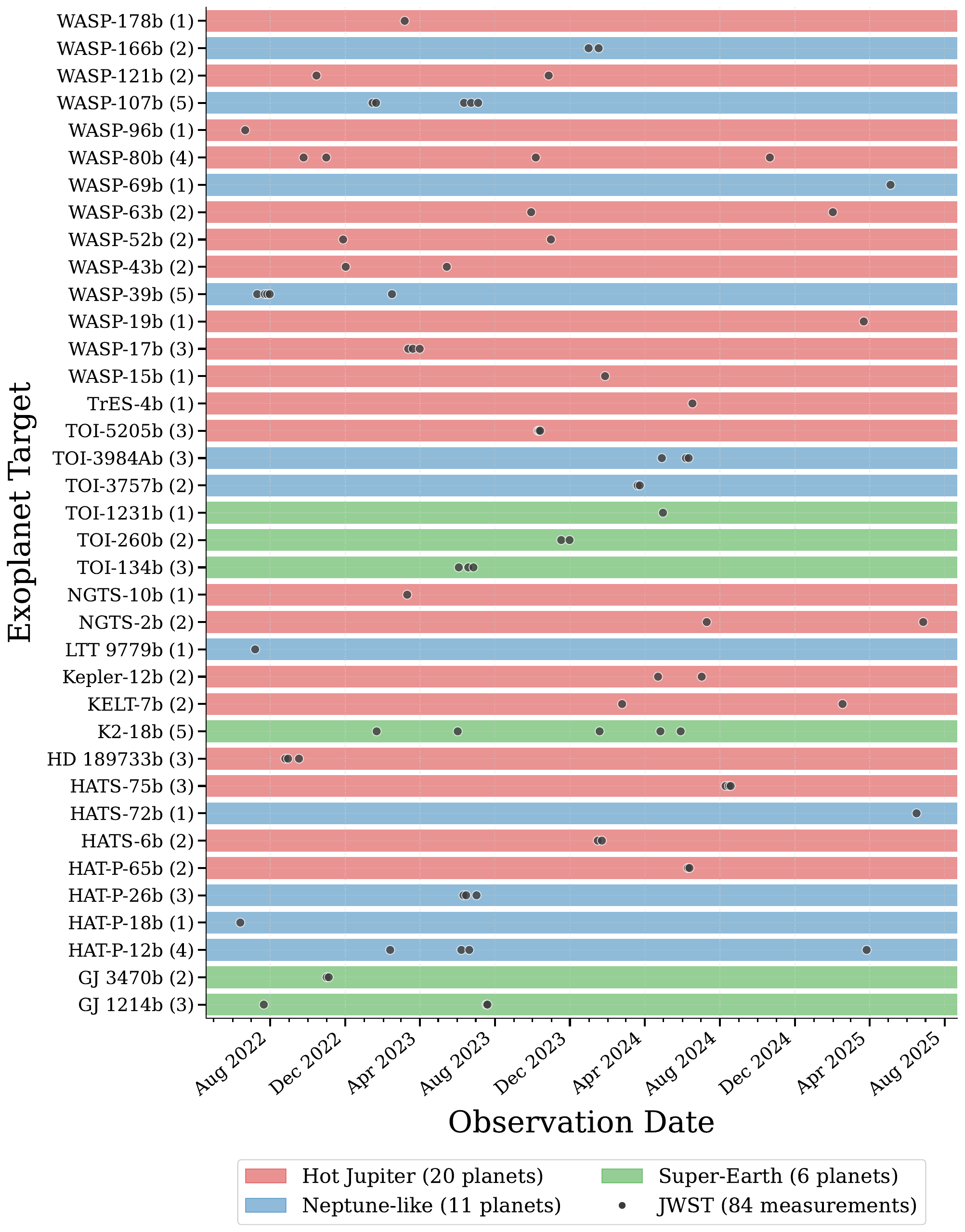}
    \caption{Reduced Transit Times from JWST Observations: Timeline showing all 84 transit times for the 37 planets in our sample. Target backgrounds are color-coded by planetary type: red for Hot Jupiters, blue for Neptune-like planets, and green for Super-Earths. Gray dots represent transit times measured in this work from JWST observations.}
    \label{fig:jwst timeline}
\end{figure}

Figure~\ref{fig:white light curve example} shows representative transit fits for the JWST white light curves of WASP-80~b, demonstrating good agreement between the model and the observations. The resulting transit times for all JWST observations are listed in Table~\ref{tab:transit times list} and displayed in Figure~\ref{fig:jwst timeline}.

\begin{deluxetable*}{cccccccccc}[!h]
\tablecaption{Physical Parameters Used in This Work \label{tab:planet parameters}}
\tabletypesize{\tiny}
\tablewidth{0pt}
\tablehead{
\colhead{Planet} & \colhead{Period[d]} & \colhead{Rp/Rs} & \colhead{a/Rs\tablenotemark{a}} & \colhead{inc[$^{\circ}$]\tablenotemark{a}} & \colhead{$M_p$[$M_J$]} & \colhead{$R_p$[$R_J$]} & \colhead{$M_s$[$M_{\bigodot}$]} & \colhead{reference\tablenotemark{b}} & \colhead{category}
}
\startdata
GJ 1214~b & 1.580404531 & 0.11589 & 14.97 & 88.98 & 0.0265 & 0.2438 & 0.182 & \cite{mahajan_using_2024} & Super-Earth \\
GJ 3470~b & 3.3366524 & 0.0764 & 13.9 & 88.9 & 0.03958 & 0.346 & 0.51 & \cite{kokori_exoclock_2023,kosiarek_bright_2019} & Super-Earth \\
HAT-P-12~b & 3.21305762 & 0.1406 & 11.77 & 89.0 & 0.211 & 0.949 & 0.74 & \cite{kokori_exoclock_2023,ozturk_new_2019,stassun_revised_2019} & Neptune-like \\
HAT-P-18~b & 5.50802941 & 0.1356 & 16.39 & 88.53 & 0.2 & 0.995 & 0.773 & \cite{kokori_exoclock_2023,bonomo_gaps_2017,stassun_revised_2019} & Neptune-like \\
HAT-P-26~b & 4.234503 & 0.0737 & 12.51 & 87.82 & 0.07 & 0.63 & 0.846 & \cite{kokori_exoclock_2023,a-thano_revisiting_2023,stassun_accurate_2017,stassun_revised_2019} & Neptune-like \\
HAT-P-65~b & 2.60544751 & 0.1006 & 5.18 & 88.3 & 0.554 & 1.611 & 1.297 & \cite{kang_simultaneous_2024} & Hot Jupiter \\
HATS-6~b & 3.32526424 & 0.1798 & 13.65 & 88.21 & 0.319 & 0.998 & 0.585471 & \cite{kokori_exoclock_2023,hartman_hats-6b_2015,stassun_revised_2019} & Hot Jupiter \\
HATS-72~b & 7.3279496 & 0.1029 & 19.821 & 89.56 & 0.1254 & 0.7224 & 0.7311 & \cite{kokori_exoclock_2023,hartman_hats-47b_2020} & Neptune-like \\
HATS-75~b & 2.7886556 & 0.1555 & 12.037 & 88.07 & 0.491 & 0.884 & 0.6017 & \cite{jordan_hats-74ab_2022} & Hot Jupiter \\
HD 189733~b & 2.218574944 & 0.15534 & 8.92 & 85.78 & 1.166 & 1.119 & 0.827851465273 & \cite{kokori_exoclock_2023,addison_minerva-australis_2019,rosenthal_california_2021} & Hot Jupiter \\
K2-18~b & 32.940045 & 0.05375 & 76.963745 & 89.522801 & 0.0272 & 0.2328 & 0.440668 &  \cite{stassun_revised_2019,benneke_water_2019} & Super-Earth \\
KELT-7~b & 2.7347656 & 0.091 & 5.53 & 83.8 & 1.39 & 1.6 & 1.45 & \cite{kokori_exoclock_2023,stassun_accurate_2017,stassun_revised_2019} & Hot Jupiter \\
Kepler-12~b & 4.43796263 & 0.11887 & 8.019 & 88.8 & 0.432 & 1.754 & 1.09 & \cite{kokori_exoclock_2023,esteves_changing_2015,stassun_revised_2019} & Hot Jupiter \\
LTT 9779~b & 0.79206425 & 0.0455 & 3.877 & 76.39 & 0.09225 & 0.421 & 1.02 & \cite{kokori_exoclock_2023,edwards_characterizing_2023} & Neptune-like \\
NGTS-10~b & 0.7668944 & 0.1765 & 4.447 & 78.950482 & 2.162 & 1.205 & 0.696 & \cite{mccormac_ngts-10b_2020} & Hot Jupiter \\
NGTS-2~b & 4.511123 & 0.0962 & 8.0 & 88.5 & 0.74 & 1.595 & 1.34 & \cite{kokori_exoclock_2023,raynard_ngts-2b_2018,stassun_revised_2019} & Hot Jupiter \\
TOI-1231~b & 24.245586 & 0.0701 & 58.1 & 89.73 & 0.0485 & 0.326 & 0.485 & \cite{burt_toi-1231_2021} & Super-Earth \\
TOI-134~b & 1.40152604 & 0.0247 & 7.4050809 & 84.27 & 0.0128 & 0.145 & 0.614 & \cite{hobson_three_2024} & Super-Earth \\
TOI-260~b & 13.475853 & 0.0258 & 33.406149 & 88.84 & 0.0133 & 0.153 & 0.616 & \cite{hobson_three_2024} & Super-Earth \\
TOI-3757~b & 3.438753 & 0.1769 & 13.26 & 86.76 & 0.268 & 1.07 & 0.64 & \cite{kanodia_toi-3757_2022} & Neptune-like \\
TOI-3984~Ab & 4.353326 & 0.1558 & 19.1 & 89.5 & 0.14 & 0.71 & 0.49 & \cite{canas_toi-3984_2023} & Neptune-like \\
TOI-5205~b & 1.630757 & 0.272 & 10.94 & 88.21 & 1.08 & 1.03 & 0.392 & \cite{kanodia_toi-5205b_2023} & Hot Jupiter \\
TrES-4~b & 3.55392889 & 0.1045 & 6.14 & 83.1 & 0.78 & 1.61 & 1.18 & \cite{kokori_exoclock_2023,stassun_accurate_2017,stassun_revised_2019} & Hot Jupiter \\
WASP-107~b & 5.72148926 & 0.1446 & 17.96 & 89.56 & 0.096 & 0.924 & 0.683 & \cite{kokori_exoclock_2023,piaulet_wasp-107bs_2021,mocnik_starspots_2017} & Neptune-like \\
WASP-121~b & 1.274924762 & 0.1245 & 3.754 & 87.6 & 1.157 & 1.753 & 1.45 & \cite{kokori_exoclock_2023,bourrier_hot_2020,stassun_revised_2019} & Hot Jupiter \\
WASP-15~b & 3.75209956 & 0.0951 & 7.3 & 85.7 & 0.54 & 1.41 & 1.176 & \cite{kokori_exoclock_2023,stassun_accurate_2017,stassun_revised_2019} & Hot Jupiter \\
WASP-166~b & 5.4435423 & 0.05301 & 11.3 & 88.0 & 0.101 & 0.63 & 1.19 & \cite{kokori_exoclock_2023,hellier_wasp-166b_2019} & Neptune-like \\
WASP-178~b & 3.3448412 & 0.11066 & 6.49 & 84.45 & 1.41 & 1.94 & 1.93 & \cite{rodriguez_martinez_kelt-25_2020} & Hot Jupiter \\
WASP-17~b & 3.73548545 & 0.1235 & 7.03 & 87.1 & 0.78 & 1.87 & 1.354 & \cite{kokori_exoclock_2023,stassun_accurate_2017,stassun_revised_2019} & Hot Jupiter \\
WASP-19~b & 0.788839092 & 0.1409 & 3.46 & 78.8 & 1.154 & 1.415 & 0.965 & \cite{kokori_exoclock_2023,cortes-zuleta_tramos_2020} & Hot Jupiter \\
WASP-39~b & 4.05528043 & 0.1457 & 11.37 & 87.75 & 0.281 & 1.279 & 0.918 & \cite{kokori_exoclock_2023,mancini_gaps_2018,stassun_revised_2019} & Neptune-like \\
WASP-43~b & 0.813474056 & 0.1594 & 4.867 & 82.11 & 2.05 & 1.036 & 0.65 & \cite{kokori_exoclock_2023,bonomo_gaps_2017,stassun_revised_2019} & Hot Jupiter \\
WASP-52~b & 1.74978117 & 0.1646 & 7.38 & 85.35 & 0.46 & 1.223 & 0.87 & \cite{kokori_exoclock_2023,ozturk_new_2019} & Hot Jupiter \\
WASP-63~b & 4.37808205 & 0.078 & 6.49 & 87.8 & 0.37 & 1.41 & 1.02 & \cite{kokori_exoclock_2023,stassun_accurate_2017,stassun_revised_2019} & Hot Jupiter \\
WASP-69~b & 3.86813888 & 0.1336 & 11.953 & 86.71 & 0.29 & 1.11 & 0.76 & \cite{kokori_exoclock_2023,stassun_accurate_2017,stassun_revised_2019} & Neptune-like \\
WASP-80~b & 3.06785251 & 0.1714 & 12.63 & 89.02 & 0.538 & 0.999 & 0.616961 & \cite{kokori_exoclock_2023,triaud_wasp-80b_2015,stassun_revised_2019} & Hot Jupiter \\
WASP-96~b & 3.42525674 & 0.1175 & 9.255 & 85.6 & 0.49 & 1.2 & 0.98 & \cite{kokori_exoclock_2023,bonomo_gaps_2017,stassun_revised_2019} & Hot Jupiter \\
\enddata
\tablecomments{
\tablenotetext{a}{Some parameters are derived from literature values, such as $a/R_s$ computed from $a$ and $R_s$, and inclination $i$ derived from $b$ and $a/R_s$ using $b = (a/R_s)\cos i$.}
\tablenotetext{b}{Parameters are primarily sourced from the \href{https://exoplanetarchive.ipac.caltech.edu/}{NASA Exoplanet Archive}.}
}
\end{deluxetable*}

\begin{deluxetable}{cccc}[!h]
\tablecaption{Full Reduced Transit Times from JWST \label{tab:transit times list}}
\tabletypesize{\scriptsize}
\tablewidth{0pt}
\tablehead{
\colhead{Planet} & \colhead{$t_0$[$BJD_{TDB}$]} & \colhead{uncertainty} & \colhead{ID\tablenotemark{a}}
}
\startdata
GJ 1214~b & 2459782.0176646 & 0.0000086 & 1803o1-MIRI \\
GJ 1214~b & 2460143.9302596 & 0.0000188 & 1185o18-NIRSpec \\
GJ 1214~b & 2460145.5106632 & 0.0000197 & 1185o19-NIRSpec \\
GJ 3470~b & 2459884.2506156 & 0.0000546 & 1185o16-NIRCam \\
GJ 3470~b & 2459887.5872886 & 0.0000529 & 1185o17-NIRCam \\
... & ... & ... & ... \\
\enddata
\tablecomments{
This table is published in its entirety in machine-readable format; a subset is shown here for reference.
\tablenotetext{a}{The ID column indicates the proposal information and instrument used. For example, GJ 1214~b with ID = 1803o1-MIRI corresponds to proposal 1803, observation 1, observed with the JWST MIRI instrument.}
}
\end{deluxetable}

\section{Transit Timing Variation Analysis\label{sec:transit timing variation analysis}}

\subsection{Transit Timing Data \label{subsec:data and method}}
To analyze the TTVs for our targets, we incorporate transit timing data from multiple sources: HST observations from \cite{ma_exoplanet_2025}, TESS data from \cite{wang_long-term_2024}, and observations specifically for WASP-107 b from \cite{wu_evidence_2026}. These three studies are our own work. We also include additional literature timings from \cite{ivshina_tess_2022}. Where the same TESS observations were processed by both \cite{wang_long-term_2024} and \cite{ivshina_tess_2022}, we used the updated and higher-precision timings from the former.

The literature transit timings are primarily drawn from the dataset of \cite{ivshina_tess_2022}. In this dataset, the authors convert the literature times into the BJD/TDB time system wherever possible and explicitly state the time system used. We exclude composite transit times derived from phase-folded light curves spanning multiple observation epochs, as well as transit times inferred from linear fits to multiple transit times under the assumption of a strictly constant orbital period. These two types of transit times are also discussed in \cite{adams_doomed_2024}. Individual epochs exhibiting anomalous deviations relative to neighboring transits are also discarded. In this regard, we discarded only one transit time observation for WASP-80~b, as the specific transit time at that epoch deviates significantly from other transit times at the same epoch. No such anomalies are present for the other targets.
In \cite{adams_doomed_2024}, the authors address the unrealistically low uncertainties of the literature transit times by rescaling all uncertainties by a common factor to achieve $\chi^2 = 1$. However, this approach is not uniformly appropriate for all data points, given the precision differences between space- and ground-based observations and the inclusion of our own transit times. Therefore, we adopt a more straightforward method, which we also employed in our previous work \citep{ma_exoplanet_2025}. For the remaining literature timings, we impose minimum uncertainty floors of $0.0005$~d and $0.0001$~d for ground-based and space-based observations, respectively, to account for unrealistically small reported errors. These floors correspond to typical uncertainties from ground- and space-based observations, ensuring a conservative treatment. For the transit timings derived from JWST data in this work, we retain their original uncertainties from the MCMC fitting without artificial inflation.

\begin{figure}[!h]
    \centering
    \includegraphics[width=0.48\textwidth]{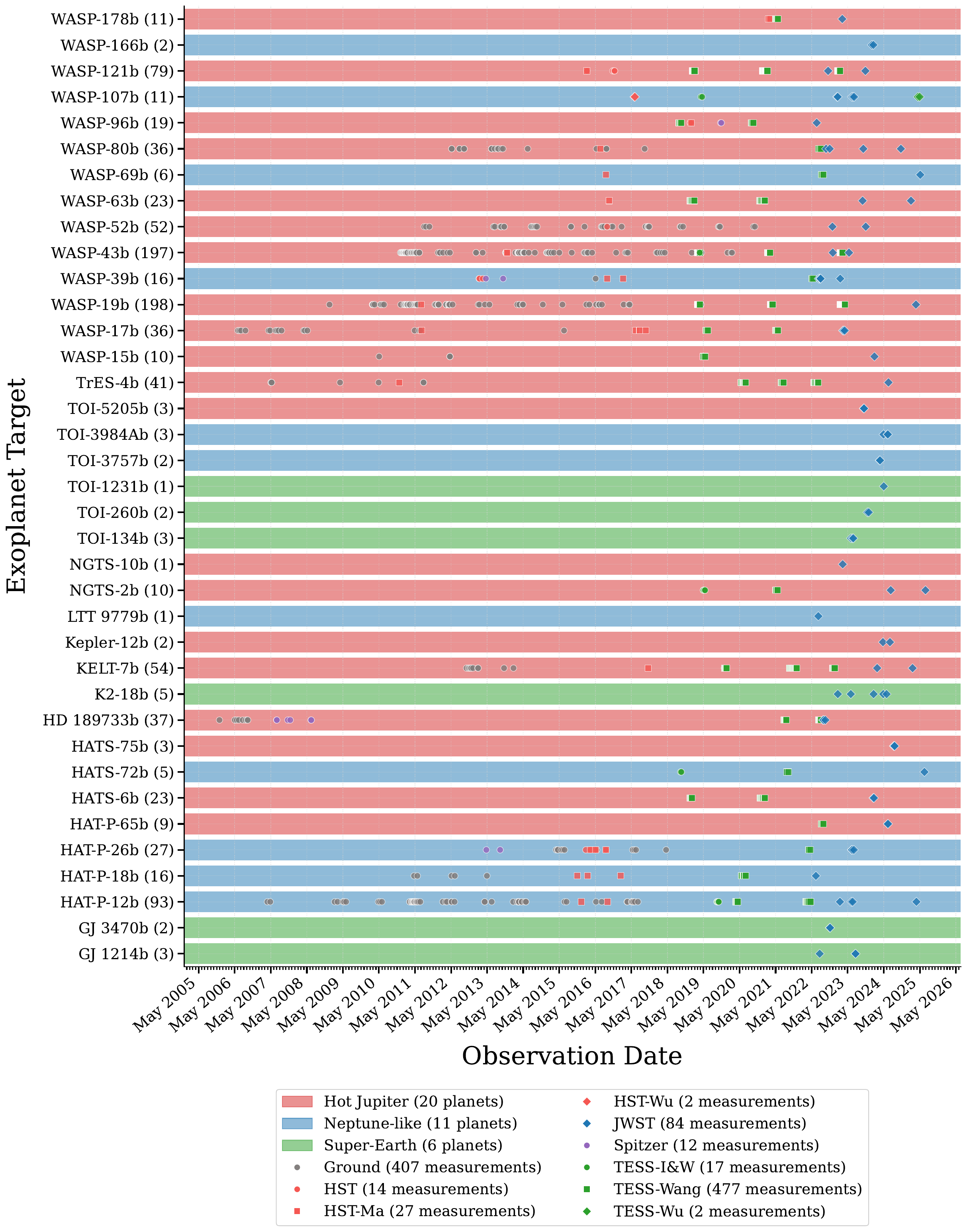}
    \caption{Transit times from all sources for the 37 planets in this work. Target backgrounds are color-coded by planetary type: red for hot Jupiters, blue for Neptune-like planets, and green for super-Earths. Points are colored by observatory (gray for ground-based, red for HST, green for TESS, purple for Spitzer, and blue for JWST) and shaped to distinguish literature data from our own measurements. Our own data comprise the JWST, HST-Ma, HST-Wu, TESS-Wu, and TESS-Wang categories.}
    \label{fig:all data timeline}
\end{figure}

All transit times used in this work are displayed in Figure~\ref{fig:all data timeline}, where points are color-coded by observatory and shaped to distinguish literature values from our own measurements (see caption for details). The categories JWST, HST-Ma, HST-Wu, TESS-Wu, and TESS-Wang represent our own data; this labeling convention applies to all subsequent figures.

\subsection{TTV Modeling}

We employed three established models for TTV fitting: a linear model for updating ephemeris, a quadratic model to detect long-term TTVs, and a sine model to identify periodical TTV signals. All three models have been extensively used in previous studies \citep{ma_exoplanet_2025,wang_long-term_2024,wang_exoplanet_2025,a-thano_revisiting_2023,parthasarathy_transit_2025,maciejewski_departure_2016,patra_apparently_2017,yee_orbit_2019,maciejewski_revisiting_2021,yang_tentative_2022}.

The linear model assumes a constant orbital period:
\begin{equation}
t_{\mathrm{mid}}(E) = t_0 + E\times P,
\end{equation}
where $t_{\mathrm{mid}}$ is the mid-transit time, $P$ is the orbital period, $E$ is the epoch number (the number of orbits since the reference transit), and $t_0$ is the reference mid-transit time.

To minimize errors over large epoch ranges, we use the cumulative form:
\begin{equation}
\label{equation:TTV linear cumulative model}
t_{\mathrm{mid}}(E) = t_{\mathrm{mid}}(E-1) + P_{E-1}.
\end{equation}

The quadratic model incorporates an uniformly changing period:
\begin{equation}
t_{\mathrm{mid}}(E) = t_0 + E\times P + \frac{1}{2}\frac{dP}{dE}E^2.
\end{equation}

Its cumulative formulation is:
\begin{equation}
\label{equation:TTV quadratic cumulative model}
\begin{aligned}
t_{\mathrm{mid}}(E) &= t_{\mathrm{mid}}(E-1) + P_{E-1}, \\
P_E &= P_{E-1} + \dot{P} \left( \frac{P_{E-1} + P_E}{2} \right),
\end{aligned}
\end{equation}
where $\dot{P} = dP/dt = P^{-1} dP/dE$ is the rate of change of the orbital period.

The sine model assumes a periodic TTV:
\begin{equation}
t_{\mathrm{mid}}(E) = t_0 + E\times P + A_{TTV}\sin(\frac{2\pi E}{P_{TTV}}-\phi_{TTV})
\end{equation}
where $A_{TTV}$, $P_{TTV}$, and $\phi_{TTV}$ are the amplitude, period, and phase of the TTV signal, respectively.

Its cumulative formulation is:
\begin{equation}
\label{equation:TTV sine cumulative model}
\begin{aligned}
t_{\mathrm{mid}}(E) &= t_{\mathrm{mid}}(E-1) + P + \frac{2\pi}{P_{TTV}}A\cos(\phi_{E-1}), \\
\phi_E &= \phi_{E-1} + \frac{2\pi}{P_{TTV}},
\end{aligned}
\end{equation}
where $\phi_E$ is the phase of the TTV signal at epoch $E$, and $\phi_0=-\phi_{TTV}$.

We evaluated model performance using the Bayesian Information Criterion (BIC) \citep{schwarz_estimating_1978}:
\begin{equation}
\mathrm{BIC} = \chi^2 + k\ln(n),
\end{equation}
where $k$ is the number of parameters (2 for the linear model, 3 for the quadratic model, and 5 for the sine model) and $n$ is the number of data points. A $\Delta\mathrm{BIC} > 10$ between models constitutes strong evidence for preferring the model with the lower BIC. Bayesian inference was performed using \texttt{dynesty} \citep{speagle_dynesty_2020,koposov_joshspeagledynesty_2022} with Dynamic Nested Sampling \citep{higson_dynamic_2019}.

\section{Results}

To ensure that our conclusions are statistically meaningful, we evaluate TTV signals only for targets with more than five measured transit times. For systems with five or fewer transit times, we restrict the analysis to updating their orbital ephemerides.

\subsection{Ephemeris Updating}
We first fitted the linear ephemeris model to all targets in the sample and derived updated orbital ephemerides, which are listed in Table~\ref{tab:linear ephemerides}. These revised ephemerides provide improved predictions of future transit events and are intended to facilitate follow-up observations with both ground- and space-based facilities. In several cases, the inclusion of newly acquired TESS and JWST transit times significantly extends the observational baseline relative to previous studies, resulting in more precise determinations of the orbital period and reference transit epoch.

\begin{deluxetable*}{ccccccc}
\tablecaption{Linear Ephemerides for Each Target \label{tab:linear ephemerides}}
\tabletypesize{\scriptsize}
\tablewidth{0pt}
\tablehead{
     \colhead{Planet\tablenotemark{a}} & \colhead{$t_0$ [$BJD_{TDB}$]} & \colhead{$1\sigma$} & \colhead{Period [d]} & \colhead{$1\sigma$} & \colhead{$N$} & \colhead{BIC}
}
\startdata
GJ 1214~b & 2459782.0176645 & 0.0000086 & 1.580404344 & 0.000000070 & 3 & 2.20 \\
GJ 3470~b & 2459884.2506157 & 0.0000545 & 3.336672304 & 0.000075071 & 2 & 1.39 \\
HAT-P-12~b & 2454187.8557142 & 0.0000535 & 3.213058158 & 0.000000031 & 93 & 251.93 \\
HAT-P-18~b & 2455673.4195636 & 0.0000785 & 5.508029543 & 0.000000115 & 16 & 101.79 \\
HAT-P-26~b & 2456405.6233967 & 0.0000658 & 4.234499799 & 0.000000089 & 27 & 668.75 \\
HAT-P-65~b & 2459799.0215141 & 0.0003767 & 2.605446601 & 0.000001454 & 9 & 5.51 \\
HATS-6~b & 2458469.3106842 & 0.0001477 & 3.325268164 & 0.000000264 & 23 & 33.50 \\
HATS-72~b & 2458373.4374542 & 0.0004938 & 7.327950634 & 0.000001464 & 5 & 3.82 \\
HATS-75~b & 2460532.4381103 & 0.0000116 & 2.788638961 & 0.000005507 & 3 & 3.32 \\
HD 189733~b & 2453702.6073468 & 0.0000439 & 2.218575184 & 0.000000016 & 37 & 93.09 \\
K2-18~b & 2459965.4694875 & 0.0001380 & 32.939958438 & 0.000011167 & 5 & 24.72 \\
KELT-7~b & 2456204.8173136 & 0.0001079 & 2.734765673 & 0.000000086 & 54 & 111.27 \\
Kepler-12~b & 2460422.7619854 & 0.0000271 & 4.437977779 & 0.000002373 & 2 & 1.39 \\
NGTS-2~b & 2458602.7063848 & 0.0001970 & 4.511123023 & 0.000000404 & 10 & 31.46 \\
TOI-134~b & 2460098.9630022 & 0.0001242 & 1.401524019 & 0.000011894 & 3 & 2.82 \\
TOI-260~b & 2460265.4394614 & 0.0000826 & 13.475962452 & 0.000111978 & 2 & 1.39 \\
TOI-3757~b & 2460389.6579820 & 0.0000085 & 3.438777795 & 0.000011693 & 2 & 1.39 \\
TOI-3984~Ab & 2460428.9699705 & 0.0000082 & 4.353331427 & 0.000001060 & 3 & 3.75 \\
TOI-5205~b & 2460227.8649322 & 0.0000029 & 1.630751332 & 0.000002217 & 3 & 10.24 \\
TrES-4~b & 2454223.7976524 & 0.0002787 & 3.553929211 & 0.000000188 & 41 & 42.22 \\
WASP-107~b & 2457904.7331773 & 0.0000321 & 5.721487610 & 0.000000087 & 11 & 31.44 \\
WASP-121~b & 2457424.8865355 & 0.0000298 & 1.274924889 & 0.000000016 & 79 & 136.18 \\
WASP-15~b & 2455320.1095083 & 0.0002496 & 3.752099374 & 0.000000206 & 10 & 9.65 \\
WASP-166~b & 2460309.8912061 & 0.0001582 & 5.443378556 & 0.000055011 & 2 & 1.39 \\
WASP-178~b & 2459265.1107418 & 0.0000944 & 3.344835851 & 0.000000550 & 11 & 55.31 \\
WASP-17~b & 2453890.5309936 & 0.0002248 & 3.735485042 & 0.000000139 & 36 & 100.18 \\
WASP-19~b & 2454817.1466501 & 0.0000480 & 0.788838983 & 0.000000007 & 198 & 201.61 \\
WASP-39~b & 2456332.4580219 & 0.0000649 & 4.055280225 & 0.000000077 & 16 & 36.62 \\
WASP-43~b & 2455537.8169330 & 0.0000376 & 0.813474049 & 0.000000007 & 197 & 287.60 \\
WASP-52~b & 2455776.1840751 & 0.0001162 & 1.749780892 & 0.000000049 & 52 & 121.31 \\
WASP-63~b & 2457650.9958847 & 0.0001049 & 4.378081604 & 0.000000216 & 23 & 25.21 \\
WASP-69~b & 2457617.1455904 & 0.0001845 & 3.868138199 & 0.000000250 & 6 & 9.25 \\
WASP-80~b & 2456054.8577135 & 0.0000836 & 3.067852056 & 0.000000066 & 36 & 40.55 \\
WASP-96~b & 2458354.3199094 & 0.0001144 & 3.425256780 & 0.000000291 & 19 & 24.48 \\
\enddata
\tablecomments{
\tablenotetext{a}{LTT 9779~b, TOI-1231~b, and NGTS-10~b each have only one measured transit time, precluding a fitted ephemeris. The corresponding transit times derived from light-curve fitting are provided in Table~\ref{tab:transit times list}.}
}
\end{deluxetable*}

\subsection{Quadratic Model}
We fitted the quadratic model to all targets with at least five transit times using three data configurations: the full dataset, the dataset excluding ground-based transit times, and the dataset containing only transit times derived in this work. For each configuration, we compared the quadratic model with the linear ephemeris model. 
A target was classified as a long-term TTV candidate if it satisfied both $\Delta \mathrm{BIC} > 10$ and $\dot{P}/\sigma_{\dot{P}} > 3$ in at least one of the three configurations.

\begin{figure}[!h]
    \centering
    \includegraphics[width=0.47\textwidth]{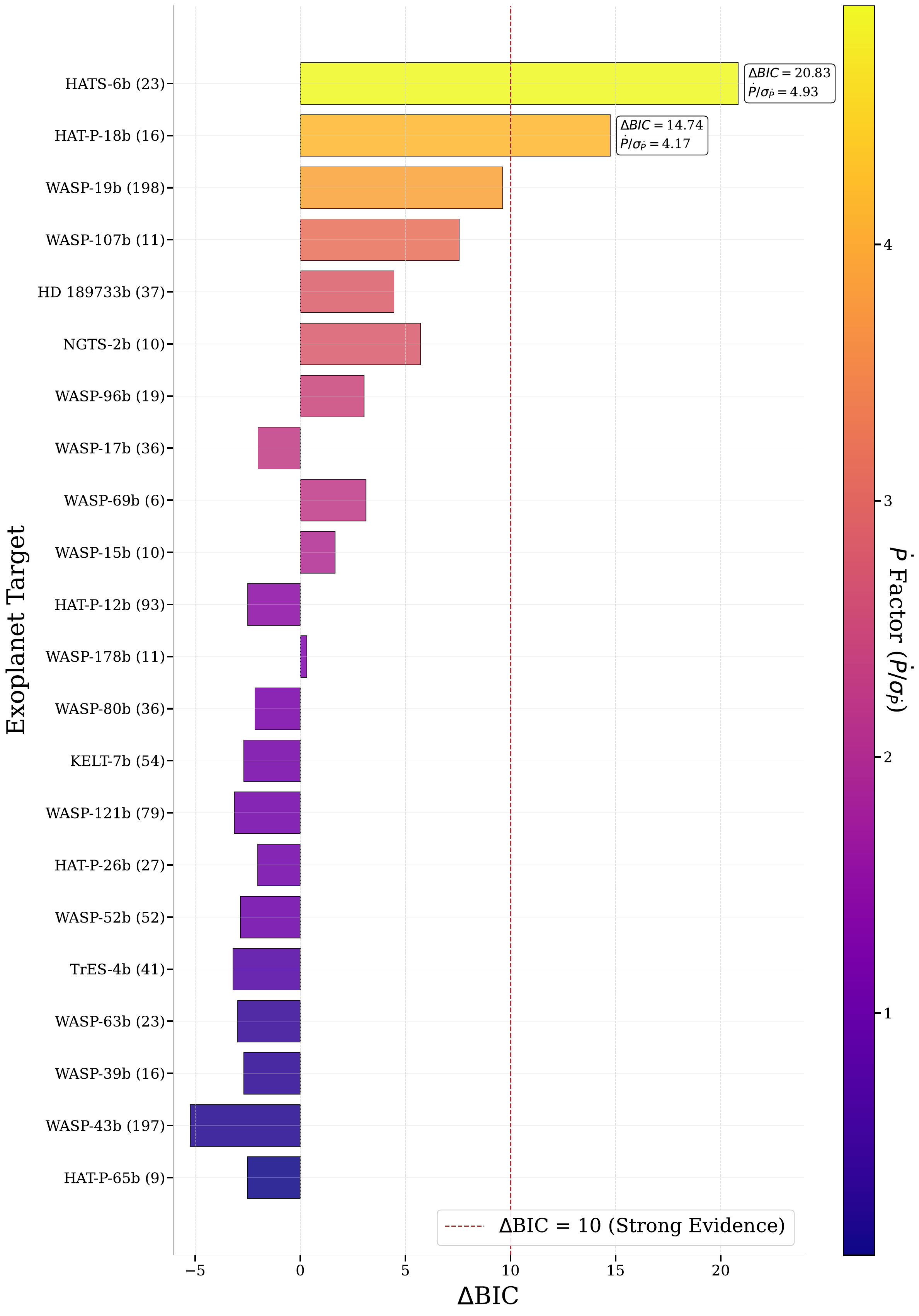}
    \caption{Long-term TTV candidates from the full dataset. Comparison of the quadratic model against the linear model for targets with more than five transit times. The red dashed line marks $\Delta \mathrm{BIC} = 10$. The color bar encodes the period-derivative significance $\dot{P}/\sigma_{\dot{P}}$. Targets satisfying $\Delta \mathrm{BIC} > 10$ and $\dot{P}/\sigma_{\dot{P}} > 3$ are annotated.}
    \label{fig:pdot candidates1}
\end{figure}

\begin{figure}[!h]
    \centering
    \includegraphics[width=0.47\textwidth]{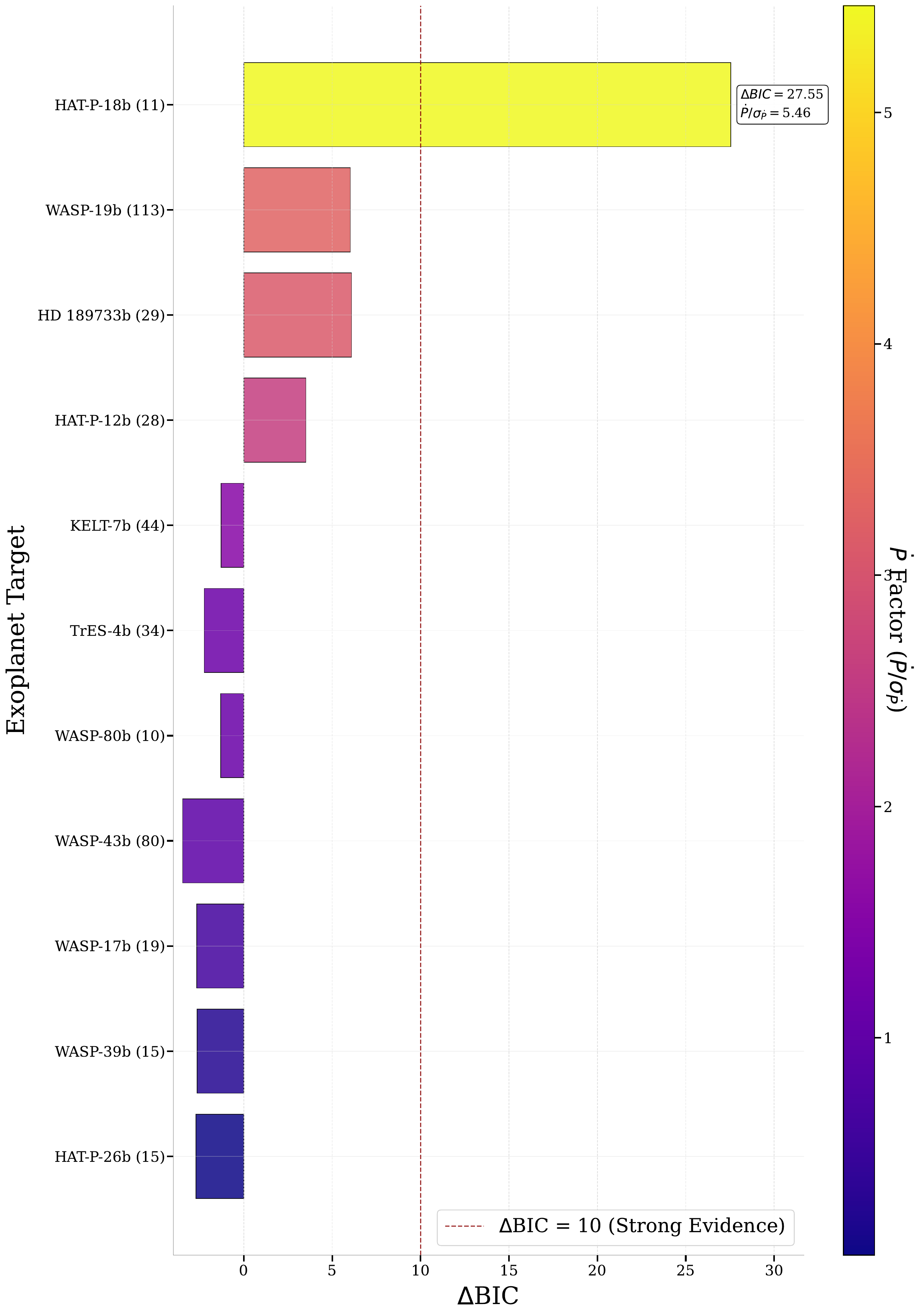}
    \caption{Same as Figure~\ref{fig:pdot candidates1}, but after excluding all ground-based transit times. Targets whose datasets remain unchanged relative to Figure~\ref{fig:pdot candidates1} are omitted to avoid redundancy.}
    \label{fig:pdot candidates2}
\end{figure}

\begin{figure}[!h]
    \centering
    \includegraphics[width=0.47\textwidth]{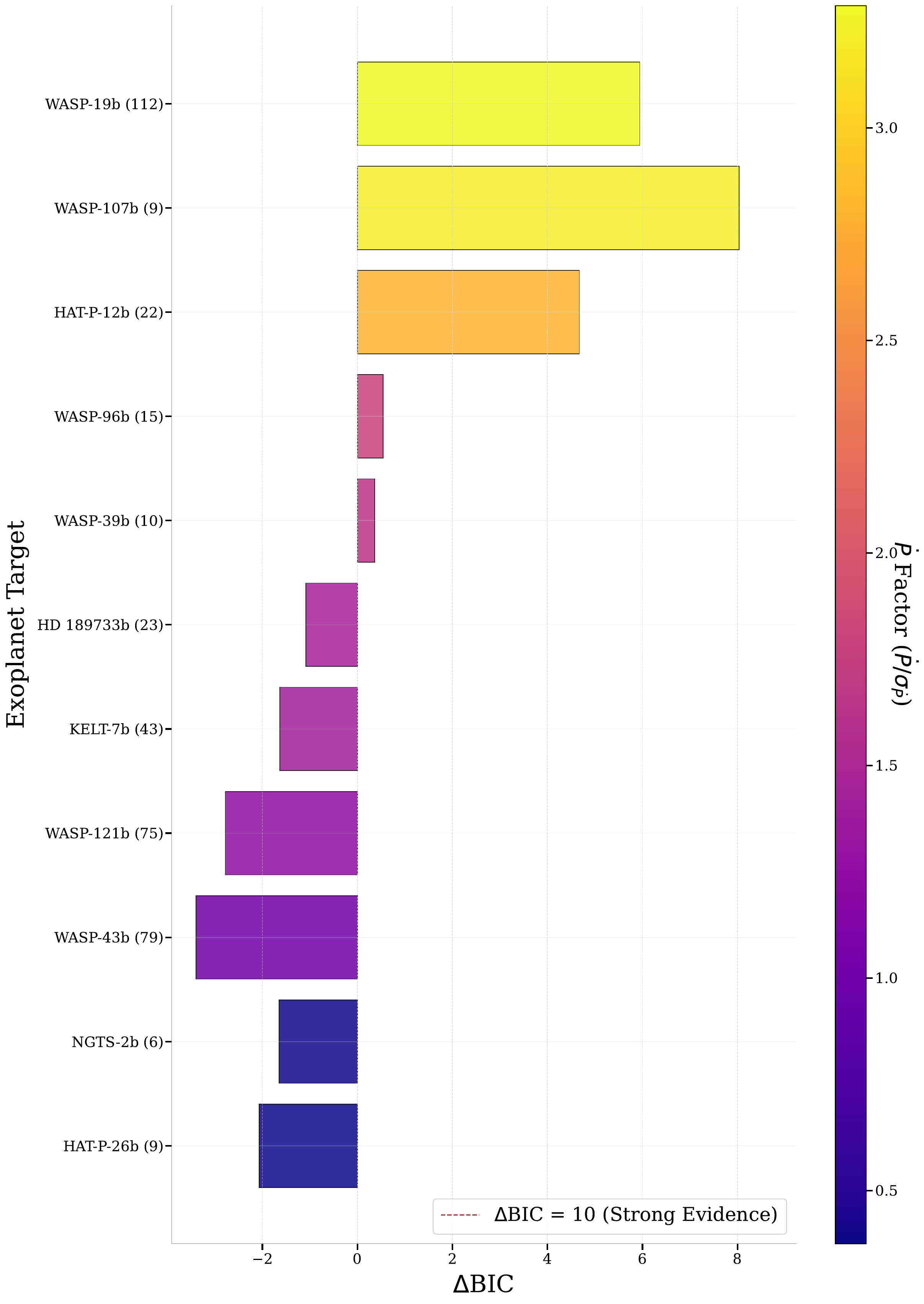}
    \caption{Same as Figure~\ref{fig:pdot candidates2}, but after further removing all literature timings not from our own work. Targets whose datasets are identical to those in Figure~\ref{fig:pdot candidates1} or Figure~\ref{fig:pdot candidates2} are omitted.}
    \label{fig:pdot candidates3}
\end{figure}

The results for the three configurations are shown in Figures~\ref{fig:pdot candidates1}--\ref{fig:pdot candidates3}. For clarity, targets whose datasets are unchanged relative to a previous configuration are omitted from subsequent figures. In the full dataset, HATS-6~b and HAT-P-18~b satisfy the candidate-selection criteria. After removing ground-based transit times, only HAT-P-18~b continues to satisfy both criteria among the targets with revised datasets. No additional candidates emerge when only transit times derived in this work are retained.

\begin{figure*}[!h]
    \centering
    \includegraphics[width=0.85\textwidth,height=0.45\textheight]{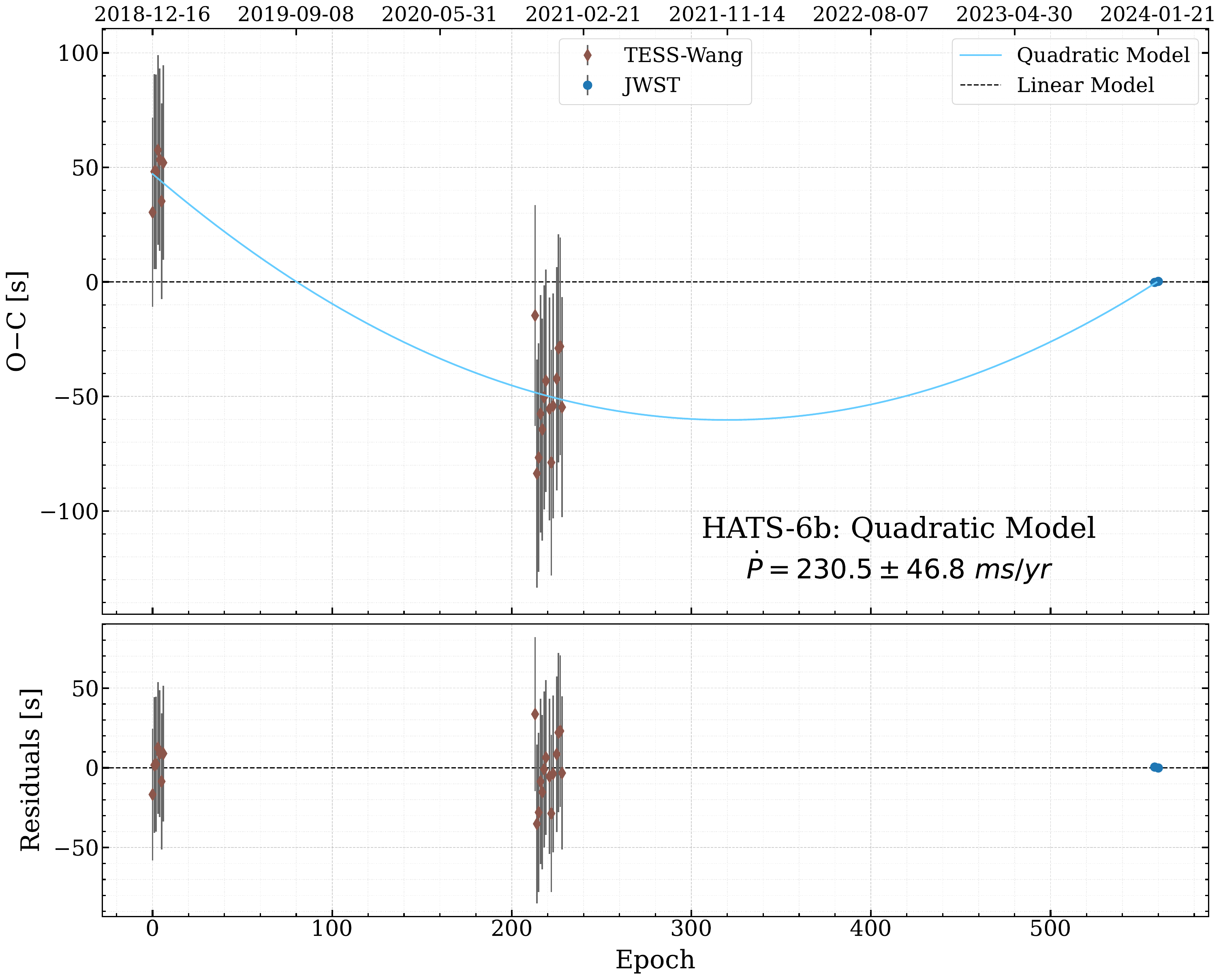}
    \caption{O--C diagram for HATS-6~b. \textit{Top:} the O-C diagram with the gray dashed line denotes the best-fit linear ephemeris, the blue curve shows the fitted quadratic model, and the annotated value gives the measured $\dot{P}$. Data points are color-coded by source. \textit{Bottom:} the residuals from the best-fit quadratic model and the observed data.}
    \label{fig:pdot TTV diagram HATS-6b}
\end{figure*}

\begin{figure*}[!h]
    \centering
    \includegraphics[width=0.8\textwidth,height=0.30\textheight]{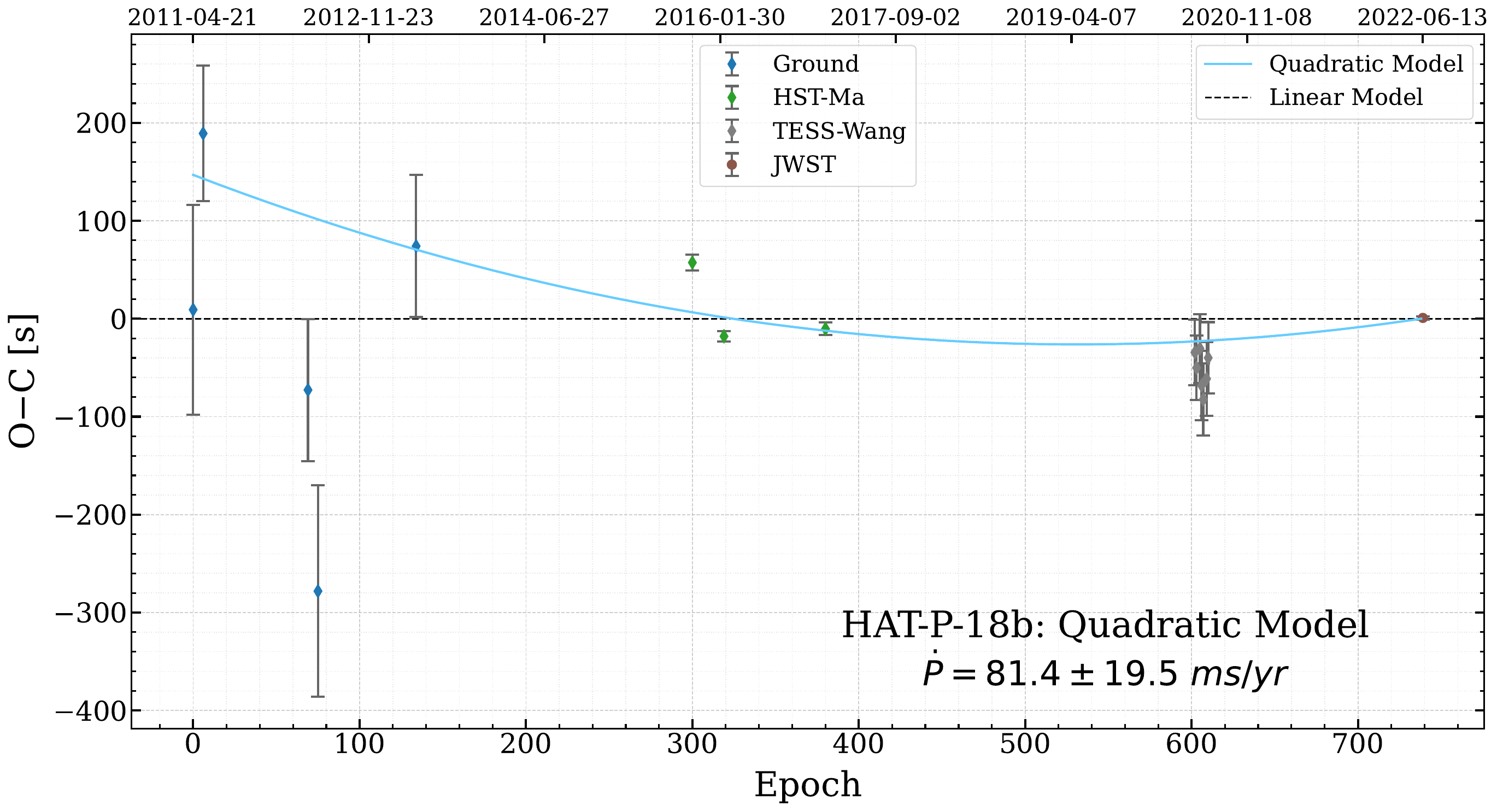}
    \includegraphics[width=0.8\textwidth,height=0.30\textheight]{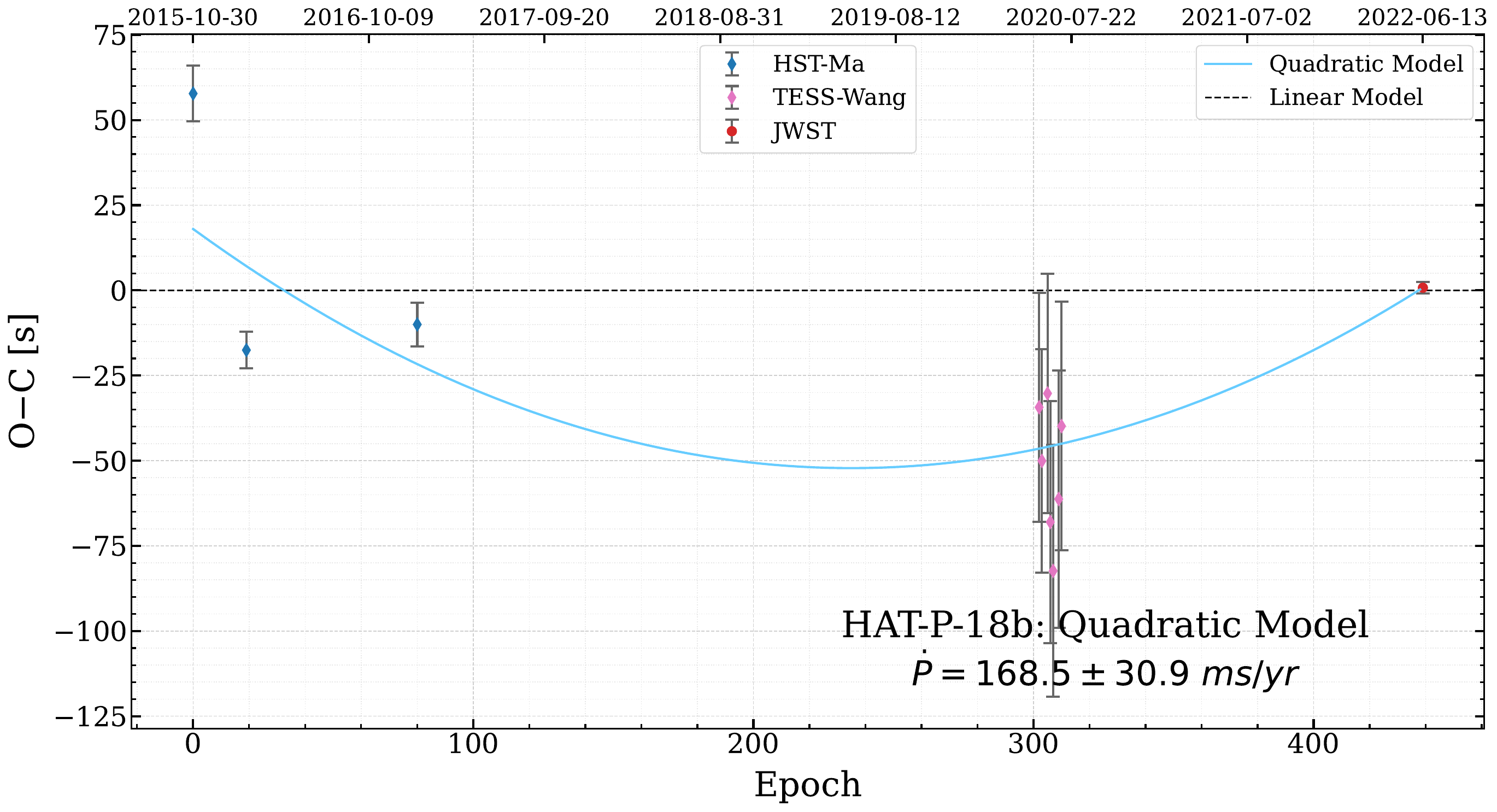}
    \caption{O--C diagrams for HAT-P-18~b. \textit{Top:} HAT-P-18~b using all available transit times. \textit{Bottom:} HAT-P-18~b after excluding ground-based transit times. In each panel, the gray dashed line denotes the best-fit linear ephemeris, the blue curve shows the fitted quadratic model, and the annotated value gives the measured $\dot{P}$. Data points are color-coded by source.}
    \label{fig:pdot TTV diagram HAT-P-18b}
\end{figure*}

Combining all three data configurations, we identify two quadratic model candidates: HATS-6~b and HAT-P-18~b. Their best-fit parameters are summarized in Table~\ref{tab:long-term ttv candidates}, and the corresponding O--C diagrams are shown in Figure~\ref{fig:pdot TTV diagram HATS-6b} and Figure~\ref{fig:pdot TTV diagram HAT-P-18b}, respectively. Figure \ref{fig:pdot TTV diagram HATS-6b} presents HATS-6~b, whose dataset is unchanged across all three configurations, while the top and bottom panels of Figure \ref{fig:pdot TTV diagram HAT-P-18b} show HAT-P-18~b using the full dataset and the dataset excluding ground-based transit times, respectively. We will discuss these two objects in Section~\ref{sec:individual}.

\begin{deluxetable*}{llccccccccccc}
\tablecaption{Long-term TTV Candidate Fitting Parameters \label{tab:long-term ttv candidates}}
\tabletypesize{\tiny}
\tablewidth{0pt}
\tablehead{
    \colhead{Planet} & 
    \colhead{Dataset\tablenotemark{a}} & 
    \colhead{$t_0$ [$BJD_{TDB}$]} & 
    \colhead{$1\sigma$} & 
    \colhead{$P$ [d]} & 
    \colhead{$1\sigma$} & 
    \colhead{$\dot{P}$} & 
    \colhead{$1\sigma$} & 
    \colhead{$N$} & 
    \colhead{BIC} & 
    \colhead{$\dot{P}/\sigma_{\dot{P}}$} & 
    \colhead{$\Delta$BIC}
}
\startdata
HATS-6~b & All & 2458469.3112296 & 1.865e-04 & 3.325260402 & 1.6081e-06 & 230.5 & 46.8 & 23 & 12.67 & 4.93 & 20.83 \\
\tableline
HAT-P-18~b & All & 2455673.4212636 & 4.170e-04 & 5.508022003 & 1.8207e-06 & 81.4 & 19.5 & 16 & 87.05 & 4.17 & 14.74 \\
 & No ground & 2457325.8286298 & 5.968e-05 & 5.508022655 & 1.2698e-06 & 168.5 & 30.9 & 11 & 57.30 & 5.46 & 27.55 \\
\enddata
\tablecomments{
\tablenotetext{a}{``All'' indicates the full dataset; ``No ground'' indicates the dataset with ground-based transit times excluded.}
}
\end{deluxetable*}

\subsection{Sine Model}
We also fitted a sinusoidal TTV model to all targets with at least five transit times using the same three data configurations. Periodic TTV candidates were identified by requiring both $\Delta \mathrm{BIC} > 10$ relative to the linear ephemeris model and an amplitude significance of $A/\sigma_A > 3$ in at least one configuration.

\begin{figure}[!h]
    \centering
    \includegraphics[width=0.48\textwidth]{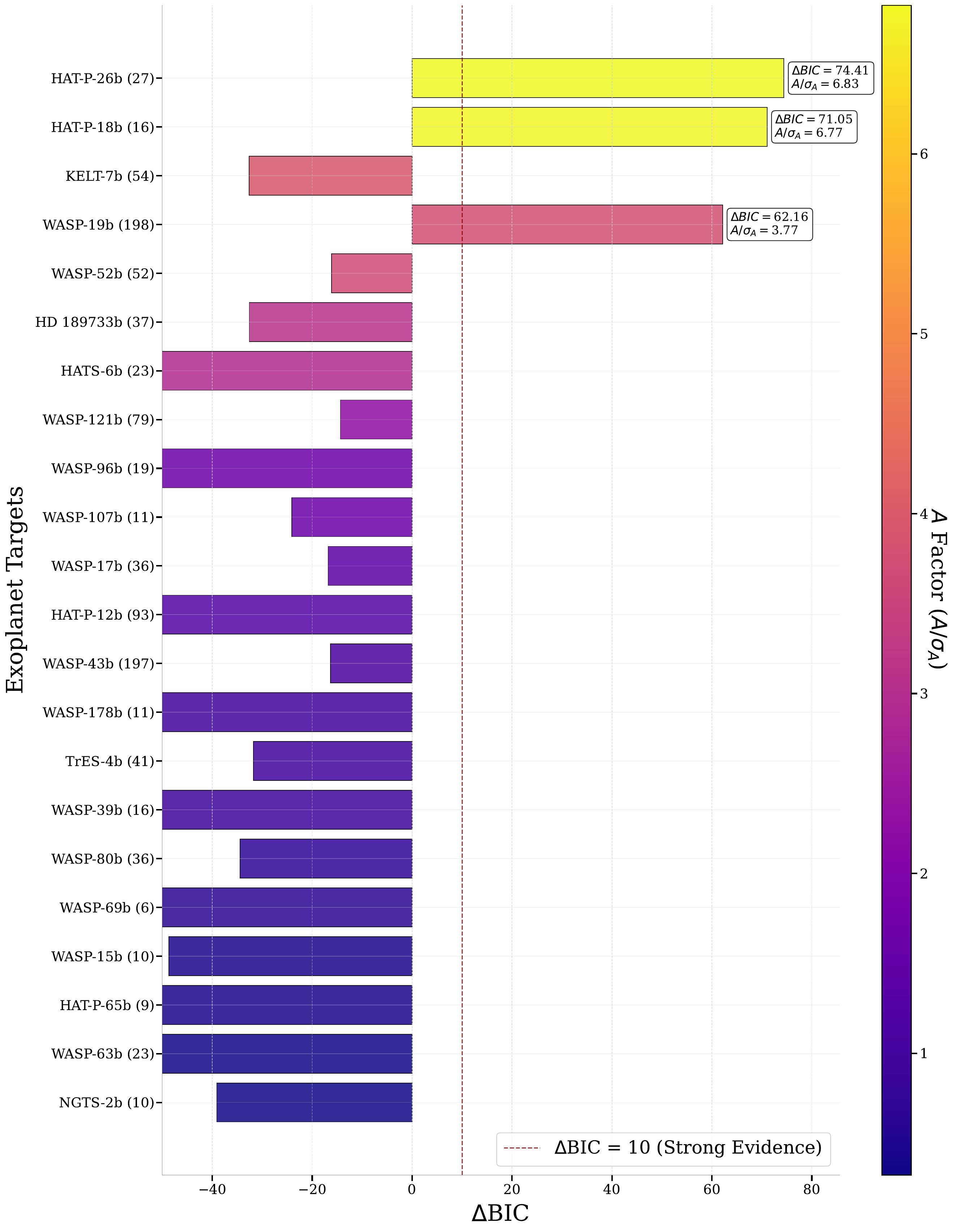}
    \caption{Short-term periodic TTV candidates from the full dataset. Comparison of the sine model against the linear model for targets with more than five transit times. The red dashed line marks $\Delta \mathrm{BIC} = 10$. The color bar encodes the amplitude significance $A/\sigma_{A}$. Targets satisfying $\Delta \mathrm{BIC} > 10$ and $A/\sigma_{A} > 3$ are annotated.  Targets with $\Delta \mathrm{BIC} < -50$ have been cut off for visibility.}
    \label{fig:sine candidates1}
\end{figure}

\begin{figure}[!h]
    \centering
    \includegraphics[width=0.48\textwidth]{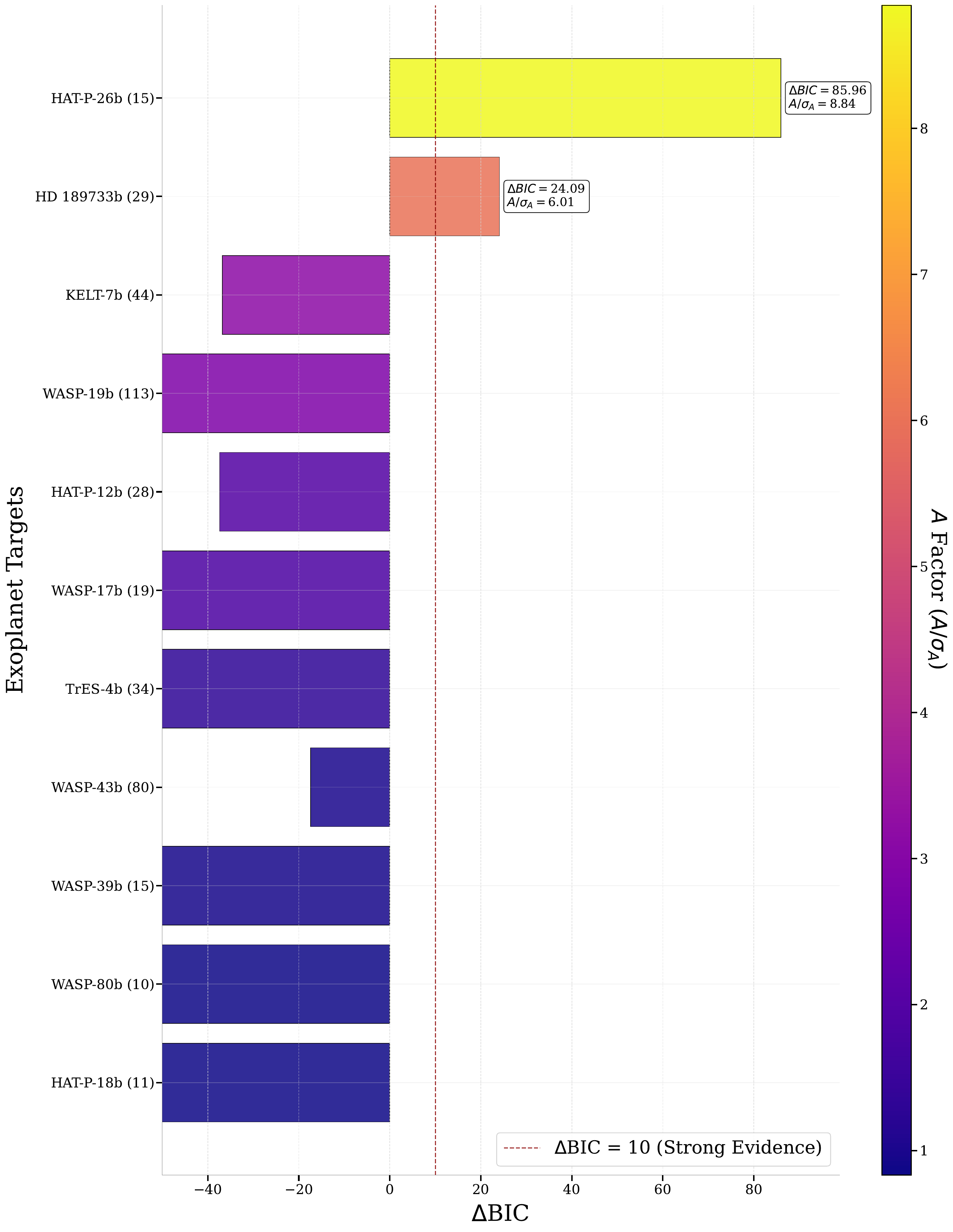}
    \caption{Same as Figure~\ref{fig:sine candidates1}, but after excluding all ground-based transit times. Targets whose datasets remain unchanged relative to Figure~\ref{fig:sine candidates1} are omitted to avoid redundancy.}
    \label{fig:sine candidates2}
\end{figure}

\begin{figure}[!h]
    \centering
    \includegraphics[width=0.48\textwidth]{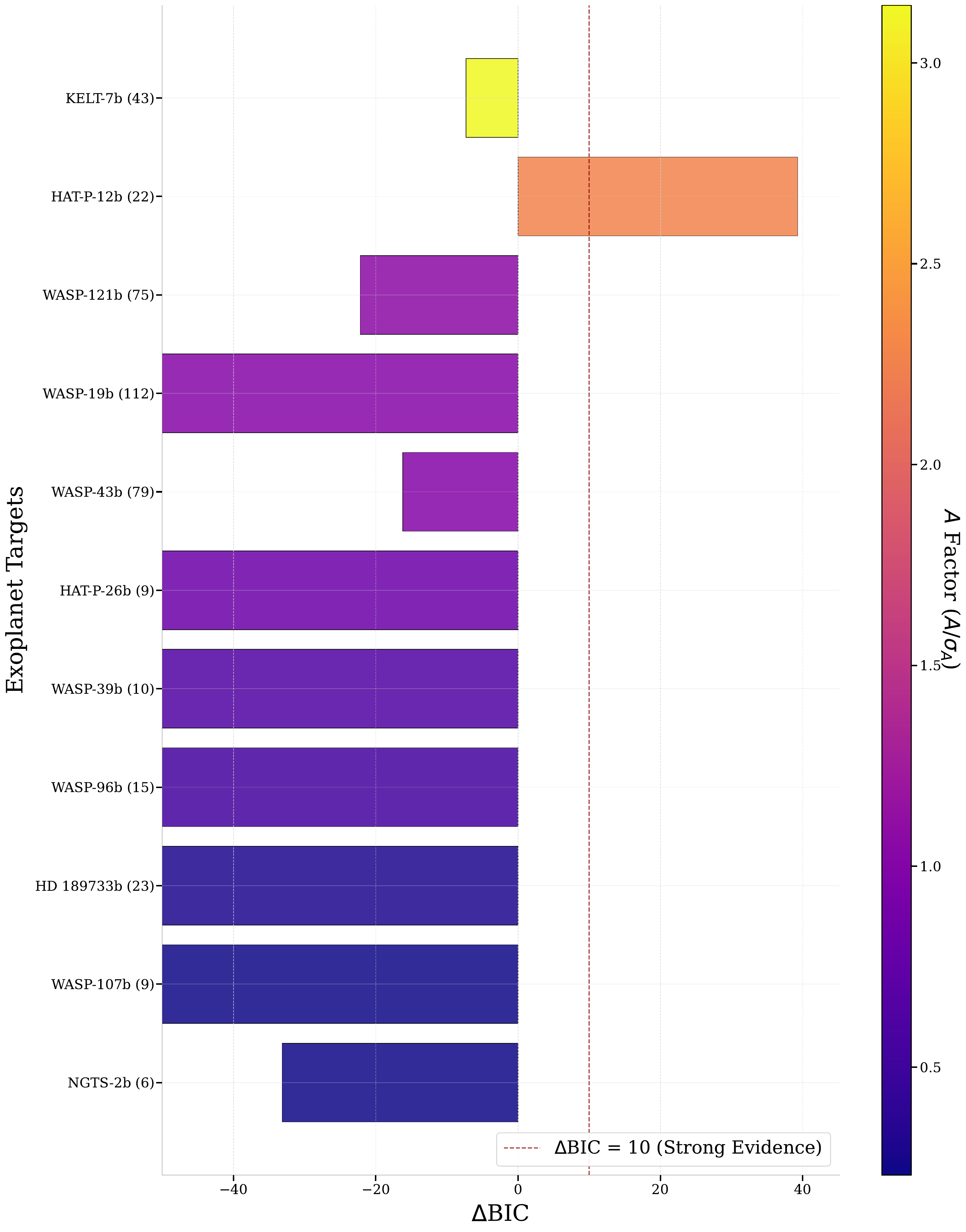}
    \caption{Same as Figure~\ref{fig:sine candidates2}, but after further removing all literature timings not from our own work. Targets whose datasets are identical to those in Figure~\ref{fig:sine candidates1} or Figure~\ref{fig:sine candidates2} are omitted.}
    \label{fig:sine candidates3}
\end{figure}

The results are presented in Figures~\ref{fig:sine candidates1}--\ref{fig:sine candidates3}. In the full dataset, HAT-P-26~b, HAT-P-18~b, and WASP-19~b satisfy the selection criteria. After excluding ground-based transit times, HAT-P-26~b and HD~189733~b remain significant among the targets with revised datasets. No targets satisfy the criteria when only transit times derived in this work are retained.

\begin{figure*}[!h]
    \centering
    \includegraphics[width=0.48\textwidth,height=0.23\textheight]{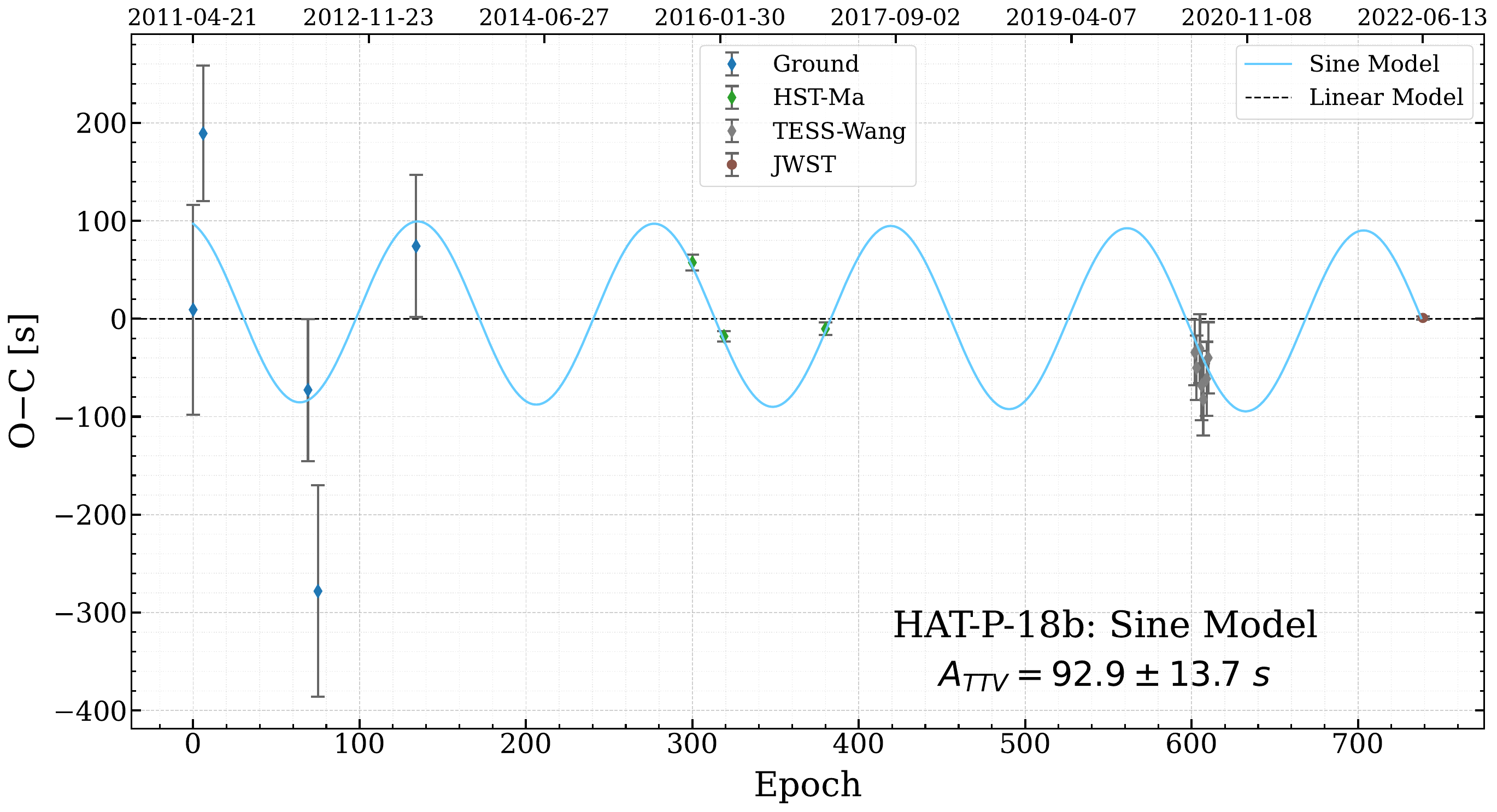}
    \includegraphics[width=0.48\textwidth,height=0.23\textheight]{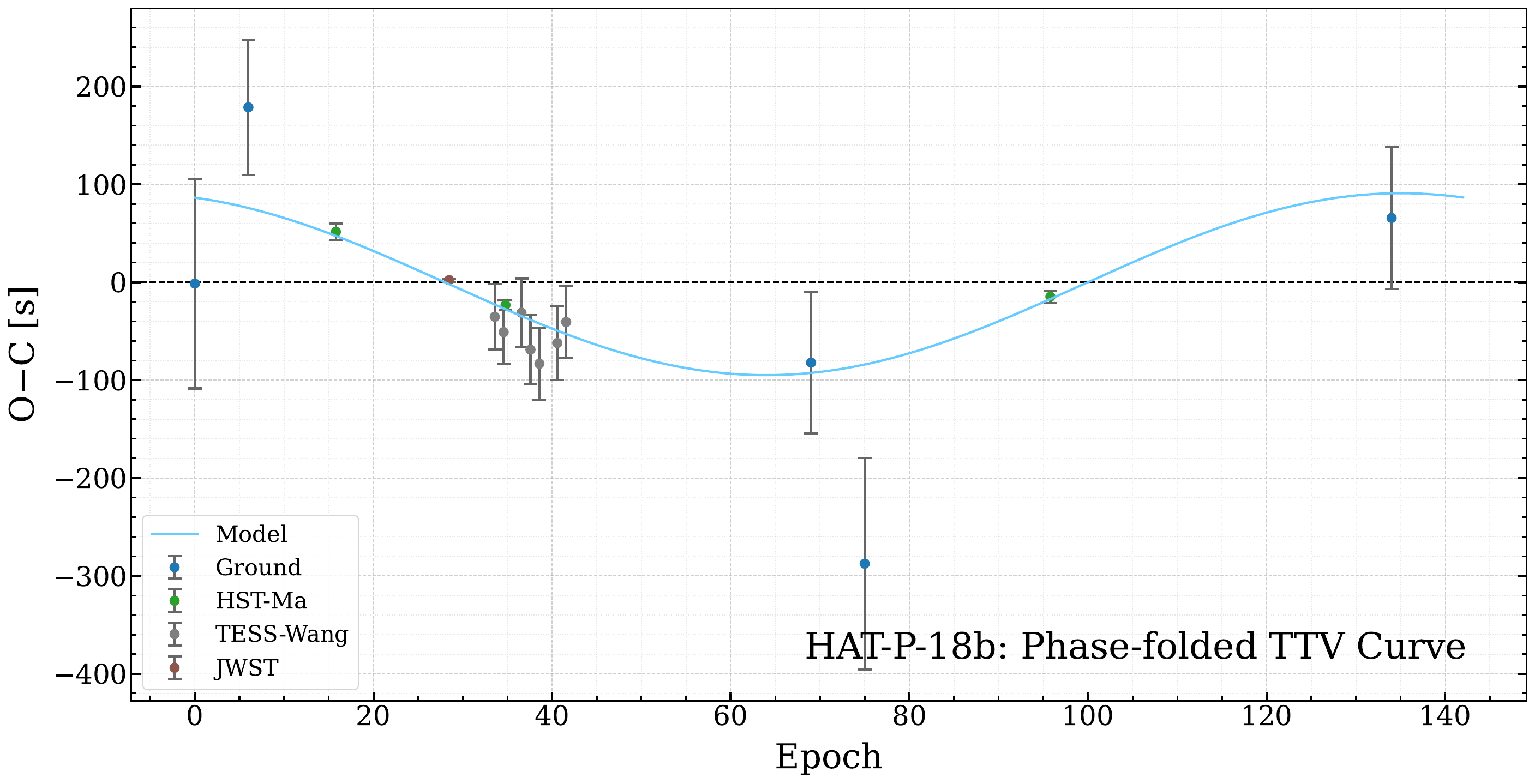}
    \includegraphics[width=0.48\textwidth,height=0.23\textheight]{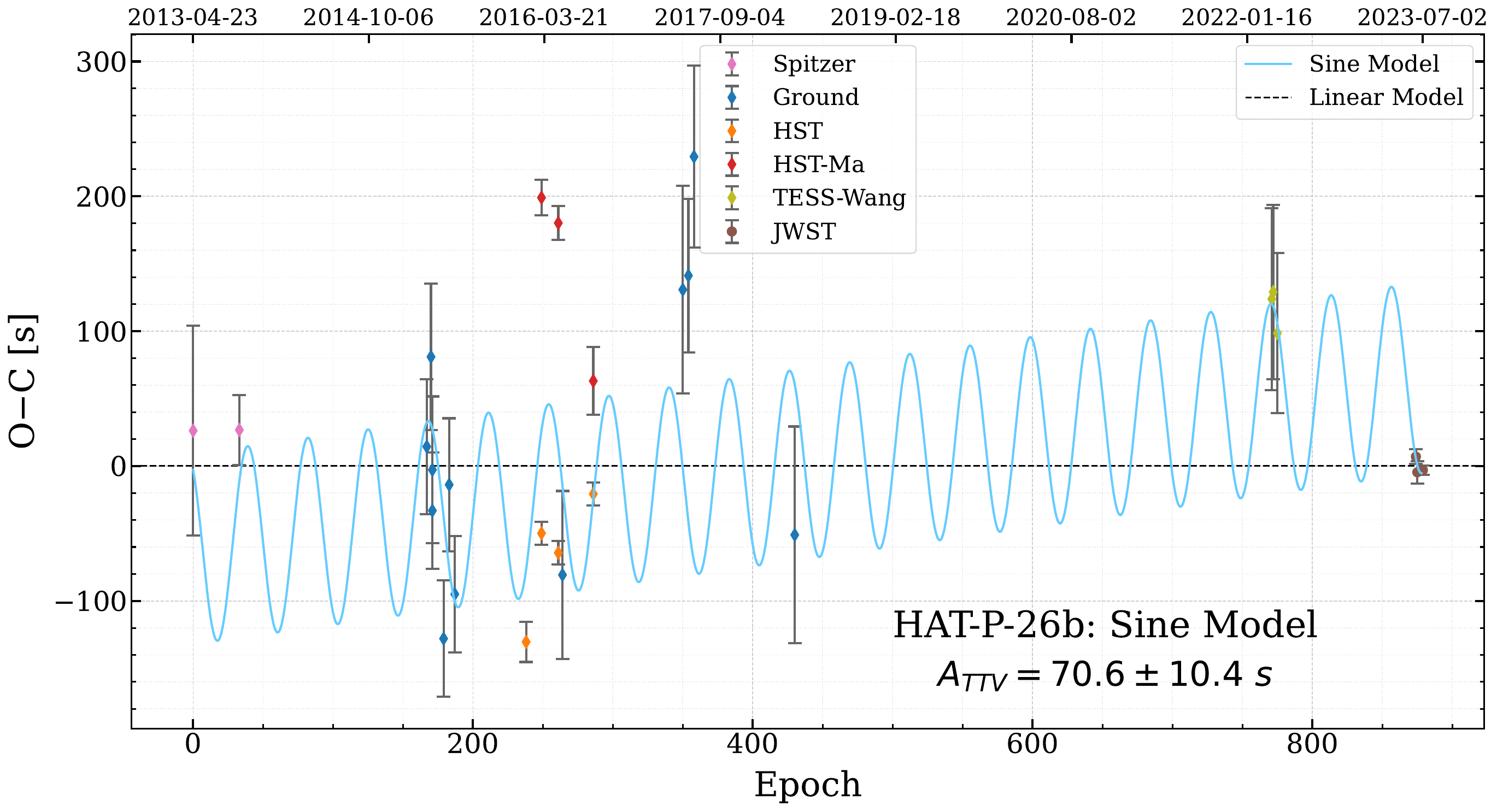}
    \includegraphics[width=0.48\textwidth,height=0.23\textheight]{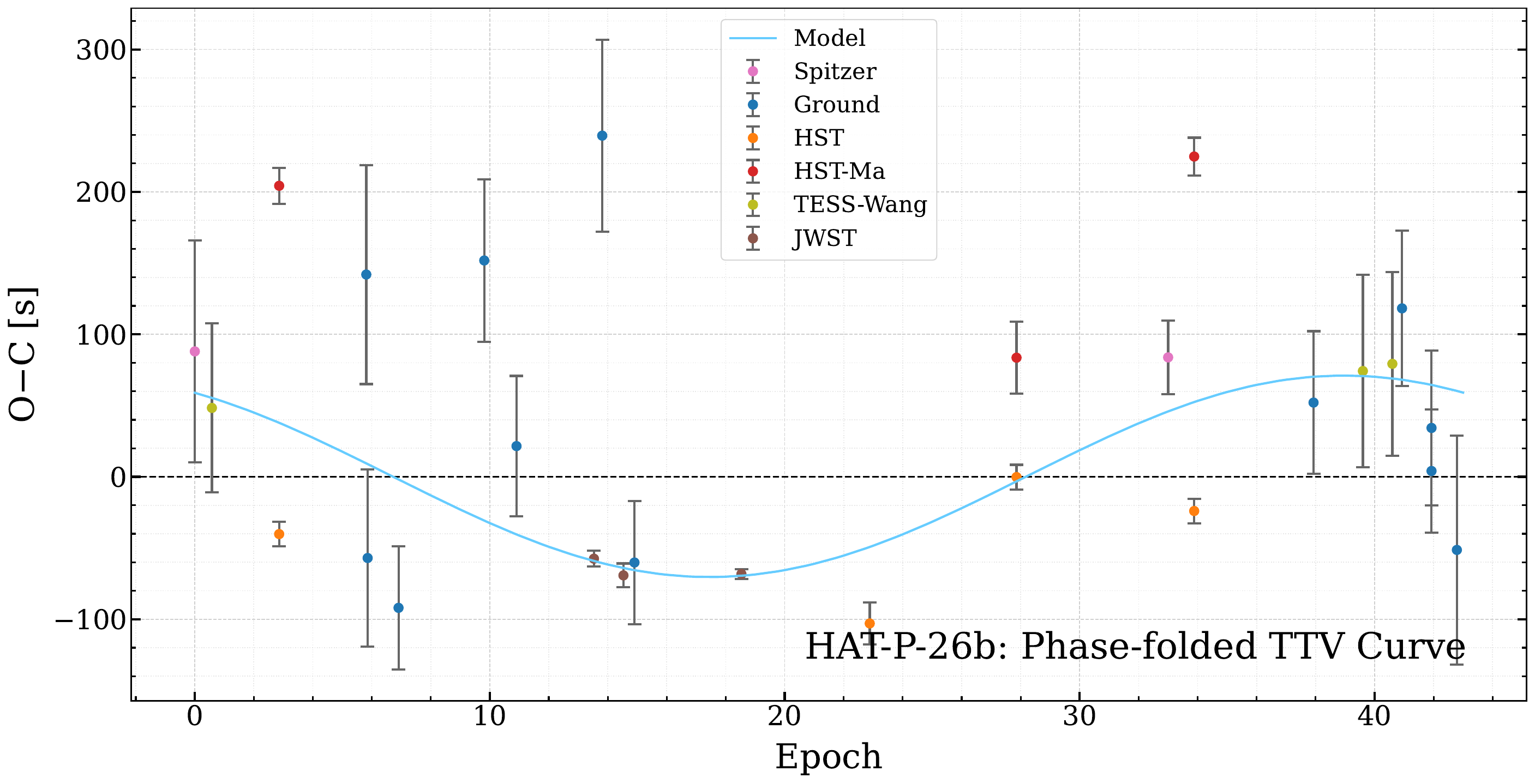}
    \includegraphics[width=0.48\textwidth,height=0.23\textheight]{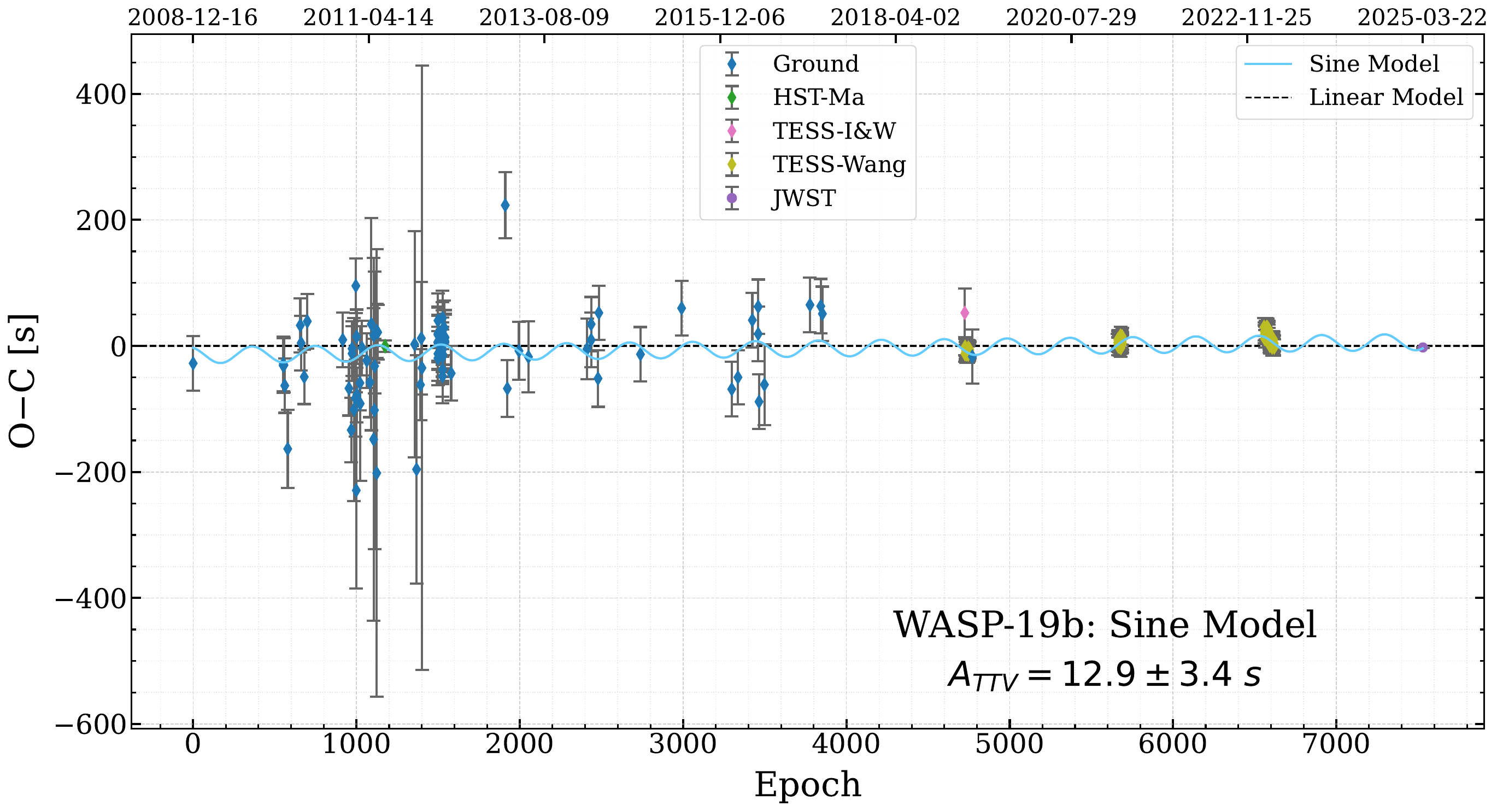}
    \includegraphics[width=0.48\textwidth,height=0.23\textheight]{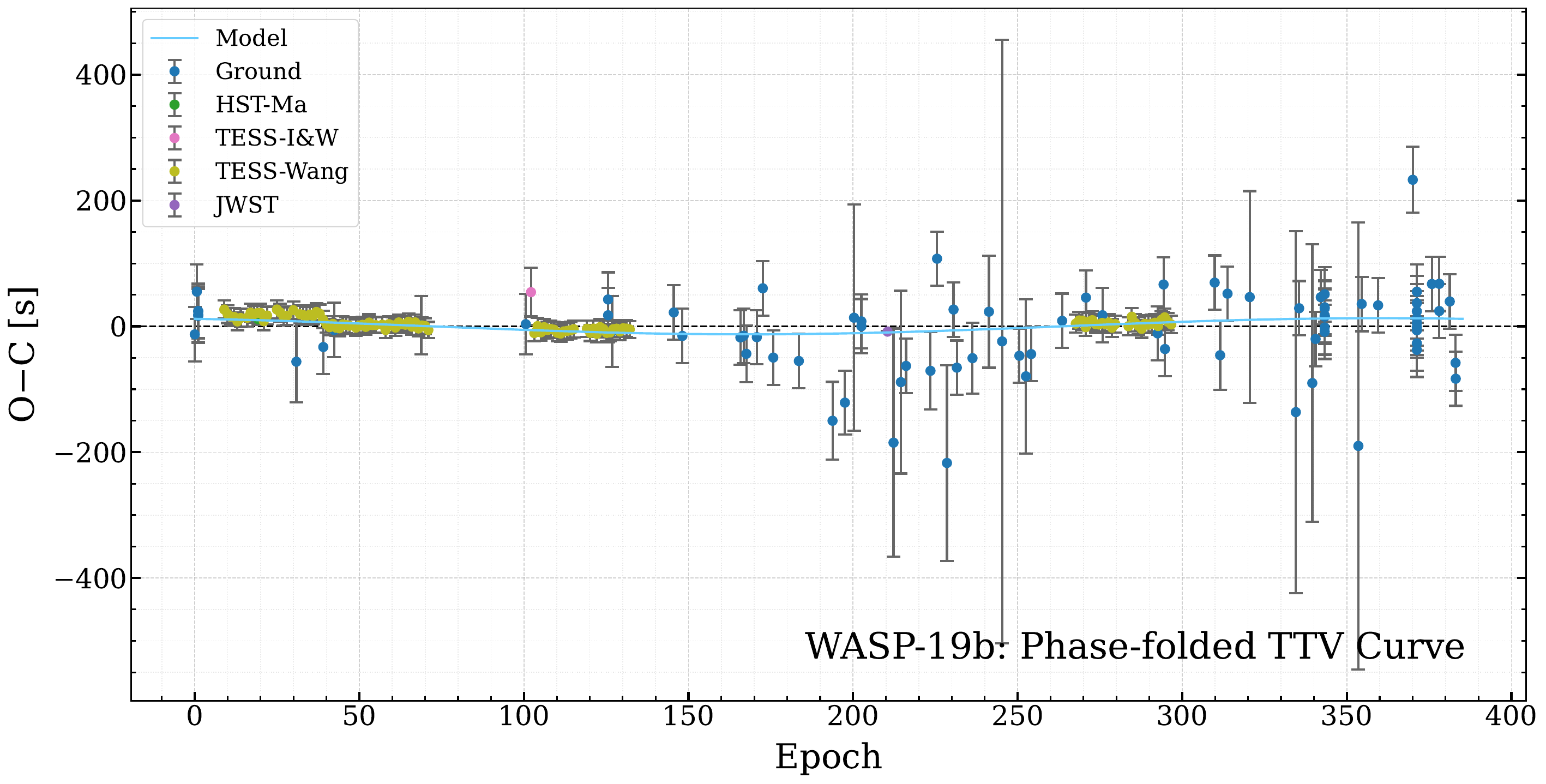}
    \caption{O--C diagrams (left panels) and phase-folded TTV curves (right panels) for the short-term periodic TTV candidates from the full dataset. The gray dashed line marks the best-fit linear ephemeris, and the blue line shows the sine model with the fitted amplitude $A_{\rm TTV}$ value annotated. Data points are color-coded by source. The linear trend in the O--C panels arises from the difference between the sine model reference epoch and the linear ephemeris given in Table~\ref{tab:linear ephemerides}.}
    \label{fig:sine TTV diagram1}
\end{figure*}

\begin{figure*}[!h]
    \centering
    \includegraphics[width=0.85\textwidth,height=0.45\textheight]{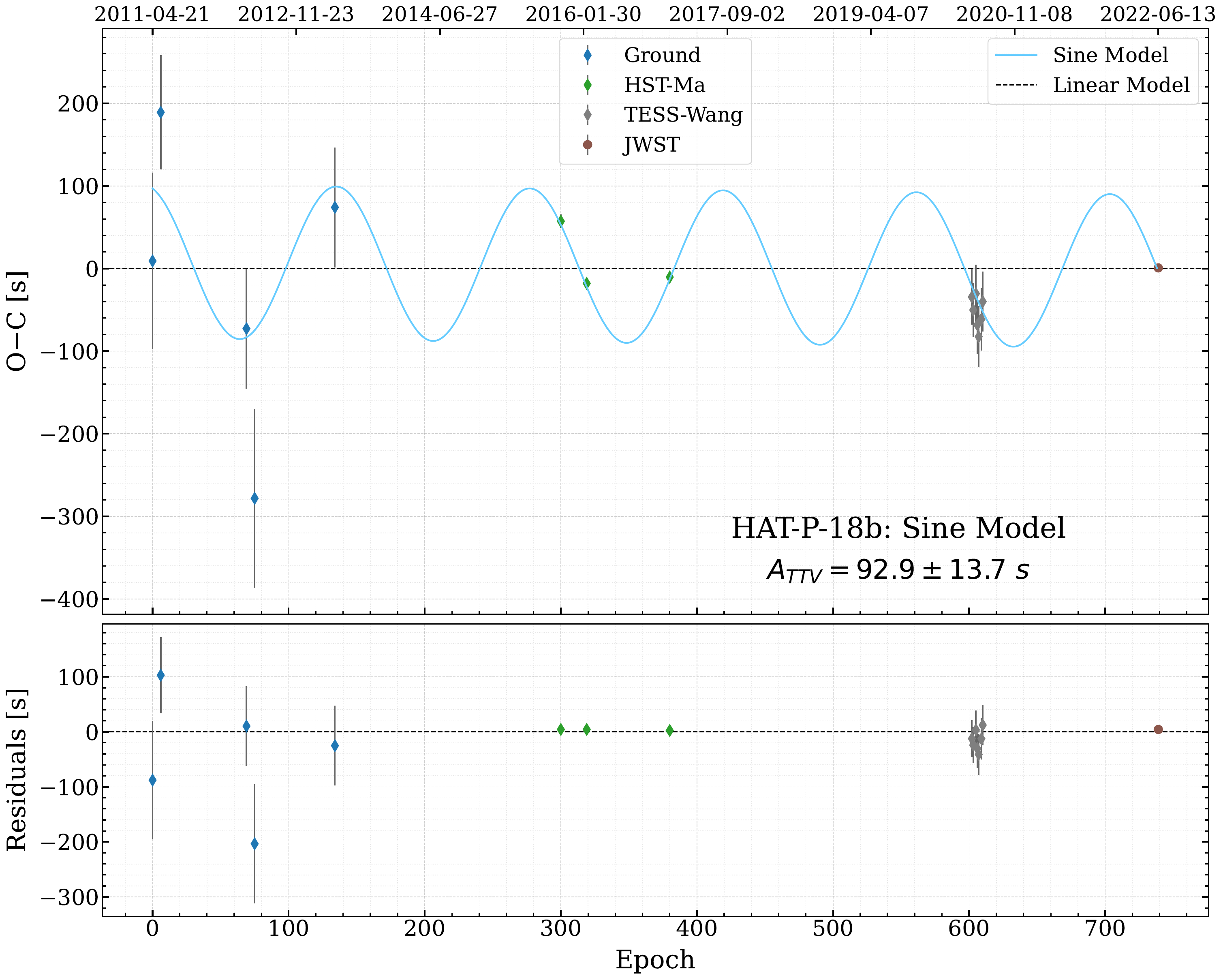}
    \caption{O--C diagram for HAT-P-18~b. \textit{Top:} the O-C diagram with the gray dashed line denotes the best-fit linear ephemeris, the blue curve shows the fitted sine model, and the annotated value gives the measured amplitude $A_{\rm TTV}$. Data points are color-coded by source. \textit{Bottom:} the residuals from the best-fit sine model and the observed data.}
    \label{fig:sine TTV diagram HAT-P-18b}
\end{figure*}

\begin{figure*}[!h]
    \centering
    \includegraphics[width=0.48\textwidth,height=0.23\textheight]{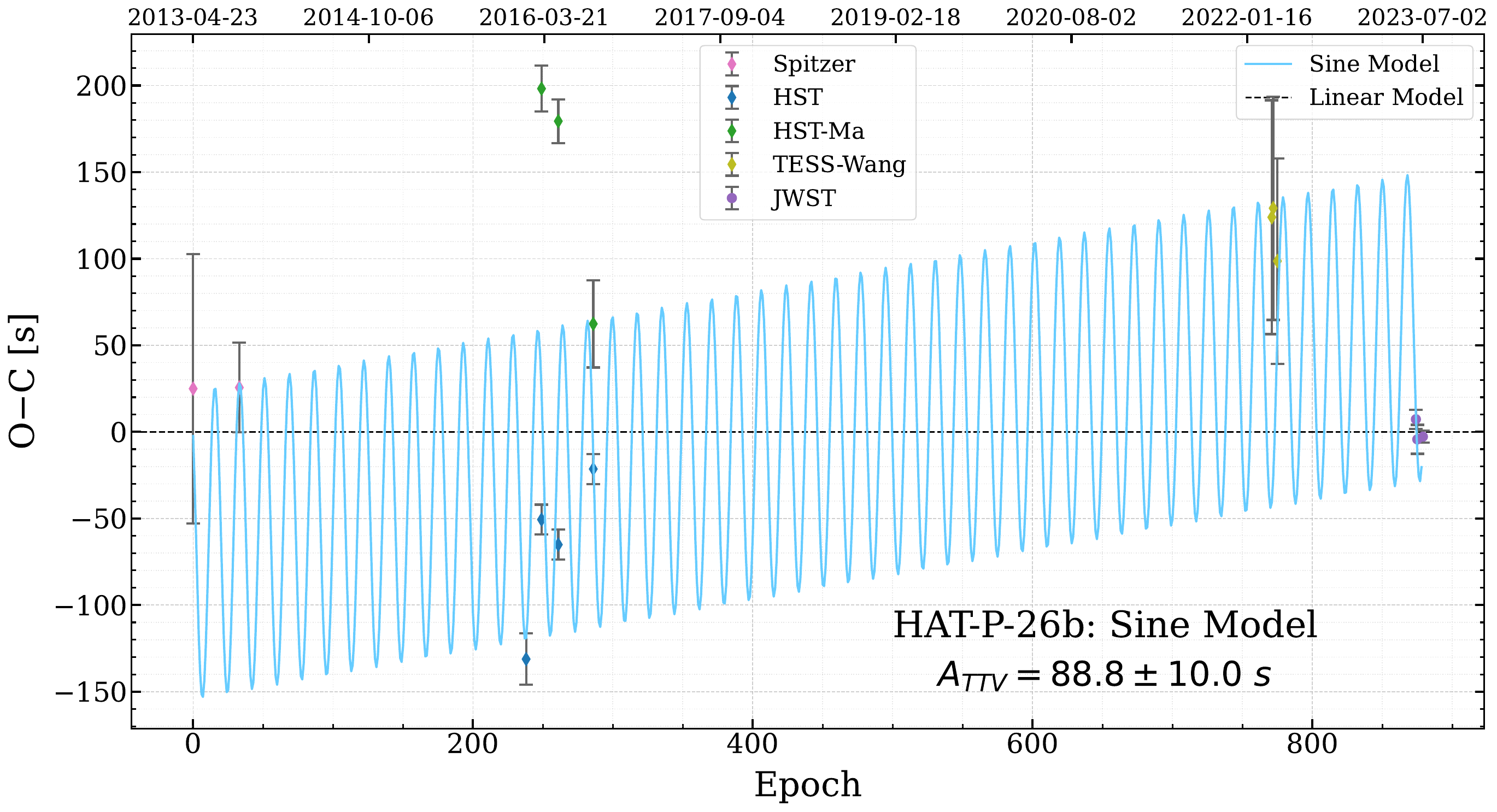}
    \includegraphics[width=0.48\textwidth,height=0.23\textheight]{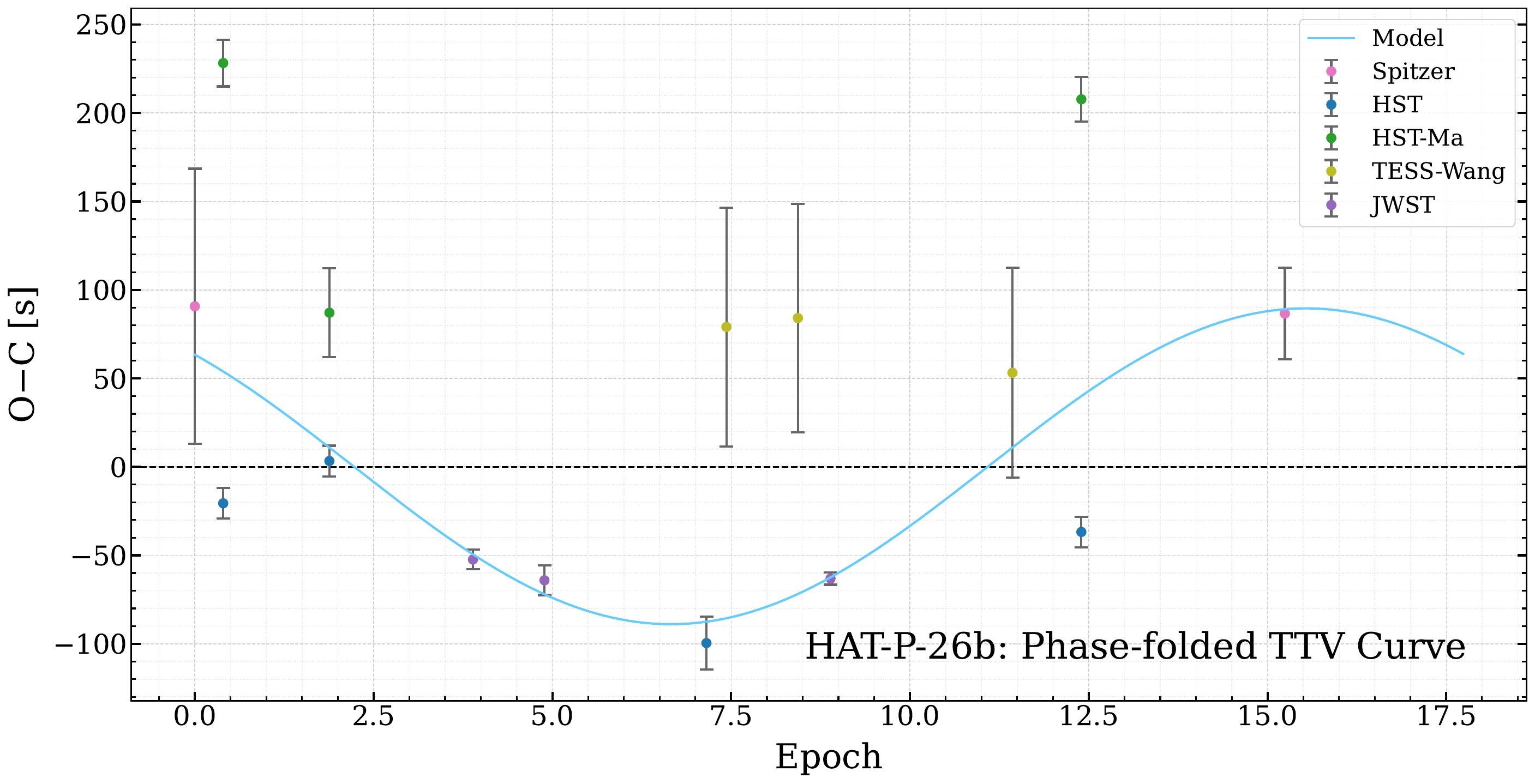}
    \includegraphics[width=0.48\textwidth,height=0.23\textheight]{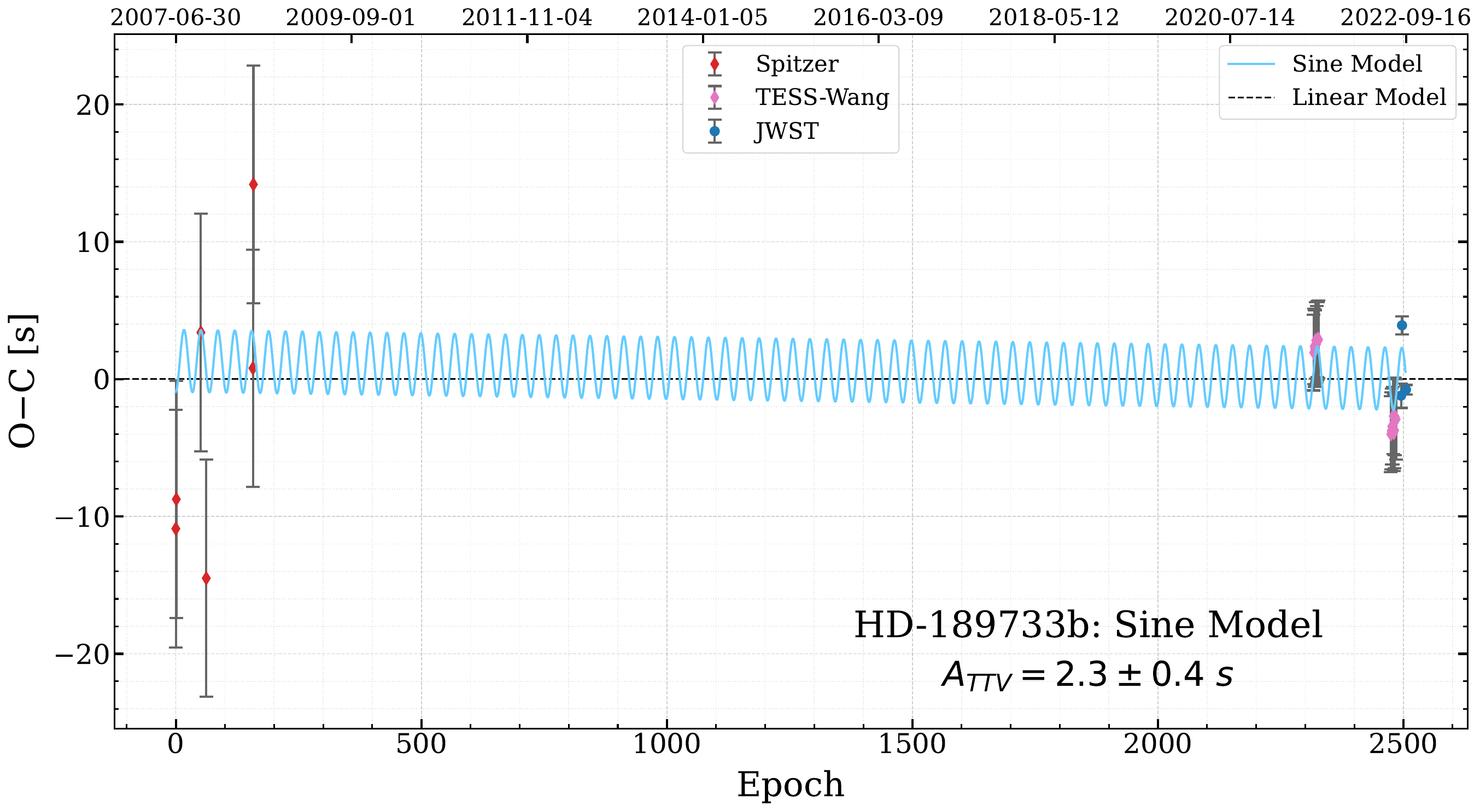}
    \includegraphics[width=0.48\textwidth,height=0.23\textheight]{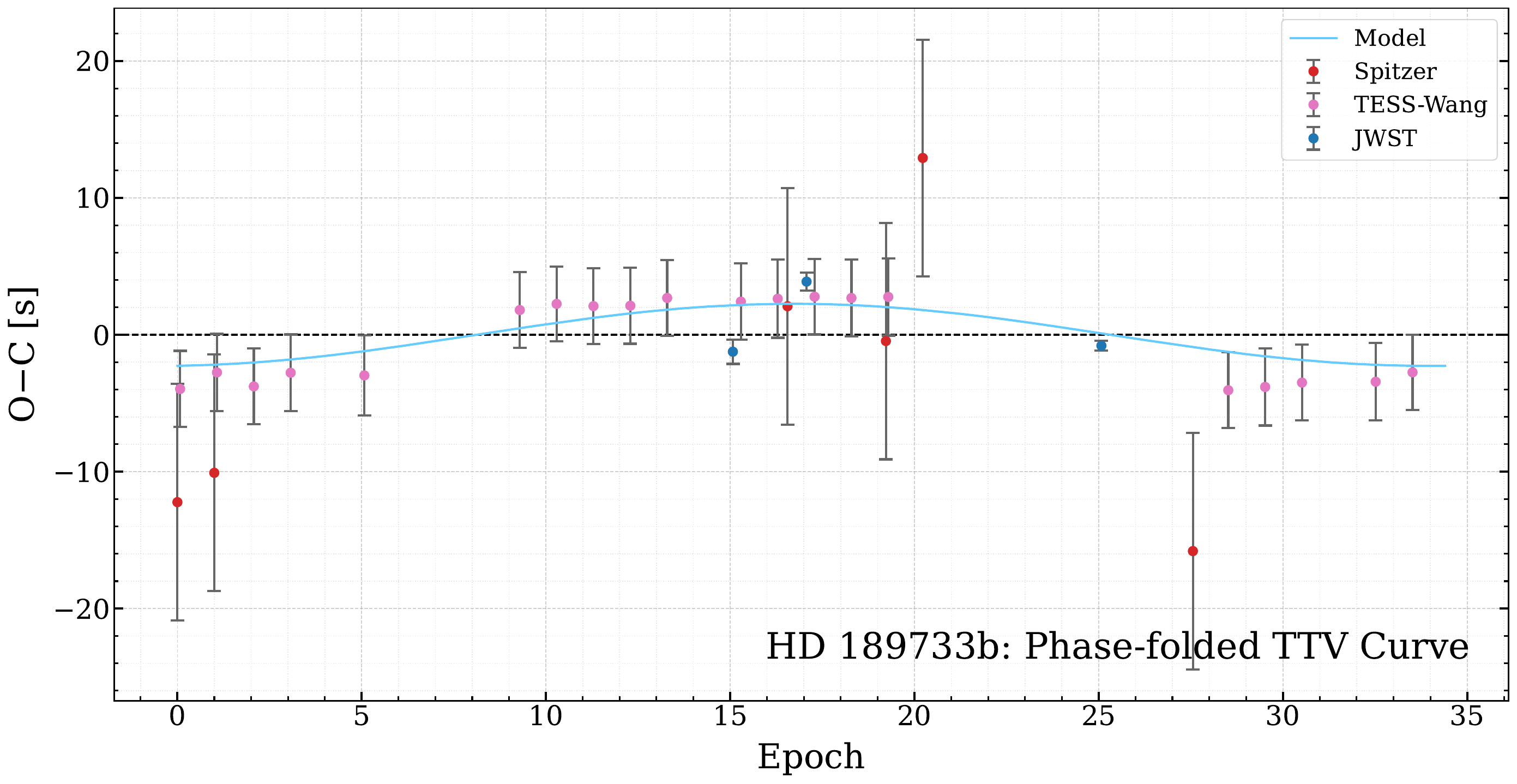}
    \caption{Same as Figure~\ref{fig:sine TTV diagram1}, but after excluding ground-based transit times. \textit{Top:} HAT-P-26~b. \textit{Bottom:} HD~189733~b.}
    \label{fig:sine TTV diagram2}
\end{figure*}

The O--C diagrams and phase-folded TTV curves for the candidates identified from the first two configurations are shown in Figures~\ref{fig:sine TTV diagram1}, \ref{fig:sine TTV diagram HAT-P-18b} and \ref{fig:sine TTV diagram2}. The apparent linear trends visible in some O--C diagrams arise because the reference epoch and orbital period inferred from the sinusoidal fit differ slightly from those obtained from the linear ephemeris model listed in Table~\ref{tab:linear ephemerides}. These offsets disappear in the phase-folded representation.

Combining the results from all three configurations, we identify four sine model candidates: HAT-P-26~b, HAT-P-18~b, WASP-19~b, and HD~189733~b. Their best-fit posterior parameters are summarized in Table~\ref{tab:sine candidates}. Individual systems are discussed in the following section.

\begin{figure*}[!h]
    \centering
    \includegraphics[width=0.48\textwidth,height=0.21\textheight]{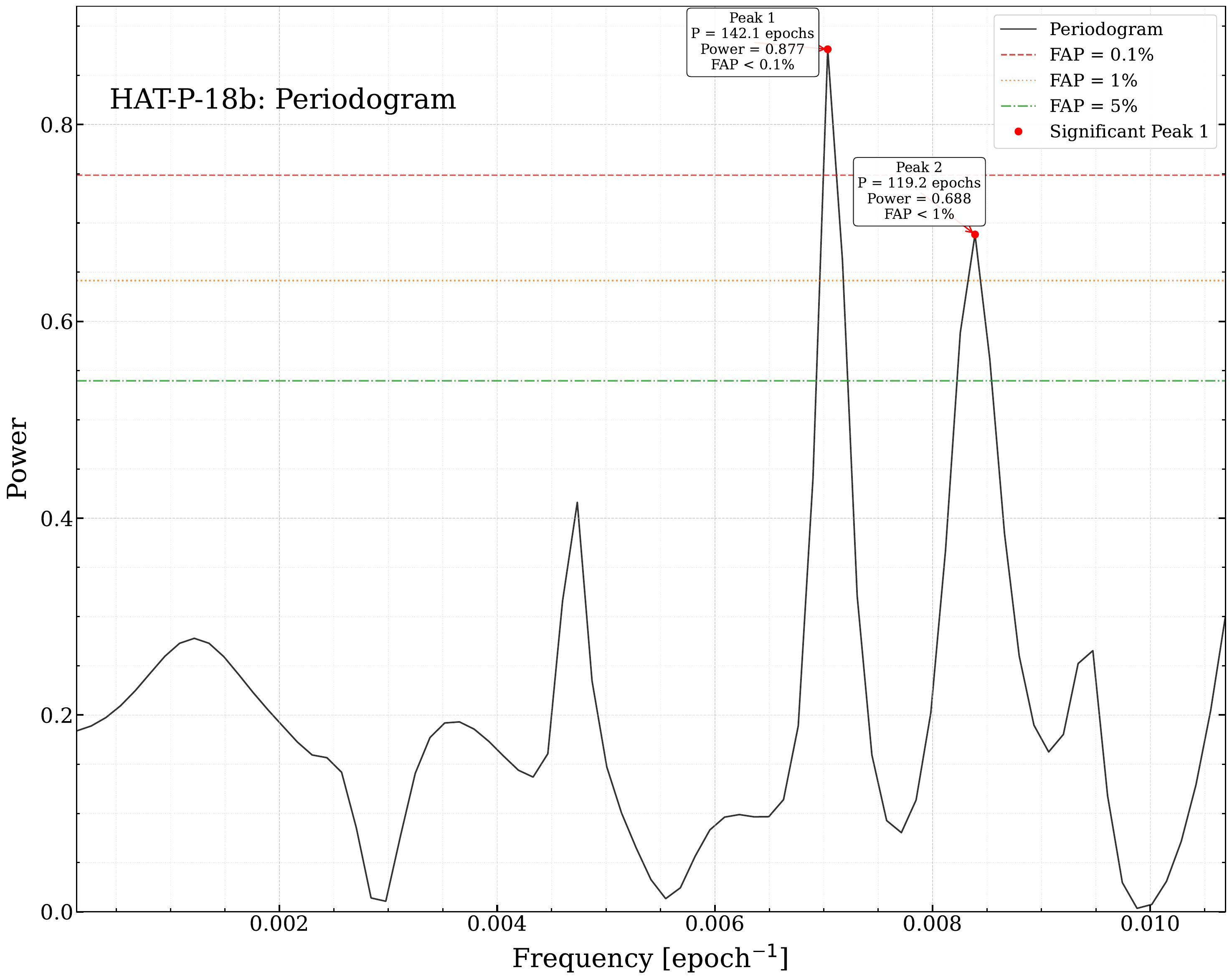}
    \includegraphics[width=0.48\textwidth,height=0.21\textheight]{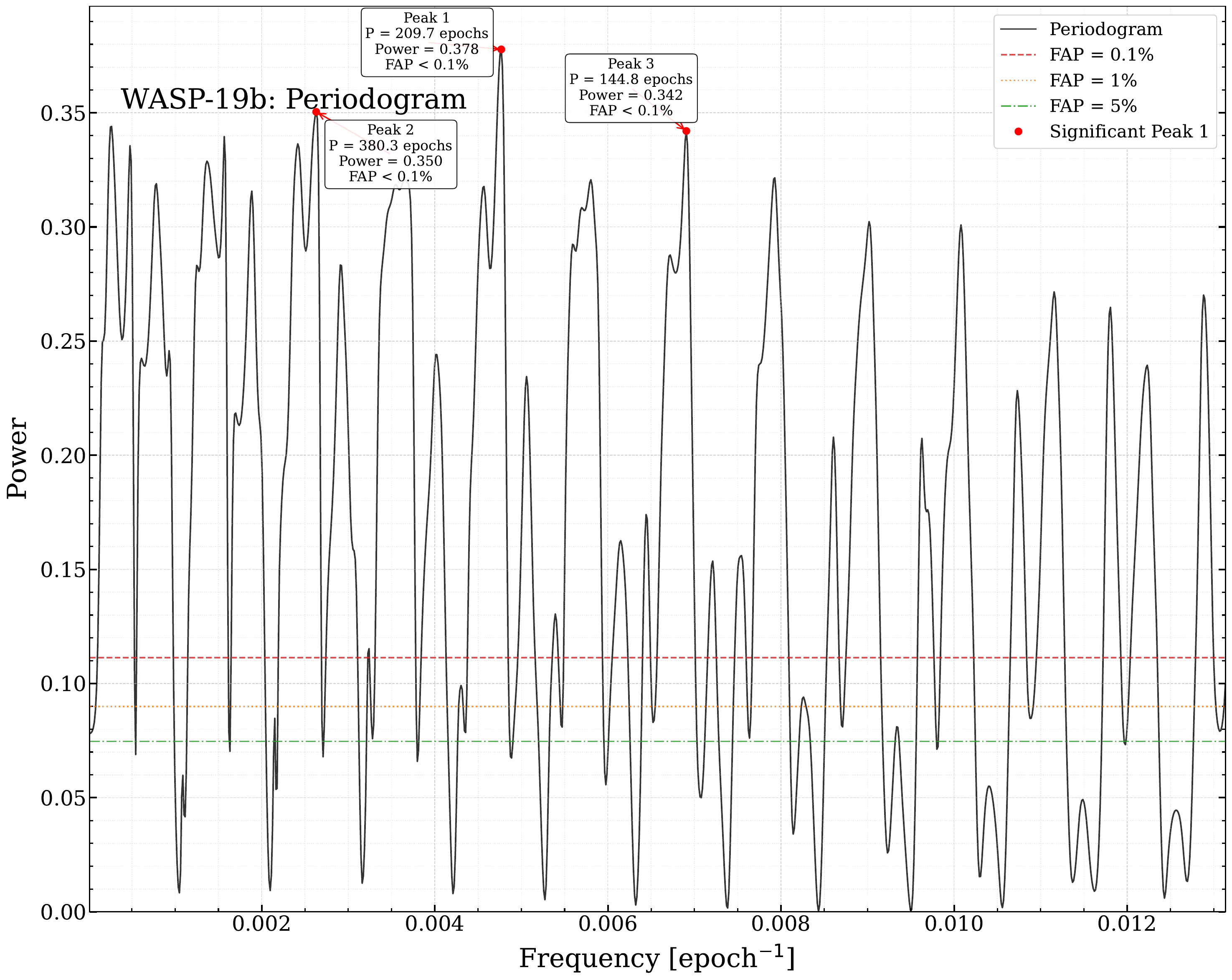}
    \includegraphics[width=0.48\textwidth,height=0.21\textheight]{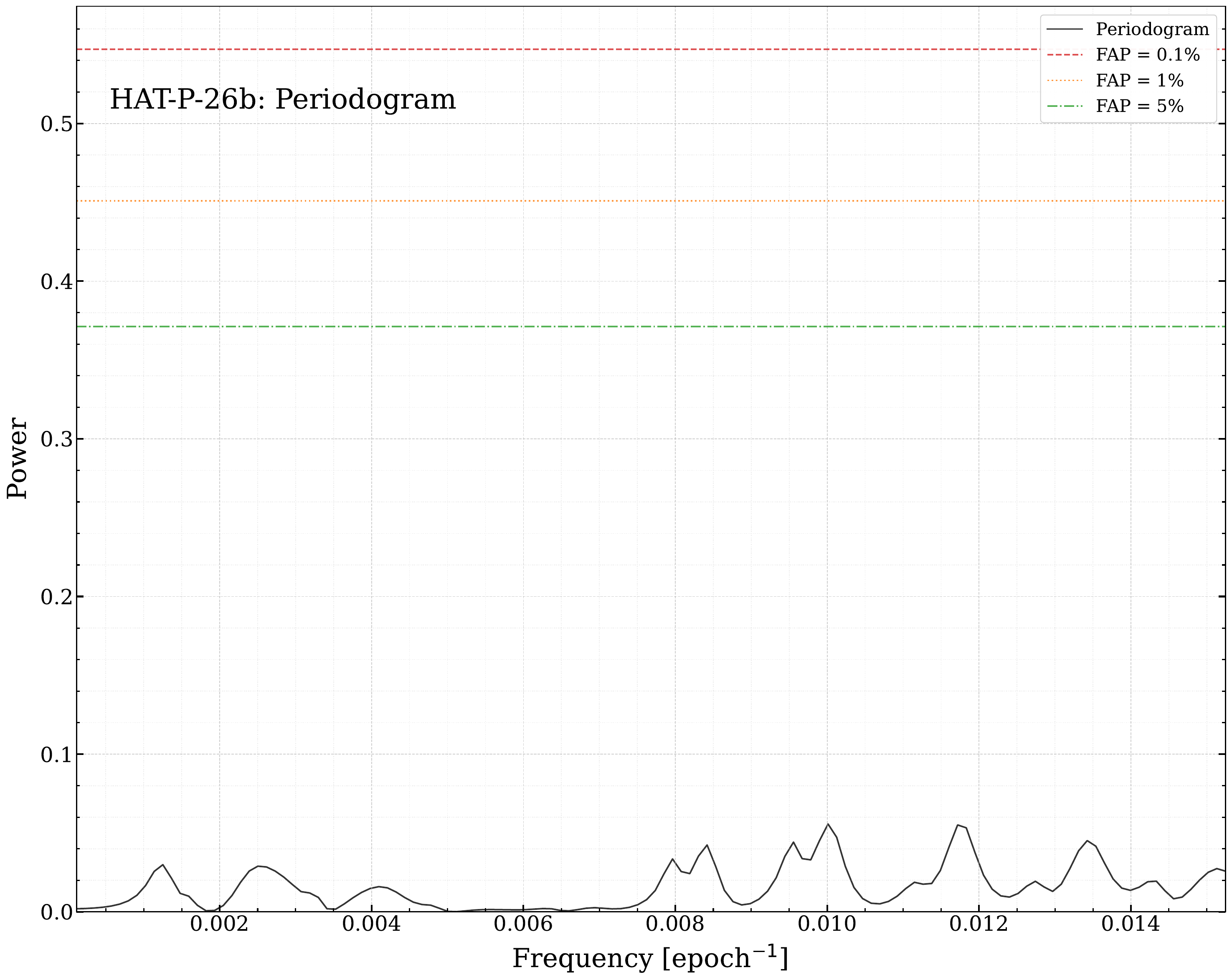}
    \includegraphics[width=0.48\textwidth,height=0.21\textheight]{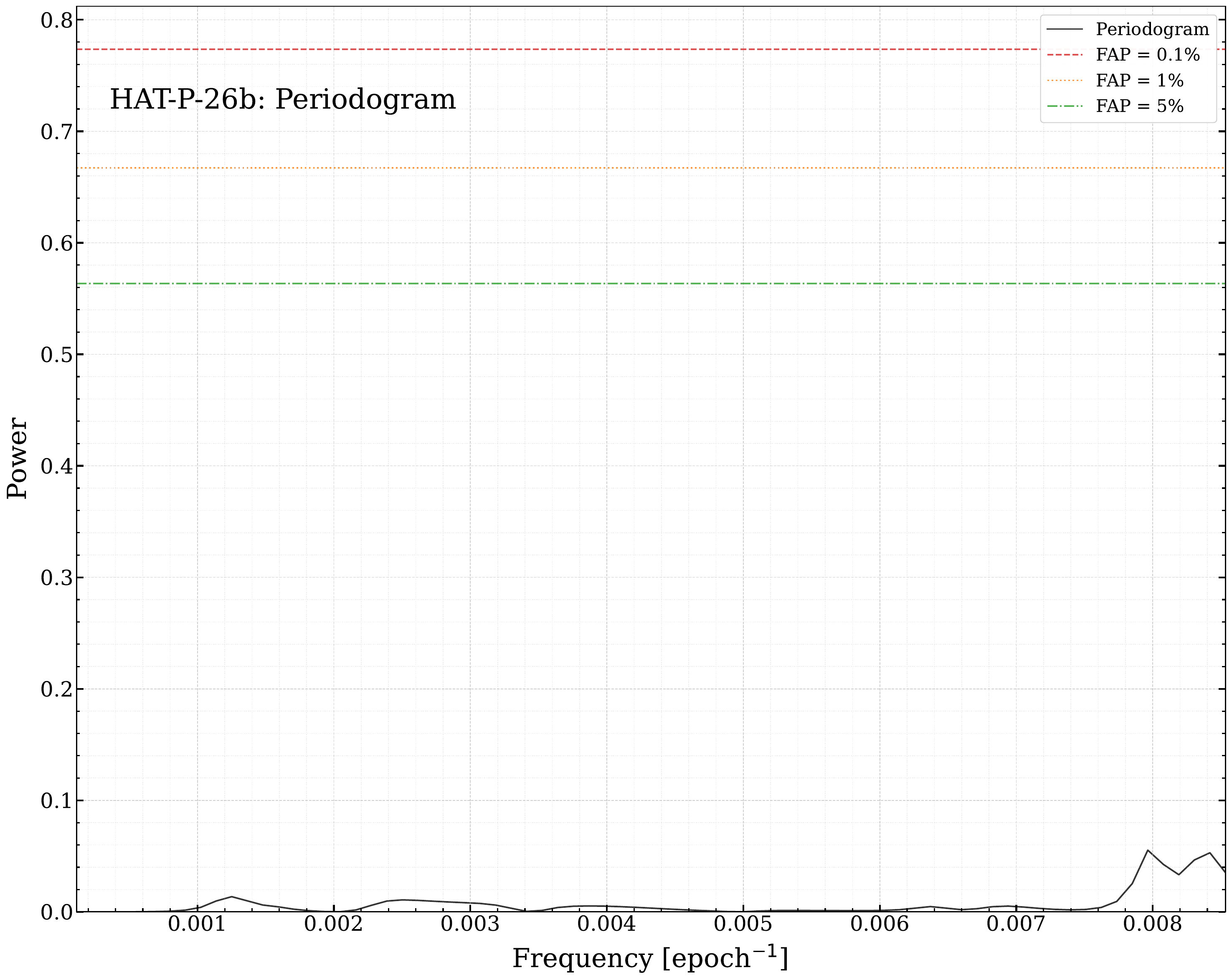}
    \includegraphics[width=0.48\textwidth,height=0.21\textheight]{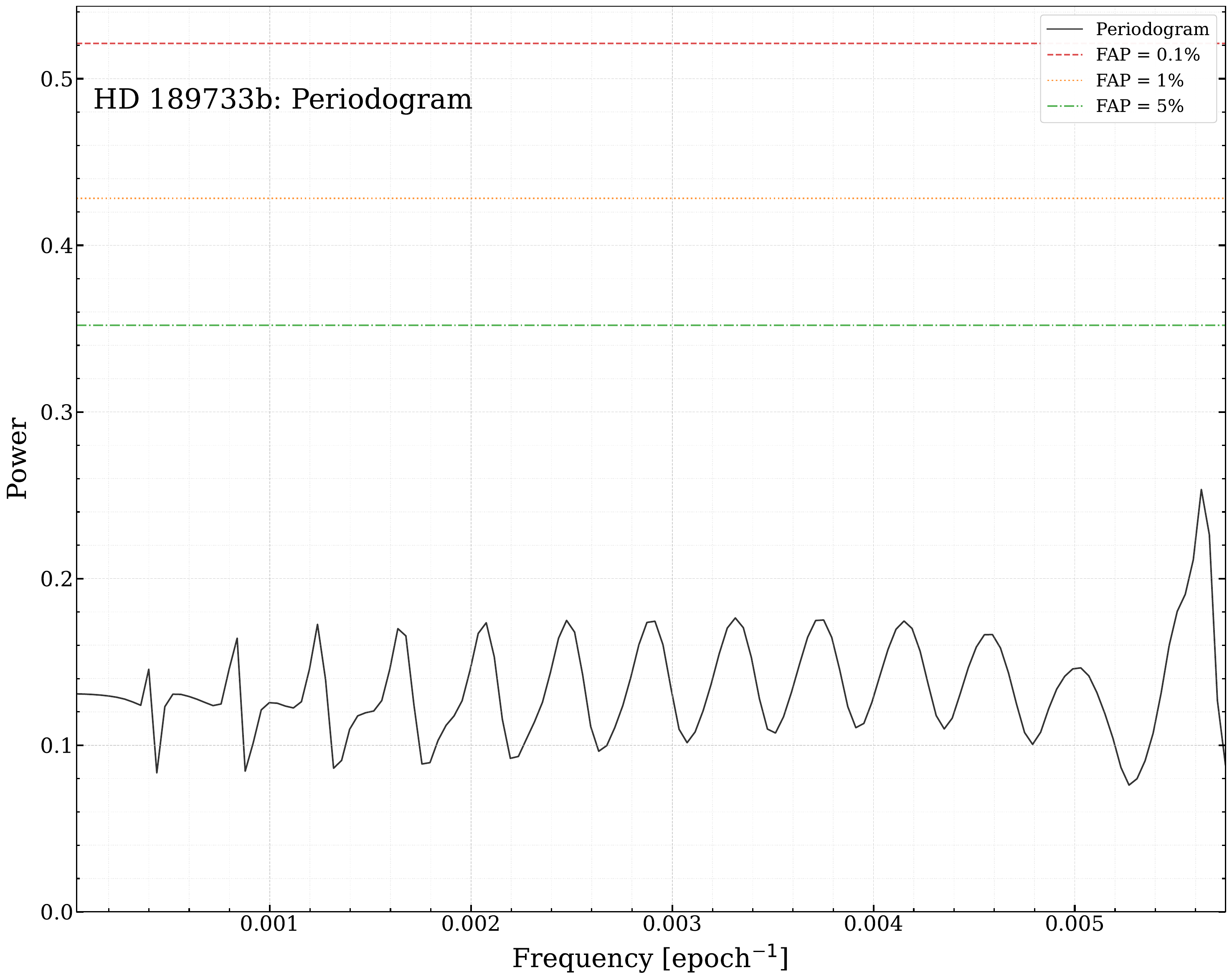}
    \caption{GLS periodograms for sine-signal candidates. \textit{Top row:} HAT-P-18~b and WASP-19~b (full dataset). \textit{Middle row:} HAT-P-26~b with (left) and without (right) ground-based data. \textit{Bottom:} HD~189733~b after excluding ground-based data. Colored dashed lines indicate false-alarm probability (FAP) thresholds, and prominent peaks are annotated.}
    \label{fig:sine periodogram}
\end{figure*}

\begin{deluxetable*}{llcccccccccccccc}
\tablecaption{Fitting Parameters for Short-term Periodic TTV Candidates\label{tab:sine candidates}}
\tabletypesize{\tiny}
\tablewidth{0pt}
\tablehead{
    \colhead{Planet} & 
    \colhead{Dataset\tablenotemark{a}} & 
    \colhead{$t_0$ [$BJD_{TDB}$]} & 
    \colhead{$1\sigma$} & 
    \colhead{$P$ [d]} & 
    \colhead{$1\sigma$} & 
    \colhead{$P_{\rm TTV}$ [epoch]} & 
    \colhead{$1\sigma$} & 
    \colhead{$\phi$ [rad]} & 
    \colhead{$1\sigma$} & 
    \colhead{$A_{\rm TTV}$ [s]} & 
    \colhead{$1\sigma$} & 
    \colhead{$N$} & 
    \colhead{BIC} & 
    \colhead{$A/\sigma_A$} & 
    \colhead{$\Delta$BIC}
}
\startdata
HAT-P-18~b & All & 2455673.4206859 & 1.672e-04 & 5.508029356 & 6.2863e-07 & 142.10448 & 2.45529 & 4.378 & 0.270 & 92.9 & 13.7 & 16 & 30.73 & 6.77 & 71.05 \\
\tableline
HAT-P-26~b & All & 2456405.6233637 & 9.801e-05 & 4.234501472 & 2.8358e-07 & 43.02339 & 0.07073 & 4.046 & 0.118 & 70.6 & 10.4 & 27 & 594.35 & 6.83 & 74.41 \\
 & No ground & 2456405.6233834 & 8.234e-05 & 4.234501441 & 2.1008e-07 & 17.75737 & 0.00589 & 3.749 & 0.082 & 88.8 & 10.0 & 15 & 542.30 & 8.84 & 85.96 \\
\tableline
WASP-19~b & All & 2454817.1466156 & 5.059e-05 & 0.788839015 & 6.8407e-09 & 385.23741 & 3.61975 & 4.298 & 0.901 & 12.9 & 3.4 & 198 & 139.46 & 3.77 & 62.16 \\
\tableline
HD 189733~b & No ground & 2454281.6554453 & 6.123e-06 & 2.218575183 & 1.1233e-08 & 34.44343 & 5.22806 & 1.387 & 1.916 & 2.3 & 0.4 & 29 & 55.29 & 6.01 & 24.09 \\
\enddata
\tablecomments{
\tablenotetext{a}{``All'' indicates the full dataset; ``No ground'' indicates the dataset with ground-based transit times excluded.}
}
\end{deluxetable*}

\subsection{Individual Object \label{sec:individual}}

We identified five systems that satisfy our statistical significance thresholds: HAT-P-18~b for both the quadratic and sine models; HAT-P-26~b, WASP-19~b, and HD~189733~b for the sine model; and HATS-6~b for the quadratic model.
In the following sections, we examine each system individually to evaluate the robustness of these signals and the extent to which they depend on the adopted timing dataset.

\subsubsection{HATS-6~b}

HATS-6~b is a hot Jupiter ($M_p = 0.319\,M_\mathrm{J}$, $R_p = 0.998\,R_\mathrm{J}$, $P = 3.3253$~days) transiting an early M dwarf star \citep{hartman_hats-6b_2015}. 
The TTV dataset for this target consists exclusively of transit mid-times from our own reduction of TESS and JWST observations. 
Our analysis identifies HATS-6~b as a promising long-term TTV candidate, with the quadratic ephemeris model favored over a linear model by $\Delta \mathrm{BIC}=20.83$ and yielding a period derivative significance of $\dot{P}/\sigma_{\dot{P}}=4.93$. 
The best-fitting model gives a period derivative of $\dot{P}=230.5\pm46.8~\mathrm{ms\,yr^{-1}}$.

As illustrated in Figure~\ref{fig:pdot TTV diagram HATS-6b}, the O--C diagram exhibits a coherent long-term trend that is well described by a quadratic ephemeris model. Based on these results, we classify HATS-6~b as a candidate system exhibiting a long-term TTV signal. 
Because this detection relies entirely on the newly measured transit times presented in this work, independent confirmation with future high-precision transit observations will be crucial for verifying the persistence of this TTV trend and refining the inferred orbital period derivative.

\subsubsection{HAT-P-18~b}\label{sec:hat-p-18b}

HAT-P-18~b is a Neptune-like planet with a mass of $0.2\,M_\mathrm{J}$ and a radius of $0.995\,R_\mathrm{J}$. For HAT-P-18~b, the timing dataset consists of both ground-based transit times collected from the literature and the transit times measured in this work. This target is a candidate for both the quadratic model and sine models.

For the quadratic model, the full dataset yields $\Delta \mathrm{BIC} = 14.74$ and $\dot{P}/\sigma_{\dot{P}} = 4.17$, while excluding the ground-based transit times increases these to $\Delta \mathrm{BIC} = 27.55$ and $\dot{P}/\sigma_{\dot{P}} = 5.46$, with corresponding period derivatives of $\dot{P} = 81.4 \pm 19.5\ \mathrm{ms\ yr^{-1}}$ and $\dot{P} = 168.5 \pm 30.9\ \mathrm{ms\ yr^{-1}}$, respectively. 
Although both datasets satisfy the adopted statistical criteria for a long-term period variation, the inferred period derivatives differ significantly between the two datasets. As shown in  Figure~\ref{fig:pdot TTV diagram HAT-P-18b}, the quadratic model does not provide a satisfactory description of the measured transit times, with significant residual discrepancies remaining among both the ground-based and HST observations. 
Therefore, despite being favored by the BIC, the quadratic model does not provide a robust description of the observed timing variations.

In contrast, the sine model provides a significantly better description of the full timing dataset, yielding an amplitude of $A_{\mathrm{TTV}}=92.9$~s with a detection significance of $A/\sigma_A=6.77$. 
The corresponding improvement in model comparison is $\Delta\mathrm{BIC}=71.05$, substantially exceeding that obtained for the quadratic model. 
As shown in Figure~\ref{fig:sine TTV diagram HAT-P-18b}, the sine model provides a good representation of both our newly measured transit times and the archival ground-based measurements.
However, when the ground-based transit times are excluded, the sinusoidal model no longer satisfies our adopted detection criteria ($\Delta\mathrm{BIC}>10$ and $A/\sigma_A>3$). 
This difference likely arises from the limited phase coverage of our new observations, which sample only a restricted portion of the inferred TTV cycle. 
Therefore, the new timings alone cannot robustly constrain the amplitude, period, and phase of the sinusoidal variation. 
The inclusion of the ground-based measurements extends the temporal baseline and provides the additional phase coverage required to identify the periodic TTV signal.

To further test this periodic interpretation, we computed a generalized Lomb--Scargle (GLS) periodogram \citep{zechmeister_generalised_2009}. As shown in Figure~\ref{fig:sine periodogram}, the strongest peak occurs at $P_{\mathrm{TTV}} = 142.1$ epochs and exceeds the $0.1\%$ false-alarm probability threshold. Moreover, this period agrees closely with the best-fitting sinusoidal solution ($P_{\mathrm{TTV}} = 142.1$ epochs), providing additional support for a coherent periodic component in the timing residuals.

\begin{figure*}[!h]
    \centering
    \includegraphics[width=0.48\textwidth]{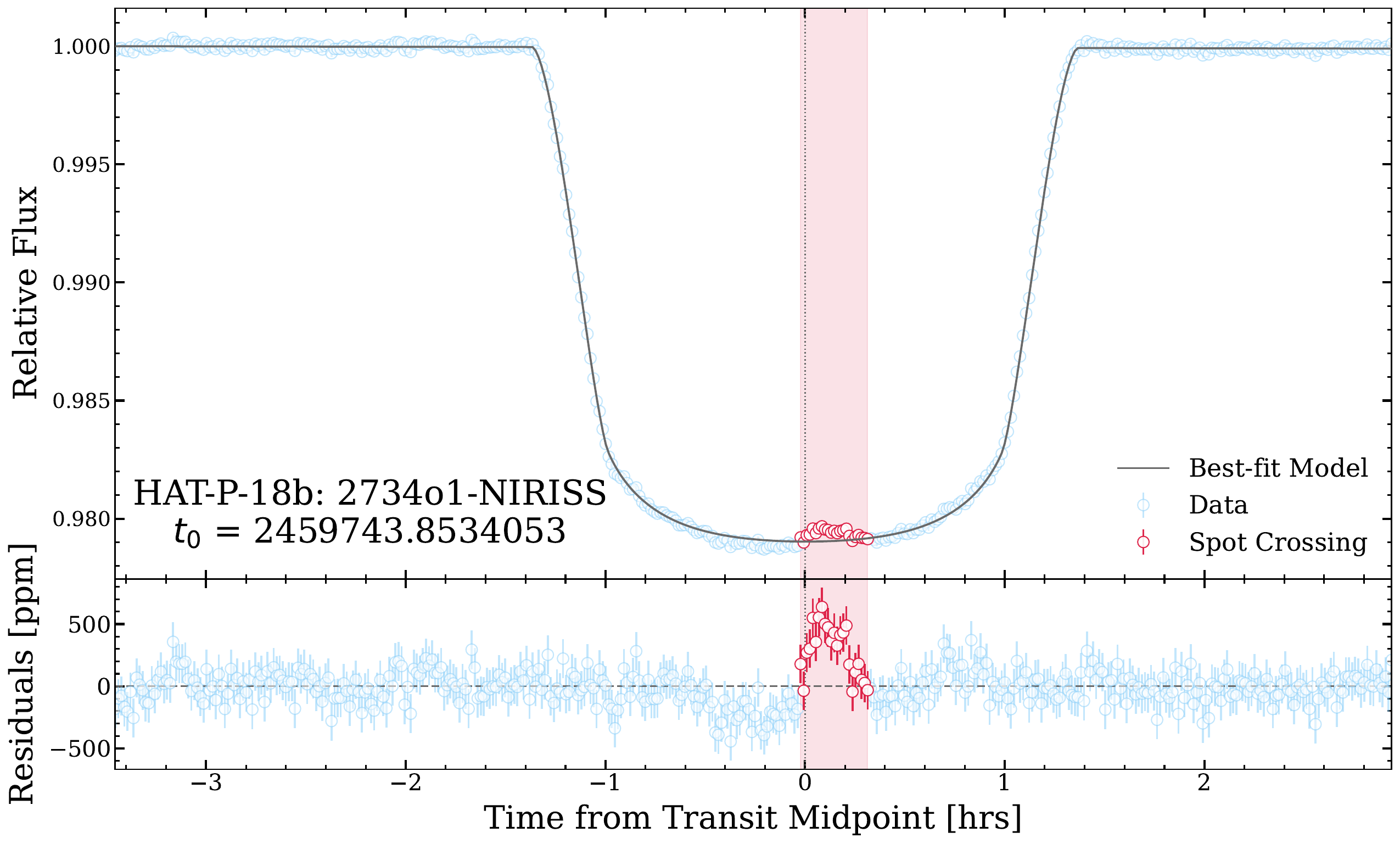}
    \includegraphics[width=0.48\textwidth]{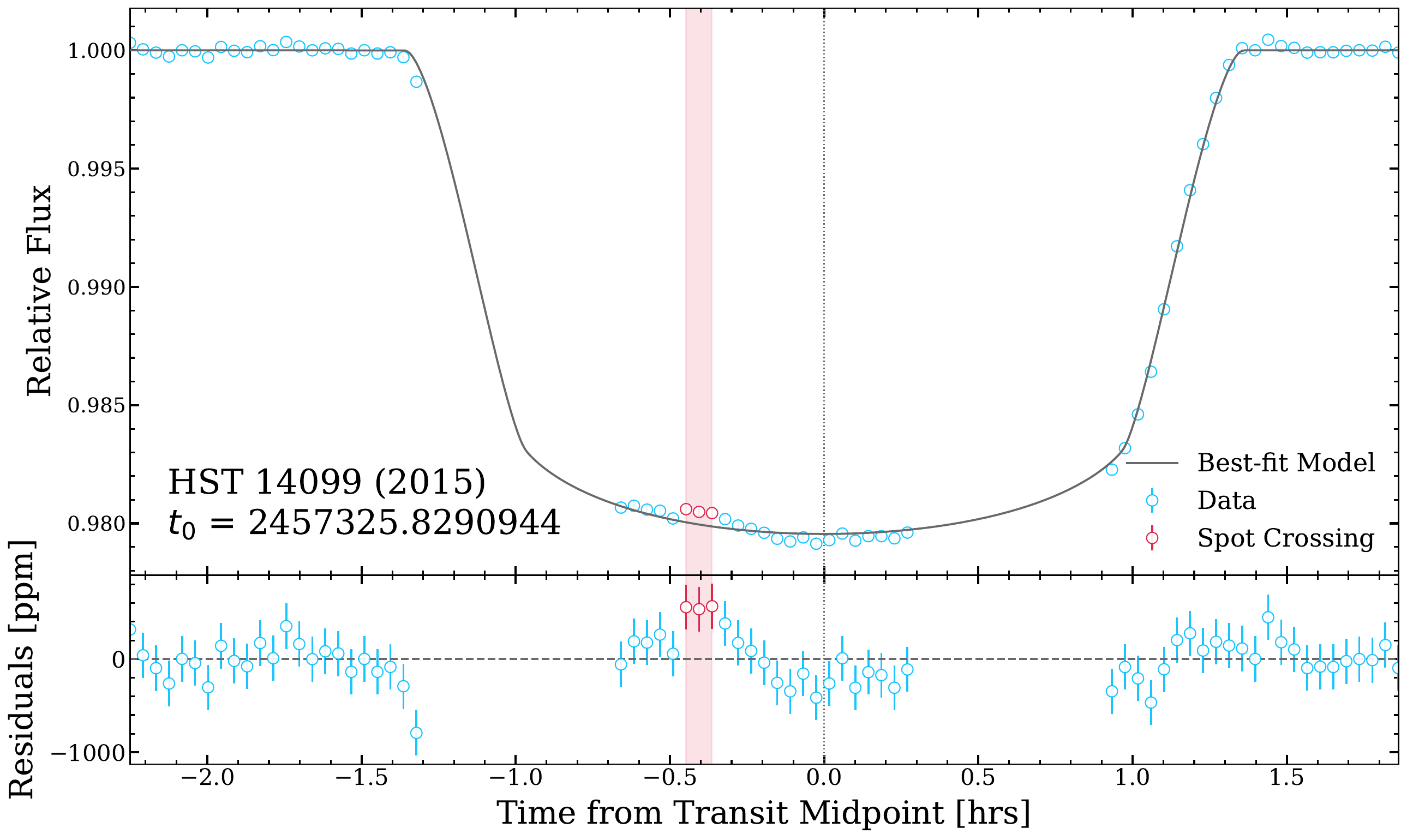}
    \includegraphics[width=0.48\textwidth]{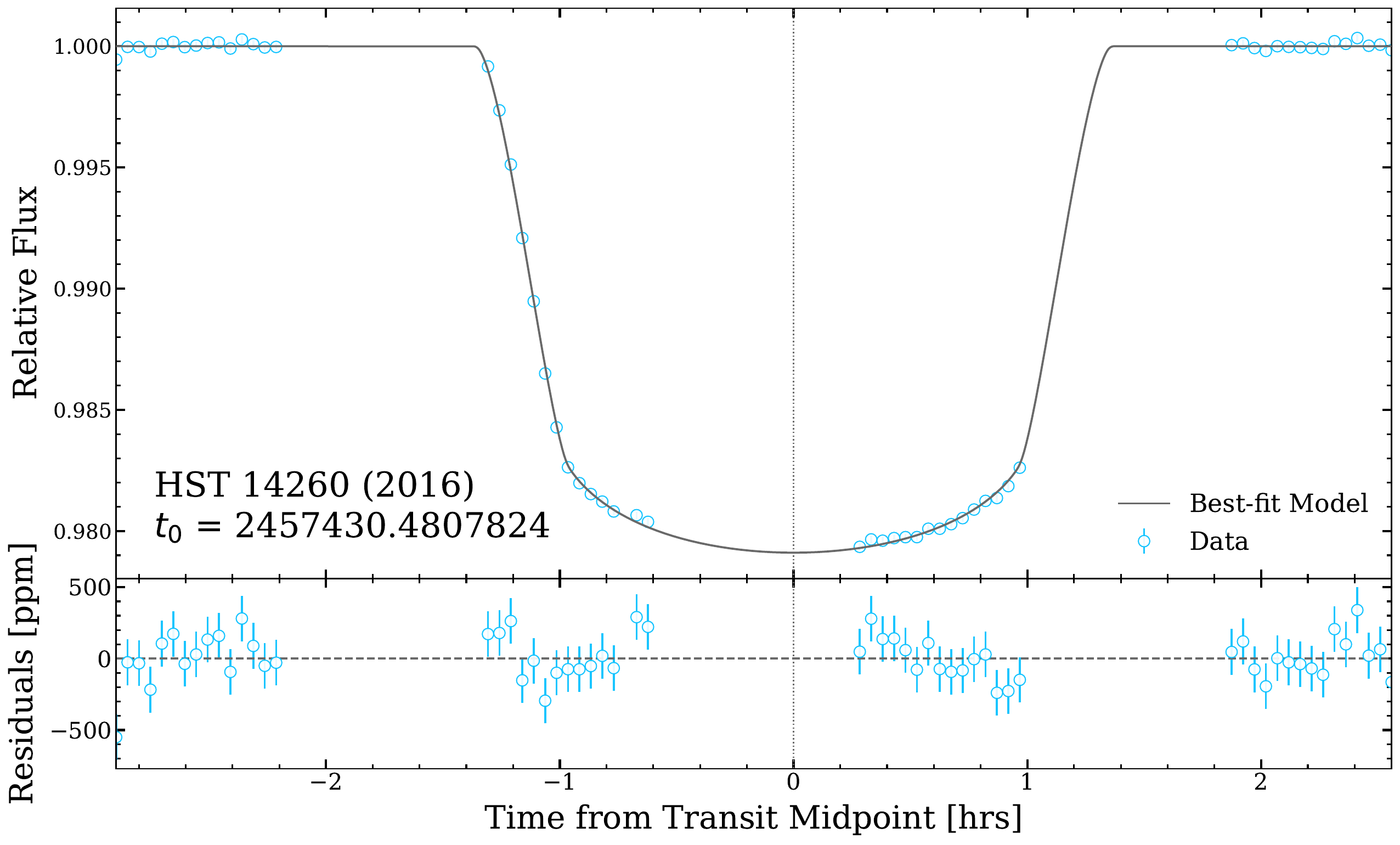}
    \includegraphics[width=0.48\textwidth]{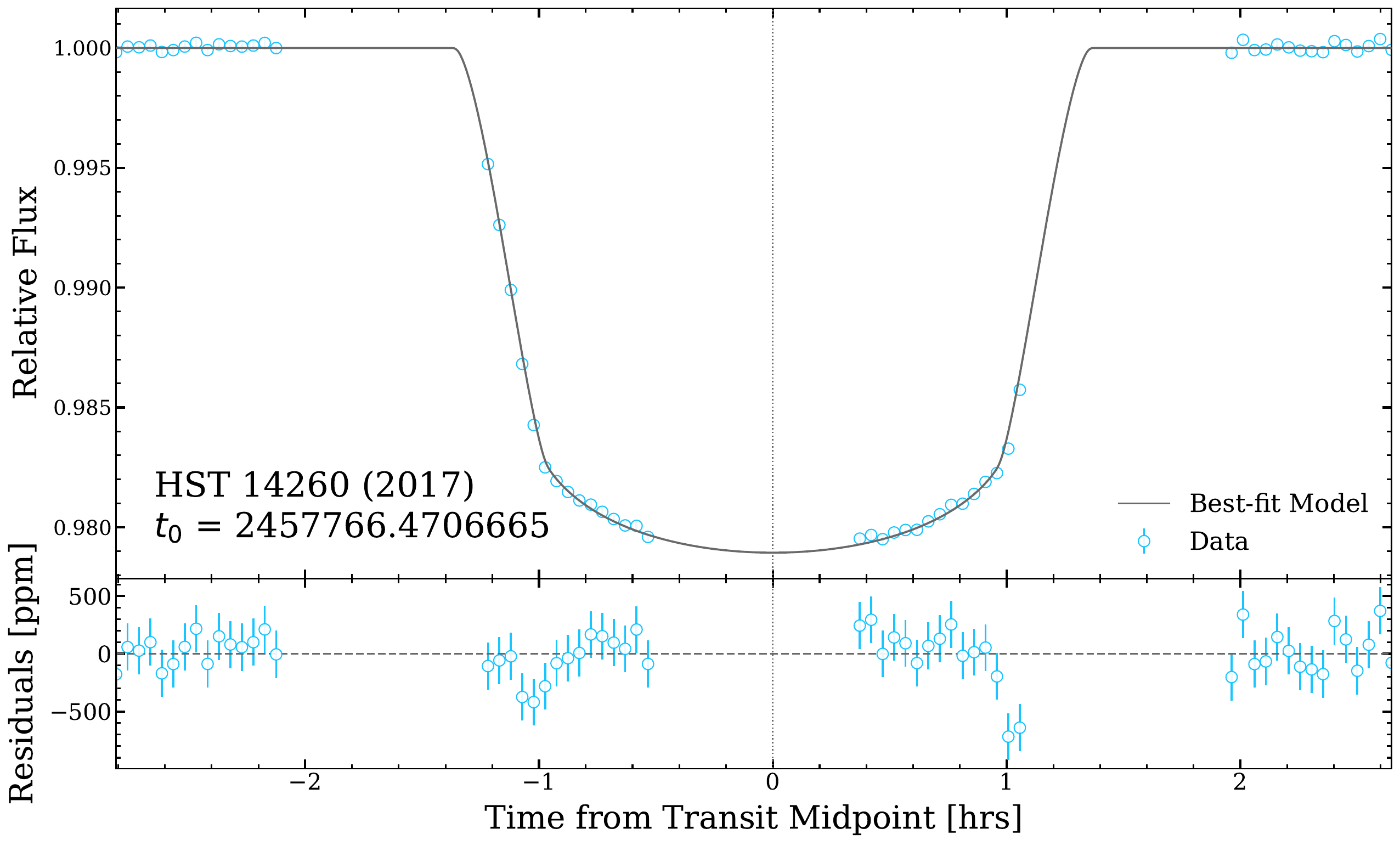}
    \caption{White light curves of HAT-P-18~b. \textit{Top left:} the JWST transit (ID: 2734o1-NIRISS, indicating proposal 2734, observation 1, observed with the JWST NIRISS instrument). \textit{Top right:} the 2015 HST/WFC3 transit (program 14099). \textit{Bottom:} the 2016 (left) and 2017 (right) HST/WFC3 transits (program 14260). In each panel, the upper subpanel shows the relative flux light curve and the lower subpanel shows the corresponding residuals; unfilled blue points represent the data and the black line indicates the best-fitting model. Red points and the shaded areas mark the spot-crossing events: the clear event near the bottom of the JWST transit, and the low-level feature near the bottom of the 2015 HST transit.}
    \label{fig:hat-p-18b spot crossing}
\end{figure*}

Finally, we investigated whether the measured transit times of HAT-P-18~b could be biased by the spot-crossing events noted in Section~\ref{sec:transit measurements} (Figure~\ref{fig:hat-p-18b spot crossing}). 
The bump in the JWST transit, first identified by \citet{fournier-tondreau_near-infrared_2024}, occurs near the transit bottom, where the derivative of the transit model with respect to the mid-transit time is close to zero. Consequently, the feature has little leverage on the inferred mid-transit time ($t_0$), which is constrained primarily by the steep ingress and egress regions.
We quantified its effect using two methods.
First, we estimated the expected timing shift to first order directly from the best-fit solution, without refitting. Because the uncertainty-weighted residuals are orthogonal to the model derivative at the best fit, masking a set of data points produces a shift in $t_0$ of 
\begin{equation}
\Delta t_0 \simeq -\frac{\sum_{i\in{\rm bump}} r_i g_i/\sigma_i^2}{\sum_{\rm all} g_i^2/\sigma_i^2},
\end{equation}
where $r_i$ and $\sigma_i$ are the fit residual and its uncertainty, respectively, and $g_i \equiv \partial F/\partial t_0$ is the derivative of the best-fit model with respect to the mid-transit time. Applying this estimate to the integrations covering the event yields a timing shift of only $0.07$~s.
Second, as a direct check, we refitted the light curve after masking the 23 integrations covering the event. The resulting mid-transit time changes by only $0.05$~s, corresponding to $0.03\sigma$ of the $1.7$~s timing uncertainty and consistent with the first-order estimate. 
For the HST transits, \citet{tsiaras_population_2018} excluded the 2015 visit (program 14099) from their sample because of a possible spot occultation. In our reduction of this visit, the only spot-like signature is a low-level bump spanning three consecutive points near the transit bottom, rather than near ingress or egress. Applying the same first-order estimate to these points yields an expected timing shift of only $0.6$~s, corresponding to $0.08\sigma$ of the $8.2$~s timing uncertainty. Even when the broader low-level in-transit feature is conservatively treated as spot contamination, the estimated shift remains below $1.3$~s ($0.15\sigma$). The other two HST transits show no comparable features. These spot-induced timing shifts are approximately two orders of magnitude smaller than the fitted TTV amplitude and therefore cannot plausibly account for the observed timing variations of HAT-P-18~b.

Overall, the quadratic model is disfavored because it fails to reproduce the observed O--C diagram, while the sinusoidal model satisfies the detection criteria with the full dataset and is corroborated by the GLS periodogram. We note that the ground-based timings on which this detection relies are taken from a single literature study and have a median per-point uncertainty ($\sim$73~s) that is smaller than the fitted amplitude ($92.9$~s); as shown in Figure~\ref{fig:sine TTV diagram HAT-P-18b}, they follow the full-cycle sinusoidal variation. This is in contrast to WASP-19~b (Section \ref{sec:WASP-19b}), whose fitted amplitude is only about one third of the median per-point uncertainty of its heterogeneous ground-based timings. 
We therefore classify HAT-P-18~b as a good sinusoidal TTV candidate.

\subsubsection{WASP-19~b\label{sec:WASP-19b}}

WASP-19~b is a hot Jupiter with a mass of $1.154\,M_\mathrm{J}$ and a radius of $1.415\,R_\mathrm{J}$. For WASP-19~b, the timing dataset similarly includes ground-based and space-based literature transit times, as well as our own measurements. This system meets our selection criteria only for the sinusoidal TTV model.

When all data are considered, the sinusoidal model yields $\Delta \mathrm{BIC} = 62.16$ and $A/\sigma_{A} = 3.77$. Excluding ground-based measurements, the reduced temporal coverage and lower data density prevent the solution from satisfying our detection thresholds  (Figures~\ref{fig:sine candidates2} and \ref{fig:sine candidates3}). As illustrated in Figure~\ref{fig:sine TTV diagram1}, the sinusoidal model reproduces the general trend in the O--C diagram, though the fit is not ideal. The fitted amplitude, $A_{\mathrm{TTV}} = 12.9$~s, is substantially smaller than that of HAT-P-18~b ($A_{\mathrm{TTV}} = 92.9$~s), indicating a much weaker timing signal.

Moreover, the fitted amplitude of the sinusoidal variation is only $\sim$30\% of the median uncertainty of the ground-based transit times ($\sim$43~s), which are compiled from multiple literature sources. 
The detection therefore rests on statistically stacking a heterogeneous dataset, whose source-dependent systematics and possibly underestimated uncertainties can generate a low-amplitude periodic signal \citep[][]{adams_doomed_2024, jackson_metrics_2023}.

To further investigate the periodic nature, the periodogram in Figure~\ref{fig:sine periodogram} exhibits multiple peaks of comparable significance, indicating a complex TTV pattern. One peak at $P_{\mathrm{TTV}} = 380.3$ epochs aligns closely with the best-fitting period of $P_{\mathrm{TTV}} = 385.237$ epochs, lending tentative support to the sinusoidal model. However, since this peak is not the highest and the periodogram is densely populated with numerous significant frequencies, it is unlikely that a true, coherent periodic TTV signal is present.

In summary, these results indicate that the sinusoidal model preference for WASP-19~b is not robust. 
The apparent signal is primarily driven by the inclusion of heterogeneous ground-based timings, has an amplitude below the typical timing uncertainty of individual measurements, and lacks a unique periodicity in the periodogram. 
We therefore do not classify WASP-19~b as a reliable candidate for a sinusoidal TTV signal.

\subsubsection{HAT-P-26~b}

HAT-P-26~b is a low-density Neptune-mass planet ($M_p = 0.07\,M_\mathrm{J}$, $R_p = 0.63\,R_\mathrm{J}$, $P = 4.2345$~days) orbiting a K1-type dwarf star ($M_\star = 0.82\,M_\odot$, $T_\mathrm{eff} \approx 5079$~K; \citealt{hartman_hat-p-26b_2011}). The system has been the subject of multiple TTV investigations, making it one of the few Neptune-like planets with a suspected periodic timing signal in the literature.

For HAT-P-26~b, the timing dataset includes ground-based transit times from the literature, space-based literature transit times (HST and Spitzer), TESS data, and the transit times measured in this work. This system satisfies our selection criteria only for the sinusoidal TTV model. Including all data, the sinusoidal model yields $\Delta \mathrm{BIC} = 74.41$ and $A/\sigma_{A} = 6.83$, with a fitted amplitude of $A_{\mathrm{TTV}} = 70.6$~s ($\sim$1.18~min). When ground-based transit times are excluded, the corresponding values increase to $\Delta \mathrm{BIC} = 85.96$ and $A/\sigma_{A} = 8.84$, with an amplitude of $A_{\mathrm{TTV}} = 88.8$~s ($\sim$1.48~min), indicating that the sinusoidal solution is favored under both data selections.

Previous studies have also reported possible TTV signals in this system. \citet{essen_indications_2019} analyzed 11 ground-based transits and found a sinusoidal TTV signal with an amplitude of $\sim$4~min and a period of $\sim$270~epochs, although they cautioned that aliasing effects from poor sampling could not be ruled out. More recently, \citet{a-thano_revisiting_2023} revisited the timing analysis using 33 transits spanning seven years and reported a TTV amplitude of $1.98 \pm 0.05$~min with a period of $\sim$222~epochs, attributing the signal to a potential additional planet of $\sim$0.02~$M_\mathrm{J}$ ($\sim$6.36~$M_\oplus$) near the 1:2 mean-motion resonance. The discrepancy in the reported TTV periods between these studies highlights the sensitivity of the HAT-P-26~b signal to the adopted dataset and analysis method.

Despite the strong statistical preference for the sinusoidal model in our analysis, Figures~\ref{fig:sine TTV diagram1} and \ref{fig:sine TTV diagram2} show that the sinusoidal model does not perfectly reproduce the observed O--C variations, suggesting that a more complex model may be required. More critically, the generalized Lomb--Scargle periodogram presented in Figure~\ref{fig:sine periodogram} shows no clearly significant peak, regardless of whether ground-based data are included. This lack of periodogram support stands in contrast to the strong $\Delta \mathrm{BIC}$ improvement, and implies that the periodic interpretation should be treated with considerable caution. The fact that the signal remains significant when only space-based data are retained does provide some reassurance against ground-based systematic effects, but the absence of a clean periodogram peak suggests that the true TTV pattern may be more complex than a simple sinusoid, or that the signal could be influenced by residual systematics in the combined dataset.

Overall, HAT-P-26~b exhibits a statistically significant improvement relative to a constant-period model, consistent with previous indications of TTVs in the literature. However, the lack of periodogram support, the discrepant TTV periods reported by different studies, and the imperfect sinusoidal fit to the O--C variations all indicate that it is not a reliable candidate for a periodic TTV signal.

\subsubsection{HD~189733~b}

HD~189733~b is a well-studied hot Jupiter ($M_p = 1.166\,M_\mathrm{J}$, $R_p = 1.119\,R_\mathrm{J}$, $P = 2.2186$~days) transiting a nearby K-type dwarf star ($M_\star = 0.828\,M_\odot$; \citealt{kokori_exoclock_2023,addison_minerva-australis_2019,rosenthal_california_2021}). It was among the first exoplanets to have its atmosphere detected and remains one of the most extensively characterized exoplanetary systems, serving as a benchmark for atmospheric studies. Despite its prominence in the exoplanet literature, robust evidence for TTVs in this system has not been previously established.

\begin{figure*}[!h]
    \centering
    \includegraphics[width=0.65\textwidth,height=0.33\textheight]{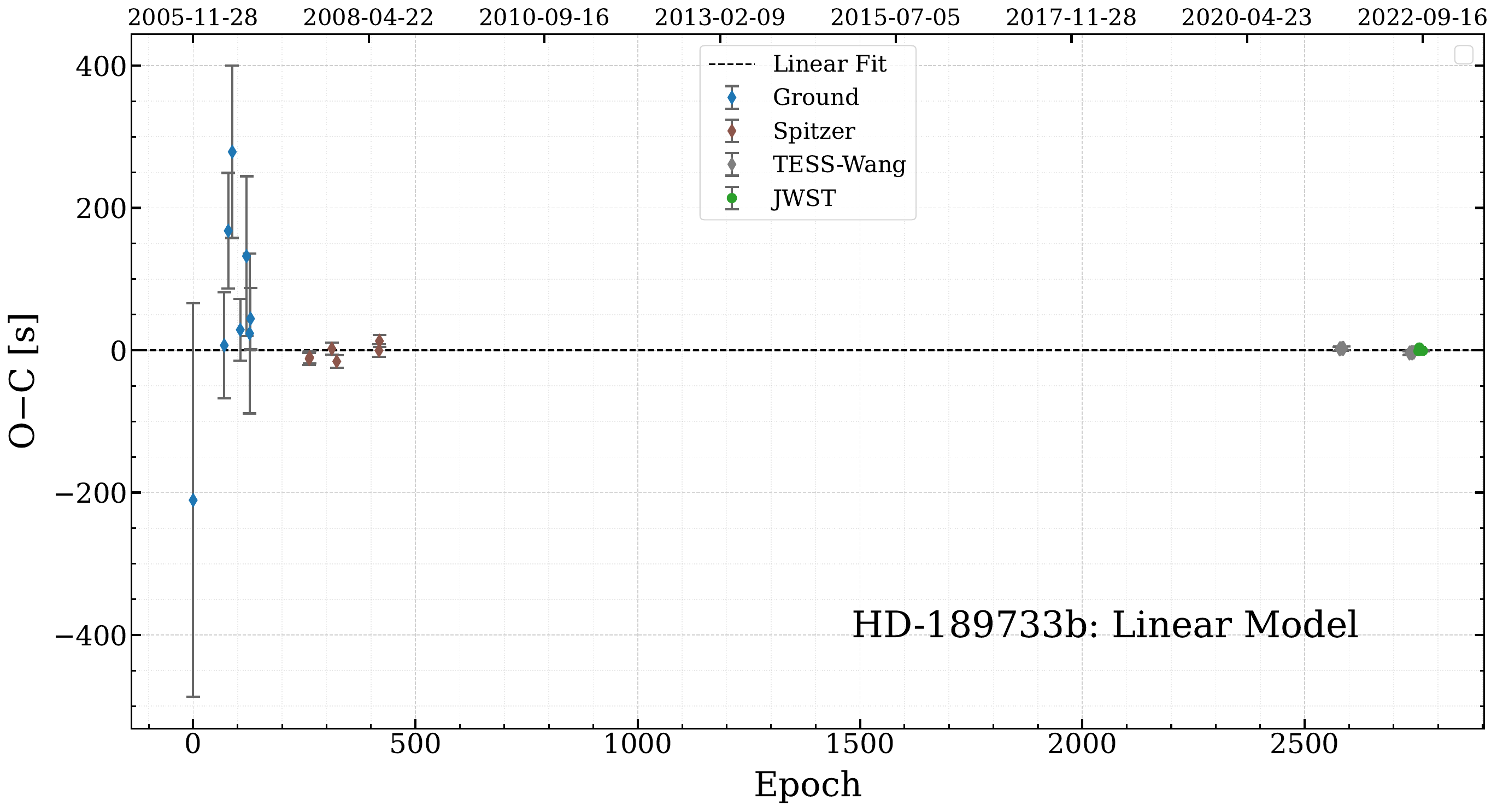}
    \caption{O--C diagram for HD~189733~b using the full dataset. The gray dashed line marks the best-fit linear ephemeris, and data points are color-coded by source.}
    \label{fig:HD 189733 all data points}
\end{figure*}

For HD~189733~b, the timing dataset combines ground-based and space-based (including Spitzer) transit times from the literature, along with the transit times measured in this work. As shown in Figure~\ref{fig:HD 189733 all data points}, a linear fit to the complete dataset produces an O--C diagram that is adequately described by a constant-period model, with no significant deviation from a linear ephemeris.

However, when the ground-based transit times are excluded, the sinusoidal model becomes statistically preferred, yielding $\Delta \mathrm{BIC} = 24$ and $A/\sigma_{A} = 6.01$. As illustrated in Figure~\ref{fig:sine TTV diagram2}, this model reproduces the observed O--C variations reasonably well, particularly in the phase-folded representation. The fitted amplitude, though, is only $A_{\mathrm{TTV}} = 2.3$~s, which is substantially smaller than those obtained for HAT-P-18~b ($A_{\mathrm{TTV}} = 92.9$~s) and WASP-19~b ($A_{\mathrm{TTV}} = 12.9$~s).

Several considerations argue against interpreting this apparent sinusoidal variation as a robust TTV detection. 
First, the inferred amplitude of $A_{\mathrm{TTV}}=2.3$~s is smaller than the typical timing uncertainties of the Spitzer and TESS measurements ($\sim$10~s and $\sim$3~s, respectively). It is also comparable to the level of possible inter-instrument timing offsets and systematic uncertainties among heterogeneous datasets, making a systematic origin difficult to exclude.
Second, the detected periodic signal is driven primarily by the TESS-Wang transit timing data, while the JWST and Spitzer measurements do not independently support the periodic TTV signal. This dependence on a specific subset of the data raises the possibility that the signal arises from subtle systematic offsets in different TESS sectors rather than from a coherent dynamical perturbation.
Finally, the GLS periodogram shown in Figure~\ref{fig:sine periodogram} does not reveal a significant peak at the best-fitting sinusoidal period, further undermining the sinusoidal interpretation. 
We therefore do not classify HD~189733~b as a candidate for a sinusoidal TTV signal.

\subsubsection{WASP-121~b}

WASP-121~b is an ultrahot Jupiter ($M_p = 1.157 \pm 0.070\,M_\mathrm{J}$, $R_p = 1.753 \pm 0.036\,R_\mathrm{J}$, $P = 1.2749$~days) orbiting an F6V-type main-sequence star ($M_\star = 1.358^{+0.075}_{-0.084}\,M_\odot$, $R_\star = 1.458 \pm 0.030\,R_\odot$, $T_\mathrm{eff} = 6460 \pm 140$~K; \citealt{Delrez2016, bourrier_hot_2020}). 

Recently, \citet{kutluay_chasing_2026} reported a positive period derivative of $\dot{P} = 15.1 \pm 0.8\ \mathrm{ms\ yr^{-1}}$ for WASP-121~b, which, if confirmed, would represent the detection of orbital expansion in an exoplanetary system. However, this measurement relies on a heterogeneous mixture of ground-based and space-based photometry, spanning observatories with widely varying photometric precision and systematic noise properties.

\begin{figure*}[!h]
    \centering
    \includegraphics[width=0.48\textwidth,height=0.22\textheight]{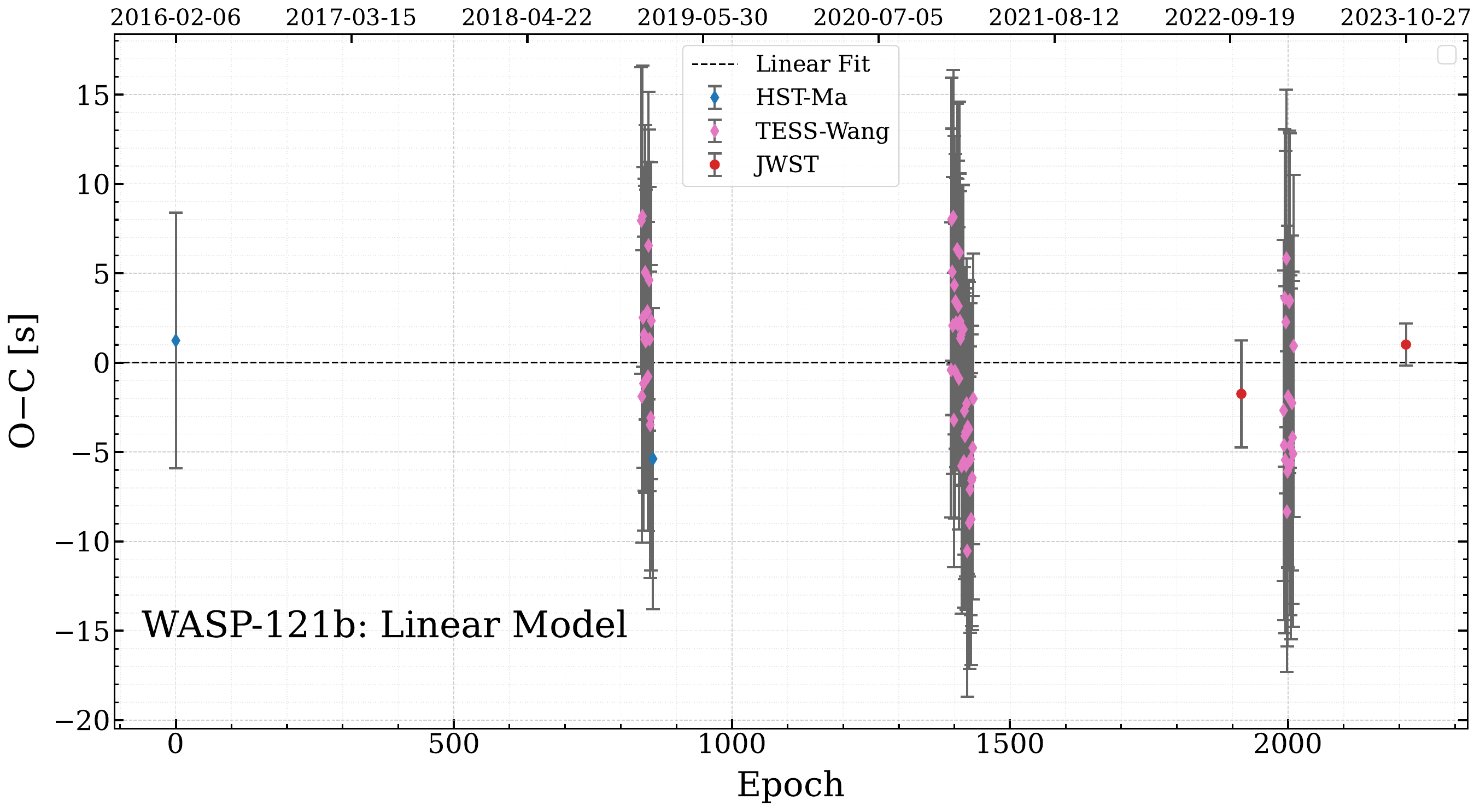}
    \caption{O--C diagram for WASP-121~b using our own data. The gray dashed line marks the best-fit linear ephemeris, and data points are color-coded by source.}
    \label{fig:WASP-121b our own data}
\end{figure*}

To independently assess this claim, we performed our own TTV analysis of WASP-121~b using exclusively our homogeneous space-based data from HST, TESS, and JWST. In contrast to the \citet{kutluay_chasing_2026} result, our analysis yields no significant detection of orbital period evolution. As shown in the O--C diagram (Figure~\ref{fig:WASP-121b our own data}), the timing residuals exhibit no coherent long-term trend. A comparison of the Bayesian Information Criterion clearly favors a constant-period ephemeris, with the linear model ($\mathrm{BIC} = 31.86$) outperforming both the quadratic model ($\mathrm{BIC} = 34.64$) and the sinusoidal model ($\mathrm{BIC} = 54.04$).

The discrepancy between our findings and those of \citet{kutluay_chasing_2026} likely stems from the heterogeneous nature of their dataset. Combining ground-based observations, which exhibit diverse systematic behaviors and potentially underestimated timing uncertainties, with space-based data can introduce spurious signals that mimic orbital evolution \citep[see also the discussion of WASP-121~b in][]{ma_exoplanet_2025}.

The WASP-121~b case serves as a powerful cautionary tale that complements the HATS-6~b detection. While HATS-6~b demonstrates what can be achieved with homogeneous, high-precision space-based photometry, WASP-121~b illustrates how heterogeneous data can lead to conclusions that may not survive scrutiny under more controlled analysis conditions. Together, these two systems underscore a central theme of this work: the reliability of TTV detections depends critically on the quality and homogeneity of the underlying timing data.

\section{Discussion\label{sec:discussion}}

Among the 37 targets in our sample, we identified two potential TTV candidates: HATS-6~b exhibiting a long-term TTV, and HAT-P-18~b showing a short-term periodic TTV. In this section, we discuss the possible physical origins of the timing variations observed in these two systems.

\subsection{Long-Term TTV}\label{subsec:long_TTV}
A possible cause for the long-term TTV signal observed in HATS-6~b is orbital expansion. 
If confirmed, this system would provide a valuable opportunity to investigate this phenomenon, which has received considerably less attention than orbital decay. 

HATS-6 consists of a hot Jupiter orbiting an early M dwarf ($M_\star = 0.57\,M_\odot$; \citealt{hartman_hats-6b_2015}), making it an interesting target for investigating tidal interaction in low-mass stellar systems. 
Orbital expansion is expected when the stellar rotation period is shorter than the planetary orbital period, i.e. $P_\mathrm{rot} < P_\mathrm{orb}$ \citep{levrard_falling_2009,matsumura_tidal_2010}.
In this regime, the tidal bulge raised on the star leads the planet and transfers angular momentum from the stellar spin to the orbit, producing an outward tidal torque \citep{hut_tidal_1981}.
Although no reliable measurement of $P_\mathrm{rot}$ is currently available for HATS-6, M dwarfs are known to retain relatively rapid rotation over several Gyrs due to inefficient angular momentum loss \citep{newton_rotation_2016}. 
Therefore, a rapidly rotating host star with $P_\mathrm{rot} < P_\mathrm{orb}$, as required for tidal orbital expansion, remains a physically plausible configuration for HATS-6.

However, we cannot exclude the possibility that the observed long-term TTV trend is a portion of a periodic variation with a timescale significantly longer than the current observational baseline. 
Thus, in addition to tidal orbital expansion, other mechanisms, including apsidal precession, the Applegate mechanism, and the Rømer effect, may also produce comparable long-term timing variations. 
Continued transit timing observations over an extended baseline will therefore be crucial for distinguishing between a genuine secular orbital evolution and alternative periodic processes, and for ultimately identifying the physical origin of the HATS-6~b TTV signal.

\subsection{Short-Term Periodic TTV}\label{subsec:short_periodic}
One possible explanation for the short-period TTV signals detected in HAT-P-18~b is gravitational perturbation from an additional, as yet unseen, planet. To explore this possibility, we estimated upper mass limits for hypothetical companions using the RMS method \citep{wang_transiting_2018, zhang_constraining_2024}. The method compares the RMS of simulated TTV signals with that of the observed timing residuals while requiring the planetary system to remain dynamically stable. Dynamical stability was evaluated using the Mean Exponential Growth factor of Nearby Orbits (MEGNO; \citealt{cincotta_phase_2003}), where $\langle Y \rangle \rightarrow 2$ corresponds to regular motion. The MEGNO analysis is intended to identify dynamically stable regions in which an additional companion could reside. We emphasize that dynamical stability alone does not imply that a given configuration can reproduce the observed TTV signal. The N-body integrations were carried out with \texttt{REBOUND} \citep{rein_rebound_2012}, assuming coplanar and circular orbits for both planets.

\begin{figure*}[!h]
    \centering
    \includegraphics[width=0.75\textwidth,height=0.32\textheight]{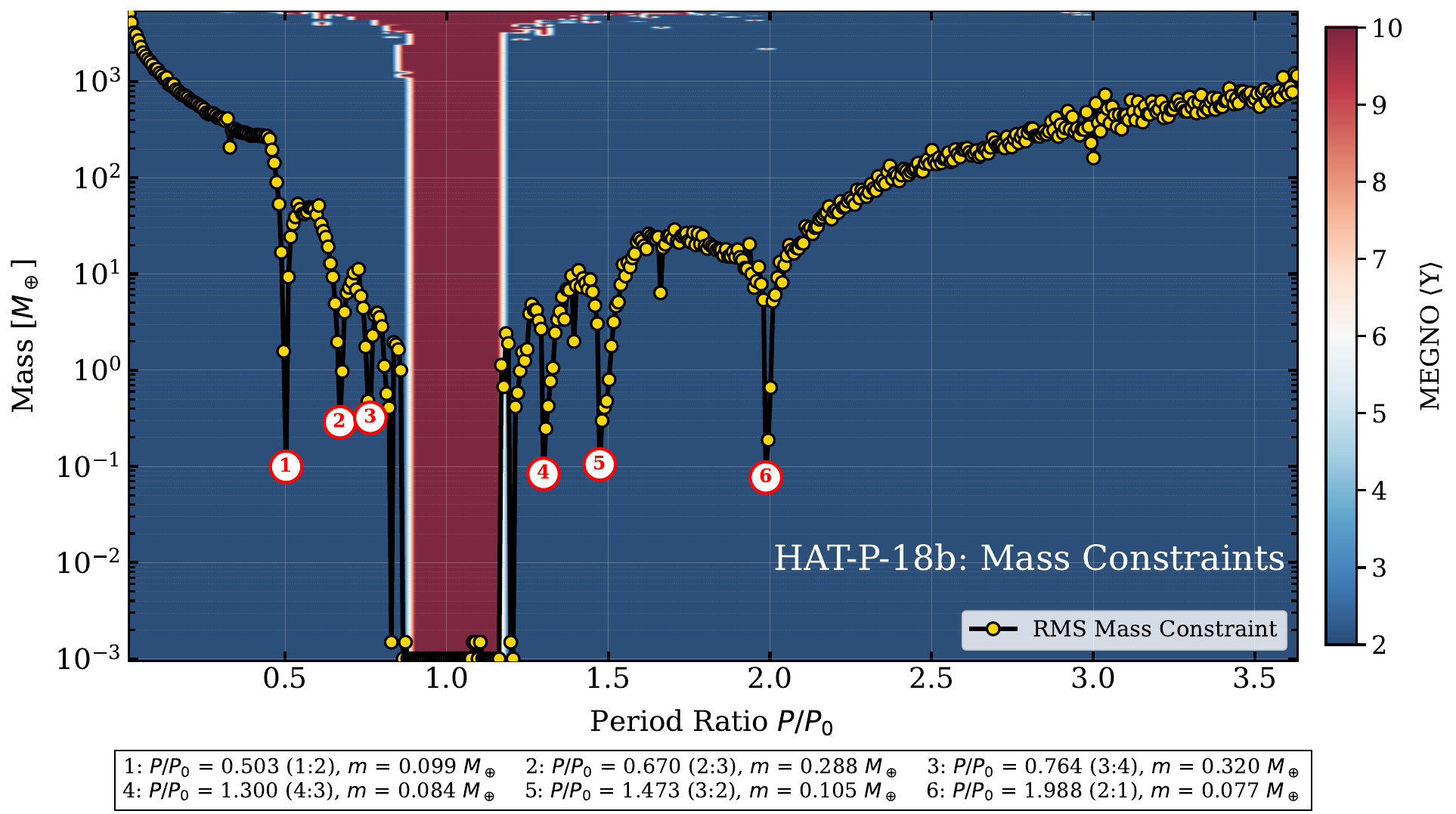}
    \caption{Mass constraints from the RMS method for a potential additional planet in the HAT-P-18~b system. The yellow line traces the RMS upper mass limit. Numbered points mark local minima (valleys) in this constraint where the period ratio $P/P_0$ of the perturber to the known transiting planet lies near a low-order ($\leq$5) mean-motion resonance (MMR); corresponding mass values are listed in the legend. The background color shows a dynamical stability map based on the MEGNO chaos indicator, with the scale provided by the color bar on the right.}
    \label{fig:rms mass constraint}
\end{figure*}

For HAT-P-18~b, the resulting constraints are shown in the top panel of Figure~\ref{fig:rms mass constraint}. Near the 1:2, 2:3, 3:4, 4:3, 3:2, and 2:1 mean-motion resonances, the companion mass is constrained to approximately $10^{-1}\,M_{\oplus}$. The region permitted by the RMS analysis lies entirely within the stable parameter space identified by MEGNO, indicating that the RMS limits are more restrictive than the stability criterion alone.

These constraints should be regarded as approximate. Determining a unique companion configuration capable of reproducing the observed TTV amplitudes and periods, while simultaneously satisfying radial-velocity constraints and long-term dynamical stability, is beyond the scope of the present work. We therefore restrict our analysis to upper mass limits derived from the RMS method. A more comprehensive dynamical investigation will be presented in a future study.

\section{Conclusion\label{sec:conclusion}}

This work extends the ExoEcho project \citep{wu_evidence_2026, wang_long-term_2024, zhang_constraining_2024, wang_exoplanet_2025, ma_exoplanet_2025} by incorporating JWST data and analyzing 84 transits across 37 exoplanets to derive precise transit timings. 
Using a linear model to update ephemerides, a quadratic model to probe long-term orbital evolution, and a sine model to search for short-periodic signals, we identify two strong TTV candidates. 
The hot Jupiter HATS-6~b exhibits a long-term trend with $\dot{P} = 230.5 \pm 46.8\ \mathrm{ms\ yr^{-1}}$.
The Neptune-like planet HAT-P-18~b shows compelling short-periodic TTVs with amplitude of $A_{\mathrm{TTV}} = 92.9 \pm 13.7\ \mathrm{s}$.

With a sample spanning 20 hot Jupiters, 11 Neptune-like planets, and 6 super-Earths, this study demonstrates the capability of JWST for high-precision transit timing across a broad range of planetary masses. By extending TTV analyses beyond the traditional hot-Jupiter population, these observations provide a valuable foundation for future studies of planetary system architectures and long-term dynamical evolution.

\section{Acknowledgments}
The work was supported by the NSFC grants 12073092 and 12103098, the National Key R$\&$D Program of China (2024YFA1611801), the science research grants from the China Manned Space Project (No. CMS-CSST2025-A16), and the Earth 2.0 Space Mission program from SHAO for their financial support. The ET mission is funded by China's Space Origins Exploration Program. We also thank Chipeng He for helpful discussion on the spot-crossing events.

\section{Data Availability\label{sec:data availability}}
All JWST data used in this work can be found in MAST: \dataset[10.17909/rbtk-qd03]{http://dx.doi.org/10.17909/rbtk-qd03}.
The configuration files, analysis notebooks, light curves, TTV data, results, and source code associated with this work are publicly available on \dataset[Zenodo]{10.5281/zenodo.20586450}\footnote{\href{https://doi.org/10.5281/zenodo.20586449}{10.5281/zenodo.20586450}} and
\href{https://github.com/bluecloud135/TTV_JWST_paper.git}{GitHub}\footnote{\href{https://github.com/bluecloud135/TTV_JWST_paper.git}{https://github.com/bluecloud135/TTV\_JWST\_paper.git}}. This research utilized the following software packages: \texttt{Eureka!} (v0.10; \citet{bell_eureka_2022}), \texttt{exoTEDRF} (v1.4.0; \citet{feinstein_early_2023,radica_awesome_2023}), \texttt{REBOUND} (v4.4.3; \citet{rein_rebound_2012}, \texttt{dynesty} (\citet{speagle_dynesty_2020,koposov_joshspeagledynesty_2022}), \texttt{emcee} (\citet{foreman-mackey_emcee_2013}), \texttt{astropy} (\citet{astropy_collaboration_astropy_2022}), \texttt{PyAstronomy} (\citet{pya}), \texttt{pandas} (\citet{team_pandas-devpandas_2025}), \texttt{numpy}/\texttt{scipy} (\citet{van_der_walt_numpy_2011}), and \texttt{matplotlib} (\citet{hunter_matplotlib_2007}).

\clearpage
\bibliographystyle{aasjournal}
\bibliography{references}

\end{document}